\documentclass[12pt]{article}
 \usepackage{graphicx} 
\usepackage{amsmath, amssymb} 
\usepackage{fancyhdr} 
\usepackage{geometry} 
\usepackage{caption}
\usepackage{natbib}
\usepackage{accents}
\usepackage{xcolor}
\usepackage{subfig}
\usepackage{float}
\usepackage{authblk}
\usepackage{stmaryrd}
\usepackage{bm}
\usepackage{amsthm}

\newtheorem{proposition}{Proposition}

\newtheorem{definition}{Definition}
\usepackage[ruled, lined, linesnumbered, commentsnumbered, longend]{algorithm2e}
\usepackage{hyperref}
\usepackage{enumitem}

\definecolor{orange1}{RGB}{255,128,0}
\definecolor{purple2}{RGB}{102,0,204}
\definecolor{blue}{RGB}{0,0,255}
\definecolor{red}{RGB}{255,0,0}

\usepackage[english]{babel}
\usepackage{csquotes}
\graphicspath{{./plots/}} 

  \newcommand{\abr}[1]{\left\|#1\right\|}
\newcommand{\cbr}[1]{\left\{#1\right\}}
\newcommand{\rbr}[1]{\left(#1\right)}
\newcommand{\sbr}[1]{\left[#1\right]}
\newcommand{\brc}[1]{\lbrace#1\rbrace}
\newcommand{\dbr}[1]{\left \llbracket #1 \right \rrbracket}
\newcommand{\ld}{,\ldots,}

 \newcommand{\cx}{\mathcal{X}}
\newcommand{\x}{x_{n,p_1 \ldots p_L}}
\newcommand{\ximp}{x^{\text{\tiny imp}}_{n,p_1 \ldots p_L}}
\newcommand{\cy}{\mathcal{Y}}
\newcommand{\y}{y_{n,q_1 \ldots q_M}}

\newcommand{\bs}{\beta_{p_1 \ldots p_L q_1 \ldots q_M}}
\newcommand{\bint}{\beta_{0, q_1 \ldots q_M}}

\newcommand{\ca}{\mathcal{A}}
\newcommand{\cb}{\mathcal{B}}

\newcommand{\core}{\mathcal{U}}

\newcommand{\ce}{\mathcal{E}}
\newcommand{\rs}{r_{n,q_1 \ldots q_M}}
\newcommand{\trs}{\tilde{r}_{n,q_1 \ldots q_M}}

\newcommand{\cl}{\mathcal{L}}

\newcommand{\cc}{\mathcal{C}^\ell}
\newcommand{\cd}{\mathcal{D}^m}
\newcommand{\cw}{\mathcal{W}}
\newcommand{\ch}{\mathcal{H}}

\newcommand{\co}{\mathcal{O}}
\newcommand{\cz}{\mathcal{Z}}

\newcommand{\bU}{\mathbf{U}}
\newcommand{\bV}{\mathbf{V}}
\newcommand{\bUbV}{\lbrace \bU_\ell \rbrace,\lbrace \bV_m \rbrace}

\newcommand{\wcase}{w_n^\text{\tiny case}}
\newcommand{\Wcase}{\cw^\text{\tiny case}}

\newcommand{\wcell}{w_{n,q_1 \ldots q_M}^\text{\tiny cell}}
\newcommand{\Wcell}{\cw^\text{\tiny cell}}

  \newcommand{\woutl}{w_n^{x}}
\newcommand{\Woutl}{\cw^{x}}

\newcommand{\ocell}{o_{n,q_1 \ldots q_M}}

 \newcommand{\sump}{\sum_{p_1 \cdots p_L}^{P_1 \cdots P_L}}
\newcommand{\sumpl}{\sum_{p_s,s \in L^{\star}_{\ell}}^{P_s}}
\newcommand{\sumq}{\sum_{q_1 \cdots q_M}^{Q_1 \cdots Q_M}}

\newcommand{\sumk}{\sum_{k_1 \cdots k_L}^{K_1 \cdots K_L}}

 \newcommand{\sgcell}{\hat{\sigma}_{1, q_1 \ldots q_M}}

\newcommand{\tsgcell}{\tilde{\sigma}_{1, q_1 \ldots q_M}}

\newcommand{\sgcase}{\hat{\sigma}_{2}}

\newcommand{\vecop}{\operatorname{vec}} 

\newcommand{\parini}{\rbr{\lbrace \bU_\ell^{0} \rbrace, \lbrace \bV_m^{0} \rbrace, \cb^{0}_0}}
\newcommand{\parcurr}{\rbr{\lbrace \bU_\ell^{k} \rbrace, \lbrace \bV_m^{k} \rbrace, \cb^{k}_0}}
\newcommand{\parnext}{\rbr{\lbrace \bU_\ell^{k+1} \rbrace, \lbrace \bV_m^{k+1} \rbrace, \cb^{k+1}_0}}
\usepackage{xr}
\begin{document}
\def\spacingset#1{\renewcommand{\baselinestretch}
{#1}\small\normalsize} \spacingset{1}

\newcommand{\blind}{0}

\if0\blind
{
\title{\bf Casewise and Cellwise Robust\\ Tensor-on-Tensor Regression}
\author[1]{Mehdi Hirari}
\author[1]{Fabio Centofanti\thanks{Corresponding author. e-mail: \texttt{fabio.centofanti@kuleuven.be}}}
\author[1]{Mia Hubert}
\author[1]{Stefan Van Aelst}
\affil[1]{Section of Statistics and Data Science, Department 
          of Mathematics, KU Leuven, Belgium}
\setcounter{Maxaffil}{0}
\renewcommand\Affilfont{\itshape\small}
\date{October 1, 2026}         
  \maketitle
} \fi

\if1\blind
{
  \bigskip
  \bigskip
  \bigskip
  \begin{center}
    {\LARGE\bf Casewise and Cellwise Robust\\\vspace{0.3cm} Tensor-on-Tensor Regression}
\end{center}
  \medskip
} \fi

\bigskip
\begin{abstract}
Tensor-on-tensor regression is an important tool for the analysis of tensor data, aiming to predict a set of response tensors from a corresponding set of predictor tensors. However, standard tensor-on-tensor regression is sensitive to outliers, which may be present in both the response and the predictor. It can be affected by casewise outliers, which are observations that deviate from the bulk of the data, as well as by cellwise outliers, which are individual anomalous cells within the tensors. The latter are particularly common due to the typically large number of cells in tensor data. This paper introduces a novel robust tensor-on-tensor regression method, named ROTOT, that can handle both types of outliers simultaneously, and can cope with missing values as well. This method uses a single loss function to reduce the influence of both casewise and cellwise outliers in the response. The outliers in the predictor are handled using a robust Multilinear Principal Component Analysis method. Graphical diagnostic tools are also proposed to visualize the different types of detected outliers. The performance of ROTOT is evaluated through extensive simulations and further illustrated using the Labeled Faces in the Wild dataset, where ROTOT is applied to predict facial attributes.
\end{abstract}

\noindent%
\vfill

\newpage
\spacingset{1.7}

\section{Introduction}
\label{sec:intro}
In many practical applications, data arise in the form of tensors which are multiway arrays in which each mode corresponds to a particular feature or characteristic of the data objects. For instance, collections of facial images can naturally be viewed as third-order tensors where the first two modes encode pixel positions, while the third mode captures the color information for each pixel. An example of such data is provided by the Labeled Faces in the Wild data discussed in Section~\ref{sec:realdata}.

A widely used strategy for analyzing tensor data is to reshape or vectorize them so that conventional matrix-based techniques can be employed. Yet, this transformation often disrupts the inherent multi-dimensional structure and inter-mode correlations, thereby discarding higher-order dependencies that may be crucial. As a result, potentially more compact and informative representations present in the original tensor form may be lost \citep{Ye:GPCA}. In contrast, multilinear tensor methods preserve and exploit this structure, allowing tensor models to capture these dependencies more effectively and thus yield improved interpretability and accuracy \citep{Bi:Tensors}.

An essential component of multilinear data analysis is tensor regression, which aims to model relationships between multiway predictors and responses. Tensor regression includes a variety of modeling frameworks depending on the structure of the inputs and outputs. Scalar-on-tensor regression models \citep{Zhao:HOPLS, Dian:TTRhyspec} estimate a scalar response from a tensor-valued predictor, whereas tensor-on-scalar approaches \citep{Yan:StructuredTReg} use sets of scalar predictors to estimate a tensor response. A natural extension of these frameworks is tensor-on-tensor  regression, which aims to predict a tensor-valued response using one or more tensor-valued predictors \citep{lock:TOT, Gahrooei:MTOT, Liu:LowTTR, Wang:BTOT}. 

As the number of parameters in the coefficient tensor increases rapidly with its dimensionality, a low-rank structure is typically assumed to substantially reduce the number of parameters to be estimated. This reduction is commonly achieved using the CANDECOMP/PARAFAC (CP) decomposition \citep{Carroll:PARanalysis, Harshman:PARAFAC}.
Tensor-on-tensor regression has been applied across several domains, including attribute prediction from images \citep{lock:TOT} and electroencephalography prediction from functional magnetic resonance imaging \citep{Lee:RTOT, Wang:BTOT}.

As for other data types, tensor data may also be affected by the presence of outliers. Robust statistical methods tackle this problem by producing estimates that are only marginally influenced by outliers \citep{Hubert:ReviewHighBreakdown, Maronna:RobustStat}. Using these robust fits, outliers can then be detected through their departures from the fitted model.

Since the 1960s, robust statistics has primarily concentrated on casewise outliers, namely observations that deviate entirely from the bulk of the data. However, with the rise of high-dimensional settings attention has increasingly shifted toward cellwise outliers \citep{Alqallaf:cell}, where the focus is on individual cells of observations that are anomalous.
Cellwise robust procedures seek to diminish their influence while retaining the useful information contained in the remaining components of the affected observations. An effective robust method should be capable of addressing both cellwise and casewise contamination at the same time.

As classical regression methods are not robust, several approaches have been developed to ensure robustness in the presence of casewise outliers, both in the response and in the predictors \citep{Rousseeuw:LMS, Yohai:MMest, Maronna:RRMM}.
Methods have also been introduced to address cellwise outliers occurring in both the response and the predictors \citep{Ollerer:ShootingS, Filzmoser:cellMreg}.

The presence of missing values in regression settings is also a common issue encountered in practice. Various methods have been proposed to address this problem, either by discarding observations that contain missing values or by proposing an imputation scheme for the missing entries \citep{Beale:Missing, Little:ReviewMissing}. From an applied perspective, missing values can be viewed as similar to cellwise outliers, with the distinction that their positions are known. This makes the problem particularly interesting, as imputation strategies can be developed similarly to those used for handling cellwise outliers \citep{Centofanti:cellpca}.

\begin{figure}[h]
\begin{center}
\includegraphics[width=0.3\textwidth]{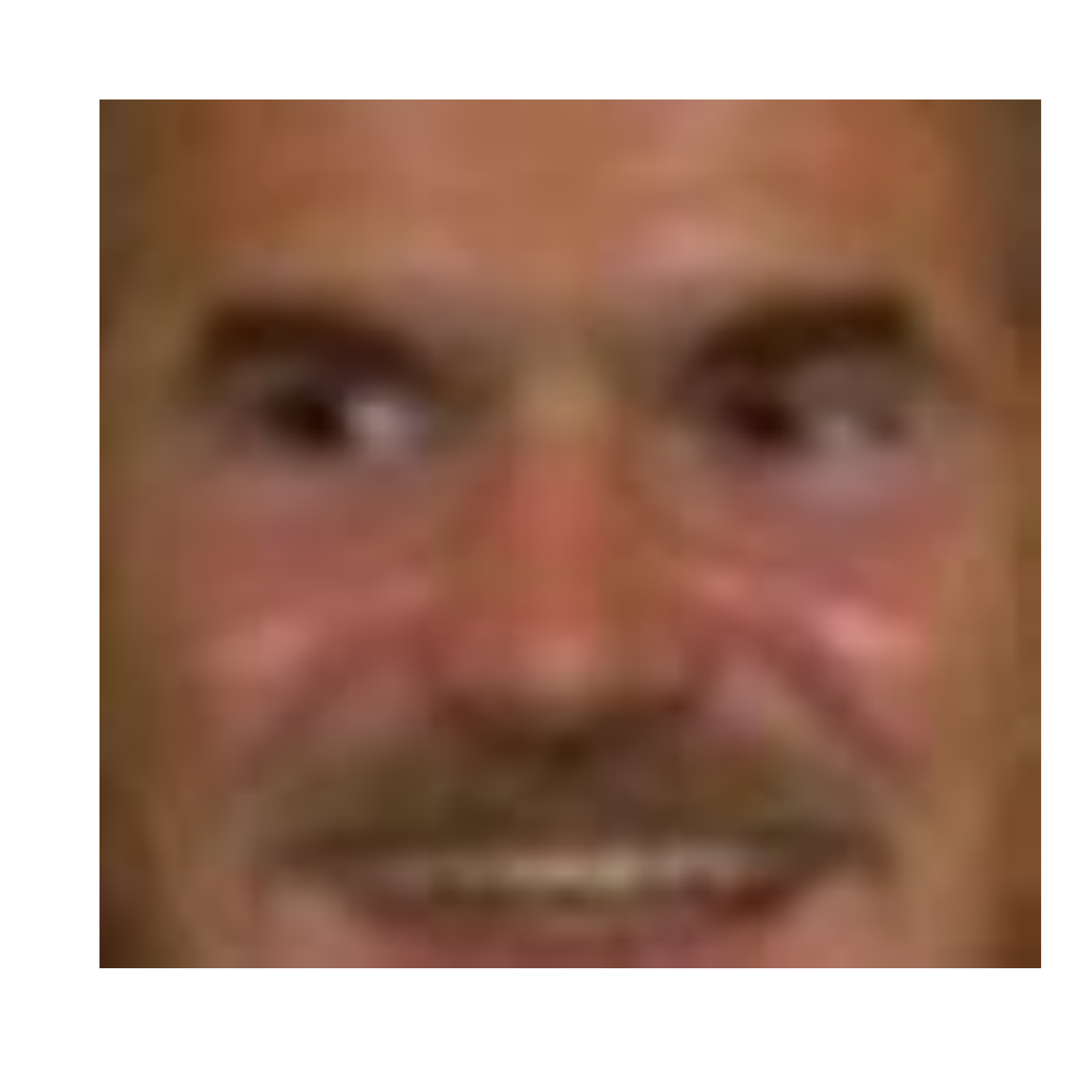}
\includegraphics[width=0.3\textwidth]{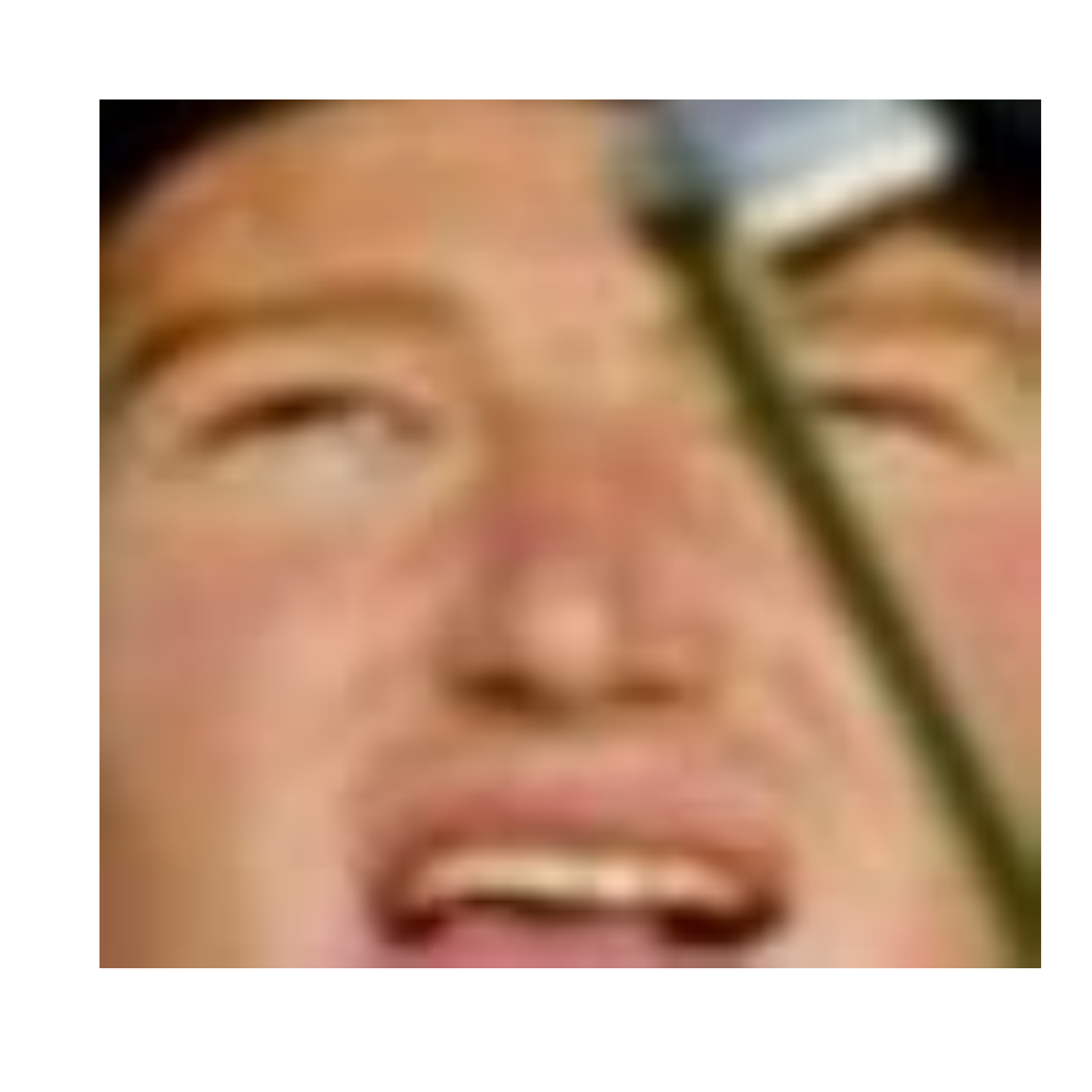}
\includegraphics[width=0.3\textwidth]{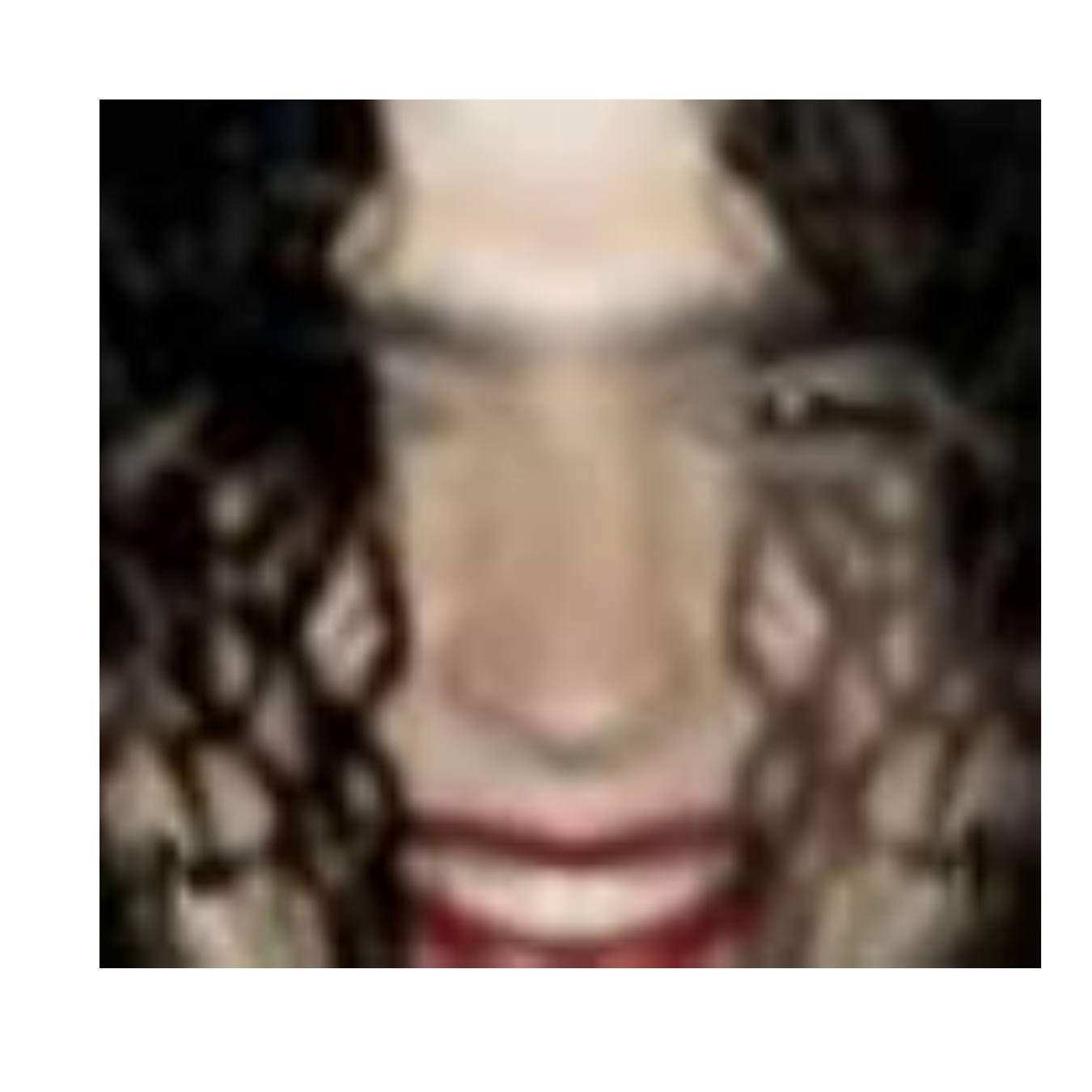}
\end{center}
\vspace{-0.5cm}
\caption{Face images without contamination (left), with cellwise contamination (middle), and a casewise outlier (right).}
\label{fig:LFW_obs_comp}
\end{figure}

Analogously, classical tensor regression is sensitive to both casewise and cellwise outliers.
As an illustration, consider the problem of estimating a collection of facial attributes from the face image of a person who exhibits various characteristics, such as different facial expressions, ethnicity, and other traits.
This analysis will be presented in Section~\ref{sec:realdata} using data from the Labeled Faces in the Wild dataset \citep{Learned:LFW}.
The attribute values for a given individual, which form the response tensor in this context, may contain unusually large values. These are examples of cellwise outliers, where only specific attributes are affected. In some cases, most of  the attribute values for an individual may be corrupted, indicating the presence of a casewise outlier. Outliers may also arise in the predictor tensor, which corresponds to the face image itself. 
Figure~\ref{fig:LFW_obs_comp} compares three face images. The left panel shows a regular image, the middle panel shows an image with cellwise outliers, and the right panel shows a casewise outlier. 

Robust methods using M-estimation have been proposed for scalar-on-tensor regression to handle casewise outliers in the response \citep{Ollila:RSOT, Konyar:RobGSTreg}.  \cite{Lee:RTOT} introduced a robust adaptation of tensor-on-tensor regression {(RTOT)} that can handle cellwise outliers in the response tensor.
{However, no existing tensor-on-tensor method jointly addresses both cellwise and casewise contamination as well as missingness in both the predictor and response tensors. 
This leaves an important methodological gap since in real-world applications contamination and missingness can occur on both sides of the regression model. To address this limitation, we propose the \textit{Robust Tensor-on-Tensor} (ROTOT) method.
Table~\ref{tab:novelty} compares the existing tensor-on-tensor regression methods in terms of missingness and types of contamination they are able to handle.}
\begin{table}[ht]
\centering
\small
\setlength{\tabcolsep}{4pt} 
\begin{tabular}{lccccc}
\hline
Method 
& Cellwise $\brc{\cy_n}$
& Casewise $\brc{\cy_n}$
& Cellwise $\brc{\cx_n}$
& Casewise $\brc{\cx_n}$
& Missing values \\
\hline
TOT \citep{lock:TOT}   & $\times$ & $\times$ & $\times$ & $\times$ & $\times$ \\
RTOT \citep{Lee:RTOT} & $\checkmark$ & $\times$ & $\times$ & $\times$ & $\times$ \\
ROTOT                  & $\checkmark$ & $\checkmark$ & $\checkmark$ & $\checkmark$ & $\checkmark$ \\
\hline
\end{tabular}
\caption{{Comparison of tensor-on-tensor regression methods in terms of the types of contamination they can handle in the predictor $\brc{\cx_n}$ and response tensors $\brc{\cy_n}$.}}
\label{tab:novelty}
\end{table}

{
The ROTOT method offers several major improvements over the existing methods.
\begin{itemize}
\item ROTOT is the first robust tensor-on-tensor regression method to minimize a single loss function that simultaneously accounts for casewise and cellwise contamination, as well as missing values, in the response tensors. Robustness is incorporated directly through bounded $\rho$-functions applied at both the cellwise and casewise levels of the tensor-on-tensor regression residuals.
\item A further novel contribution is the ability to deal with contamination in the predictor tensors. 
To this end, residuals in the loss function are computed using imputed predictor tensors to accommodate cellwise outliers and missing values, while casewise weights are used to downweight observations with outlying predictors. The imputed predictors and their corresponding casewise weights are obtained using the robust multilinear PCA method ROMPCA~\citep{Hirari:ROMPCA}.
\item A robust prediction scheme based on imputed predictor tensors is introduced, enabling predictions for new observations affected by cellwise outliers and missing values.
\item New diagnostic visualization tools are developed that distinguish between casewise and cellwise outlyingness in both the predictor and response tensors, providing insight into the type, size and location of the contamination in the data.
\end{itemize}
}
{It is important to emphasize that ROTOT is not a straightforward extension of either ROMPCA~\citep{Hirari:ROMPCA} or the robust PCA method cellPCA~\citep{Centofanti:cellpca}. ROMPCA is designed for robust tensor reconstruction, whereas tensor-on-tensor regression aims to model the relationship between predictor and response tensors. CellPCA is formulated for multivariate data and therefore does not preserve the inherent multiway structure of tensor data. Consequently, these methods cannot be directly applied to model the relationship between the response and predictor tensors while simultaneously handling contamination. ROTOT builds on ideas from ROMPCA and cellPCA, but integrates and extends them to address the distinct challenges posed by outliers in tensor-on-tensor regression.
}

Section~\ref{sec:meth} introduces basic multilinear concepts and our newly proposed ROTOT method. To solve the ROTOT minimization problem, we introduce an iteratively reweighted least squares algorithm that uses alternating least squares in each iteration step. Section~\ref{sec:simu} evaluates its performance through extensive simulations while Section~\ref{sec:outdetect} presents several diagnostics to detect and visualize outlying cells and cases. Finally, we apply our method to a real dataset in Section~\ref{sec:realdata} and Section~\ref{sec:disc} concludes the paper.

\section{Methodology}
\label{sec:meth}
\subsection{Preliminaries and Notation}
\label{sec:notations}
Denote an $L$th-order tensor $\ca = \sbr{a_{p_1 \ldots p_L}} \in \mathbb{R}^{P_1 \times \cdots \times P_L}$, where $p_\ell \in \{1,\ldots,P_\ell\}$ for $\ell = 1,\ldots,L$.
Let the sets $I_r=\{r_1,\ldots,r_{|I_r|}\}$ and $I_c=\{c_1,\ldots,c_{|I_c|}\}$ denote a partition of the modes $\cbr{1,\ldots,L}$ into row modes and column modes, with $r_1 < r_2 < \ldots < r_{|I_r|}$ and $c_1 < \ldots < c_{|I_c|}$. 
The matricization of the tensor $\ca$ with respect to the row set $I_r$ and the column set $I_c$ is defined by $\mathbf{A}_{(I_r \times I_c)} = \sbr{a_{j k}} \in \mathbb{R}^{J \times K}$, where
$J = \prod_{\ell=1}^{|I_r|} P_{r_\ell}$ and $K = \prod_{\ell=1}^{|I_c|} P_{c_\ell}$.
The entry $a_{p_1 \ldots p_L}$ maps to the $(j,k)$ element of the matrix $\mathbf{A}_{(I_r \times I_c)}$, that is, 
$a_{p_1\ldots p_L} = a_{jk}$, 
where $j = 1 + \sum_{\ell=1}^{|I_r|} \sbr{(p_{r_\ell} - 1) \prod_{\ell' = 1}^{\ell-1} P_{r_{\ell'}}}$ and $k = 1 + \sum_{m=1}^{|I_c|} \sbr{(p_{c_m} - 1) \prod_{m' = 1}^{m-1} P_{c_{m'}}}$.
Vectorization corresponds to the special case in which $I_c = \varnothing$. In this case, $\vecop(\ca)$ is a vector of length $\prod_{\ell=1}^L P_\ell$, whose $j$th entry is given by $\vecop(\ca)_j = a_{p_1 \ldots p_L}$, where the index $j$ is defined above.
The Frobenius norm of $\ca$ is defined as
\begin{equation*}
\abr{\ca}_F=\sqrt{\sum_{p_1=1}^{P_1} \sum_{p_2=1}^{P_2} \cdots \sum_{p_L=1}^{P_L} a_{p_1 \ldots p_L}^2} :=\sqrt{\sump  a_{p_1 \ldots p_L}^2}\,.
\end{equation*}
The Hadamard product of two tensors $\ca = \sbr{a_{p_1 \ldots p_L}} \in \mathbb{R}^{P_1 \times \cdots \times P_L}$ and $\cb = \sbr{b_{p_1 \ldots p_L}} \in \mathbb{R}^{P_1 \times \cdots \times P_L}$ multiplies their entries elementwise, $\ca \odot \cb = \sbr{a_{p_1 \ldots p_L} b_{p_1 \ldots p_L}}$.
For two tensors, $\ca = \sbr{a_{p_1 \ldots p_L j_1 \ldots j_K}} \in \mathbb{R}^{P_1 \times \cdots \times P_L \times J_1 \times \cdots \times J_K}$ and $\mathcal{B} = \sbr{b_{j_1 \cdots j_K q_1 \ldots q_M}} \in \mathbb{R}^{J_1 \times \cdots \times J_K \times Q_1 \times \cdots \times Q_M}$, the contracted tensor product $\langle \ca, \cb \rangle_{\cbr{J_k}}= \sbr{c_{p_1 \ldots p_L q_1 \ldots q_M}} \in \mathbb{R}^{P_1 \times \cdots \times P_L \times Q_1 \times \cdots \times Q_M}$,
with $
c_{p_1 \ldots p_L q_1 \ldots q_M}=\sum_{j_1 \ldots j_K}^{J_1 \cdots J_K} a_{p_1 \ldots p_L j_1 \ldots j_K} b_{j_1 \cdots j_K q_1 \ldots q_M}
$,
where $\cbr{J_k}$ is the shortened notation for the collection of dimensions $\cbr{J_1 \ld  J_K}$ for which the contraction is done.
Note that for matrices $\mathbf{A} \in \mathbb{R}^{P \times J}$ and $\mathbf{B} \in \mathbb{R}^{J \times Q}$,
$\langle \mathbf{A}, \mathbf{B} \rangle_{\cbr{J}} = \mathbf{A} \mathbf{B}$.
       
Consider a collection of matrices denoted by $\bV_\ell = \sbr{v_{p_\ell k_\ell}^{(\ell)}} \in \mathbb{R}^{P_\ell \times K_\ell}$, $\ell=1,\dots,L$, and a  tensor $\core = [u_{k_1\ldots k_L}] \in \mathbb{R}^{K_1 \times \cdots \times K_L}$. The notation  $ \dbr{\, \core; \bV_1, \dots, \bV_L}$ stands for a tensor of size $P_1 \times \cdots \times P_L$ whose $(p_1p_2\dots p_L)$th element is $\sumk {u}_{k_1 \ldots k_L} v_{p_1 k_1}^{(1)} \cdots v_{p_L k_L}^{(L)}$. When $K_1 = \cdots = K_L =K$ the tensor
 $\dbr{\bV_1 \ld \bV_L} := \dbr{\mathcal{I}_K ; \bV_1 \ld, \bV_L}$ where $\mathcal{I}_K \in \mathbb{R}^{K \times \cdots \times K}$ is the $L$th order  identity tensor which has ones along the superdiagonal and zeros elsewhere \citep{BallardKolda:bookTensor}. Its $(p_1p_2\dots p_L)$th element is given by $\sum_{k=1}^K v_{p_1 k}^{(1)} \cdots v_{p_L k}^{(L)}$.

Let $\cbr{\cx_n=\sbr{x_{n , p_1 \ldots p_L}} \in \mathbb{R}^{P_1 \times \cdots \times P_L}}_{n=1}^N$, be a given set of $N$ independent tensors. The objective of Multilinear Principal Component Analysis (MPCA) is to find a collection of  mode-$\ell$ projection matrices $\bV_\ell^x = \sbr{v_{p_\ell k_\ell}^{x(\ell)}} \in \mathbb{R}^{P_\ell \times K_\ell}$ of rank $K_\ell \leqslant P_\ell$,  together with a center $\mathcal{C}^x=\sbr{c_{p_1 \ldots p_L}^x} \in \mathbb{R}^{P_1\times \cdots \times P_L} $ and core tensors $\core_n^x = [u_{n,k_1\ldots k_L}^x] \in \mathbb{R}^{K_1 \times \cdots \times K_L}$ such that the reconstructed tensors 
\begin{equation} 
\label{eq:MPCA}
\widehat{\cx}_n = \mathcal{C}^x + \dbr{\, \core_n^x; \bV_1^x, \dots, \bV_L^x}
\end{equation}
are a good approximation of the original tensors $\cx_n$ \citep{Lu:MPCA}.

\subsection{Tensor-on-tensor Regression}
\label{sec:Framework}
Tensor-on-tensor regression aims to model the relation between a set of $N$ independent {continuous} predictor tensors $\cbr{\cx_n=\sbr{x_{n, p_1 \cdots p_L}} \in \mathbb{R}^{P_1 \times \cdots \times P_L}}_{n=1}^N$ and $N$ {continuous }response tensors $\cbr{\cy_n=\sbr{y_{n, q_1 \cdots q_M}} \in \mathbb{R}^{Q_1 \times \cdots \times Q_M}}_{n=1}^N$.
The tensor-on-tensor regression model introduced by \cite{lock:TOT} assumes that 
\begin{equation}
\label{eq:TOTmodel}
\cy_n = \cb_0 + \langle \cx_n, \cb \rangle_{\{P_\ell\}} + \ce_n,
\end{equation}
where $\cb_0 = \sbr{\bint} \in \mathbb{R}^{Q_1 \times \cdots \times Q_M}$ is the intercept, $\cb{=\sbr{\bs}} \in \mathbb{R}^{P_1 \times \cdots \times P_L \times Q_1\times \cdots \times Q_M}$ is the slope tensor and $\cbr{\ce_n = \sbr{\epsilon_{n,q_1 \ldots q_M}} \in \mathbb{R}^{Q_1 \times \cdots \times Q_M}}_{n=1}^N$ are the error tensors. 
{The error tensor is assumed to satisfy $E(\ce_n \mid \cx_n)=\mathcal{O}_{Q_1 \times \cdots \times Q_M}=[0_{q_1 \ldots q_M}]$, and to have components that are symmetrically distributed.}
Equivalently, model~\eqref{eq:TOTmodel} postulates that 
\begin{equation*}
\y = \bint + \sump \x \bs + \epsilon_{n,q_1 \ldots q_M}\,. 
\end{equation*}
One could estimate $\cb_0$ and  $\cb$ by minimizing the 
least squares objective
\begin{equation}
\label{eq:LScent}
\sum_{n=1}^N \abr{\cy_n - \cb_0 - \langle \cx_n, \cb \rangle_{\{P_\ell\}}}_F^2,
\end{equation}
but this would be ill-defined or lead to overfitting  due to the large number of parameters to be estimated.
To avoid these problems, a penalized LS objective can be considered, as in ridge regression \citep{lock:TOT, Liu:LowTTR}:
\begin{equation}
\label{eq:RR}
\sum_{n=1}^N \abr{\cy_n - \cb_0 - \langle \cx_n, \cb \rangle_{\{P_\ell\}}}_F^2 + \lambda\abr{\cb}_F^2.
\end{equation}
Since the slope tensor $\cb$ is of significantly higher order than the predictor tensors, this can still lead to a substantial computational burden and potential instability in minimizing~\eqref{eq:RR}.  To deal with this problem, it is reasonable to assume that there exists a low-rank structure in the slope array $\cb$. Following~\cite{lock:TOT}, $\cb$ is represented using a CP decomposition of rank $R$, that is
\begin{equation}\label{eq:Bapp}
\cb = \dbr{\bU_1 \ld \bU_L, \bV_1 \ld \bV_M},
\end{equation}
where $\bU_\ell = \sbr{u_{p_\ell r}^{(\ell)}} \in \mathbb{R}^{P_\ell \times R}$ for $\ell = 1 \ld L$ and $\bV_m = \sbr{v_{q_m r}^{(m)}} \in \mathbb{R}^{Q_m \times R}$ for $m = 1 \ld M$. Estimates for $\cb_0, \{\bU_\ell\}$ and$\{\bV_m\}$ are obtained via an alternating LS algorithm, whereas $R$ and $\lambda$ are selected via cross-validation.

\subsection{{Casewise and Cellwise Outliers in Tensor-on-Tensor Regression}}
\label{sec:DefOutl}
{Traditionally, the term outlier refers to an entire outlying case. 
In tensor-on-tensor regression, casewise contamination may occur in the predictor tensors, the response tensors, or both.  Moreover, contamination may affect only a subset of the entries of a tensor, leading to cellwise outliers.  Such cellwise outliers can occur anywhere within either the predictor or response tensors.  In practice, casewise and cellwise contamination may occur simultaneously, while some entries of the predictor and response tensors may additionally be missing.}

{Formally, following~\cite{Hirari:ROMPCA} we extend the mixed contaminated and partially observed tensor contamination model to the tensor-on-tensor regression setting. 
For each $n=1,\ldots,N$, the observed predictor and response tensors are given by
\begin{align*}
\cx_{n,\varepsilon} &= \ca_{\cx_n} \odot \cx_n  + \left(\mathbf{1}_{P_1 \times \cdots \times P_L} - \ca_{\cx_n}\right) \odot \cz_{\cx_n},\\
\cy_{n,\varepsilon} &= \ca_{\cy_n} \odot \cy_n  + \left(\mathbf{1}_{Q_1 \times \cdots \times Q_M} - \ca_{\cy_n}\right) \odot \cz_{\cy_n},
\end{align*}
where $\cx_n$ and $\cy_n$ satisfy the tensor-on-tensor regression model in~\eqref{eq:TOTmodel}, while $\cz_{\cx_n} \sim H_{\cz_{\cx}}$ and $\cz_{\cy_n} \sim H_{\cz_{\cy}}$ denote tensors generated from unspecified outlier-generating distributions.
The tensors $\mathbf{1}_{P_1 \times \cdots \times P_L} \in \mathbb{R}^{P_1 \times \cdots \times P_L}$ and $\mathbf{1}_{Q_1 \times \cdots \times Q_M} \in \mathbb{R}^{Q_1 \times \cdots \times Q_M}$  contain only ones.}

{For the response tensor, we define $\ca_{\cy_n} = \ca_{\cy_n}^{\text{\tiny case}} \odot \ca_{\cy_n}^{\text{\tiny cell}} \odot \ca_{\cy_n}^{\text{\tiny obs}}$. The variable $\ca_{\cy_n}^{\text{\tiny case}}$ has Bernoulli-distributed marginals 
$A_{\cy_n,q_1 \ldots q_M}^{\text{\tiny case}}$
with success parameter $1-\varepsilon_\cy^{\text{\tiny case}}$, where $0 \leqslant \varepsilon_\cy^{\text{\tiny case}} \leqslant 0.5$. Its components are fully dependent, such that all elements of $\ca_{\cy_n}^{\text{\tiny case}}$ are simultaneously equal to either 1 or 0. The $(Q_1 \times \cdots \times Q_M)$-variate variable $\ca_{\cy_n}^{\text{\tiny cell}}$ has Bernoulli components $A_{\cy_n,q_1 \ldots q_M}^{\text{\tiny cell}}$ with success probabilities $1-\varepsilon_{\cy,q_1 \ldots q_M}^{\text{\tiny cell}}$. Finally, the components $A_{\cy_n,q_1 \ldots q_M}^{\text{\tiny obs}}$ of $\ca_{\cy_n}^{\text{\tiny obs}}$ are binary variables taking values in $\{1,\text{NA}\}$, with $P\rbr{A_{\cy_n,q_1 \ldots q_M}^{\text{\tiny obs}} = 1} = 1-\varepsilon_{\cy,q_1 \ldots q_M}^{\text{obs}}$.}

{For the predictor tensor, the tensors $\ca_{\cx_n}^{\text{\tiny case}}$, $\ca_{\cx_n}^{\text{cell}}$, and $\ca_{\cx_n}^{\text{\tiny obs}}$, and hence $\ca_{\cx_n} = \ca_{\cx_n}^{\text{\tiny case}} \odot \ca_{\cx_n}^{\text{\tiny cell}} \odot \ca_{\cx_n}^{\text{obs}}$,  are defined analogously, using the tensor indices $p_1,\ldots,p_L$.}

{Different assumptions on the dependence structure of $\cx_n$, $\cy_n$,  $\cz_{\cx_n}$, $\cz_{\cy_n}$, $\ca_{\cx_n}^{\text{\tiny case}}$, $\ca_{\cx_n}^{\text{\tiny cell}}$, $\ca_{\cx_n}^{\text{\tiny obs}}$, $\ca_{\cy_n}^{\text{\tiny case}}$, $\ca_{\cy_n}^{\text{\tiny cell}}$, and $\ca_{\cy_n}^{\text{\tiny obs}}$ lead to different contamination models. In particular, the model allows casewise and cellwise contamination to occur separately or simultaneously in the predictor and response tensors. When the observation indicators  are independent of the underlying tensor values, the outlier-generating variables, and the other contamination indicators, the corresponding values are missing completely at random.}

\subsection{Robust Tensor-on-Tensor Regression}
\label{sec:robust}
Since Equation~\eqref{eq:RR} relies on the Frobenius norm, which is known to be sensitive to both casewise and cellwise outliers, the resulting solution lacks robustness to outliers. Therefore, we introduce our ROTOT method which can cope with both casewise and cellwise outliers, as well as missing values.

To address the possible presence of contamination in $\cbr{\cx_n}$ and $\cbr{\cy_n}$, we adopt a weighted M-estimation approach~\citep{Maronna:RobustStat}.
First, outliers and missing entries in the predictors $\cbr{\cx_n}$ are addressed through Robust Multilinear Principal Component Analysis (ROMPCA), proposed by \cite{Hirari:ROMPCA}. ROMPCA is a robust extension of the MPCA decomposition introduced in~\eqref{eq:MPCA}, and is fully described in Section~\ref{app:Def_ROMPCA} of the Supplementary Material. For each predictor 
$\cx_n$ it yields a core tensor $\widehat{\core}^x_n$ of  dimension $K_1 \times K_2 \times \cdots \times K_L$ (with each $K_\ell \leqslant P_\ell)$, and an imputed tensor $\cx_n^{\text{\tiny imp}}$ in which the cellwise outliers and the missing values of $\cx_n$ have been replaced by regular values. 
 It also produces a weight $w_n^x$ that takes values between 0 and 1 to quantify the degree of casewise outlyingness of $\cx_n$.

To indicate missing values in each response tensor $\cy_n$, the tensor $\co_n = \sbr{\ocell} \in \mathbb{R}^{Q_1 \times \cdots \times Q_M}$ has value $\ocell = 1$ if $\y$ is observed, and 0 otherwise. The scalar $o_n = \sumq \ocell$ yields the number of observed cells in $\cy_n$ and $o = \sum_{n=1}^N o_n$ the total number of observed cells in the set $\cbr{\cy_n}$.
ROTOT estimates the slope array $\cb$, which is parameterized as in \eqref{eq:Bapp} by $(\bUbV)$, along with the intercept $\cb_0$, by minimizing
\begin{multline}
\label{eq:robustWLS}
\cl \rbr{\cbr{\cx_n}, \cbr{\cy_n}, \bUbV, \cb_0}:=\\\frac{\sgcase^2}{o} \sum_{n=1}^N o_n w_n^x \rho_2\rbr{\frac{1}{\sgcase}\sqrt{\frac{1}{o_n}\sumq \ocell \sgcell^2 \rho_1\rbr{\frac{\rs}{\sgcell}}}} + \lambda \abr{\cb}^2_F,
\end{multline}
where
\vspace{-3mm}
\begin{align}
\rs &= \y - \bint - \sump \ximp\bs, \label{eq:cellres}\\
\bs & = \sum_{r=1}^{R} u_{p_1r}^{(1)} \cdots u_{p_Lr}^{(L)} v_{q_1r}^{(1)} \cdots v_{q_Mr}^{(M)}\, . \label{eq:cp}\end{align}

Each of the scales $\sgcell$  standardizes the cellwise residuals $\rs$ and the scale $\hat{\sigma}_2$  standardizes the casewise deviations
\begin{equation}
\label{eq:caseres}
t_{n} := \sqrt{ \frac{1}{o_n} \sumq \ocell \sgcell^2 \rho_1\left(\frac{\rs}{\sgcell}\right)}.
\end{equation}

If $\rho_1(z) = \rho_2(z) = z^2$ and there are no missing entries nor outlying predictor tensors, the objective function~\eqref{eq:robustWLS} reduces to the classical tensor-on-tensor regression minimization problem~\eqref{eq:RR}. To simultaneously address both cellwise and casewise outliers in ROTOT, \textit{valid} $\rho$-functions  are employed \citep{Centofanti:cellpca}. 
\begin{definition}\label{def:validrho}
A function $\rho:\mathbb{R}\rightarrow\mathbb{R}$ is called a valid $\rho$-function if,
for any  $z\in\mathbb{R}$, it is continuous and differentiable, even,
nondecreasing in $|z|$, satisfies $\rho(0)=0$, is bounded for large $|z|$, and the
mapping $z \mapsto \rho(\sqrt{z})$ is concave for $z \geqslant 0$.
\end{definition}
Specifically, the function $\rho_1$ is designed to limit the influence of cellwise outliers in~\eqref{eq:robustWLS}. A cellwise outlier in the $n$th response tensor $\cy_n$ results in a large absolute residual $\rs$, but its contribution to the objective function is tempered thanks to the boundedness of $\rho_1$.
Similarly, $\rho_2$ mitigates the effect of casewise outliers with a large deviation $t_{n}$.
 Note that the presence of $\rho_1$ in $t_{n}$ reduces the influence of outlying cells and avoids that a single cellwise outlier would always result in a large casewise deviation. Specifically, the $\rho_1$ and $\rho_2$ functions are set to be the hyperbolic tangent $(tanh)$ function \citep{Hampel:CVC}, which is described in Section \ref{app:rho_func} of the Supplementary Material. As shown in \cite{Centofanti:cellpca}, the $tanh$ is a valid 
$\rho$-function. 
{Note that whether an observation is classified as a casewise outlier is determined not by the number of corresponding outlying cells, but by the magnitude of $t_n$.}

It is important that the scale estimates $\sgcell$ and $\sgcase$ in~\eqref{eq:robustWLS} are robust as well. Therefore, we use M-scale estimators based on an initial fit, which is described in  Section~\ref{sec:setup}. An M-scale estimator of a univariate sample $(z_1 , \ldots, z_n)$ is the solution $\hat{\sigma}$ of the equation
\begin{equation}
\frac{1}{n} \sum_{i=1}^{n} \rho \left( \frac{z_i}{a\sigma} \right) = \delta,
\label{eq:Mscale}
\end{equation}
where the $\rho$-function  is the \textit{tanh} function. 
 {We set $\delta=1.88$ to achieve maximal robustness and $a=0.3431$ to ensure consistency under the standard Gaussian distribution. See Section~\ref{app:rho_func} of the Supplementary Material for further details on the choice of these constants. Moreover, Section~\ref{app:simres} of the Supplementary Material further investigates the impact of  $\delta$ and $a$ on the ROTOT performance.} 
 
The minimization of the ROTOT objective function is performed via an iterative LS algorithm, detailed in Section~\ref{sec:alg}. It starts from a robust initial fit described in Section~\ref{sec:setup}. Selection of the tuning parameters $\lambda$ and $R$ is discussed in Section~\ref{sec:tuning}. The resulting estimates are $\widehat{\cb}_0$ and $\widehat{\cb}=\dbr{\widehat{\bU}_1 \ld \widehat{\bU}_L, \widehat{\bV}_1 \ld \widehat{\bV}_M}$. 

 {An important objective of regression is to provide accurate predictions of the response for a new predictor. This task is straightforward when the new predictor is uncontaminated. However, it becomes substantially more challenging when it contains cellwise outliers or missing values, since these can directly affect the resulting prediction. To address this issue, we propose a robust prediction scheme based on the imputed predictor tensor. For a given $\cx_n$, its prediction is given by}
\begin{equation} 
\label{eq:fitted}
\widehat{\cy}_n =\widehat{\cb}_0 + \langle \cx_n^{\text{\tiny imp}}, \widehat{\cb} \rangle_{\{P_\ell\}}\,.
\end{equation}
When a new predictor $\cx_* = \sbr{x_{*,p_1 \ldots p_L}}$ arrives, we first construct its imputed version $\cx_*^{\text{\tiny imp}}$ as outlined in Section~\ref{app:Def_ROMPCA} of the Supplementary Material. 
Then, the ROTOT prediction  of the response associated with $\cx_*$ is given by
$\widehat{\cy}_* = \widehat{\cb}_0 + \langle \cx_*^{ \text{\tiny imp}}, \widehat{\cb} \rangle_{\{P_\ell\}}$.

\subsection{The IRLS Algorithm}
\label{sec:alg}
Since the loss function in \eqref{eq:robustWLS} is continuously differentiable, we show in Section~\ref{app:firstorder} of the Supplementary Material that its solution needs to satisfy the following first-order conditions:
\begin{align}
\label{eq:condi1}
\sum_{n=1}^N \left\langle \rbr{\cy_n - \cb_0 - \langle \cx_n^{\text{\tiny imp}}, \cb \rangle_{\{P_\ell\}}} \odot \cw_n, \cc_{n} \right\rangle_{\{Q_m\}} - (4\lambda \, o) \bU_\ell \,\mathbf{T}_\bU^{(-\ell)} & = \mathbf{0}_{P_\ell \times R}, \quad \ell= 1 \ld L\,, \\
\label{eq:condi2}
\sum_{n=1}^N \left\langle \rbr{\cy_n - \cb_0 - \langle \cx_n^{\text{\tiny imp}}, \cb \rangle_{\{P_\ell\}}}  \odot \cw_n, \cd_{n} \right\rangle_{Q_m^{\star}} - (4\lambda \, o) \bV_m \,\mathbf{T}_\bV^{(-m)} & = \mathbf{0}_{Q_m \times R}, \;\; m=1 \ld M\,,\\
\label{eq:condi3}
\sum_{n=1}^N \rbr{\cy_n - \cb_0 - \langle \cx_n^{\text{\tiny imp}}, \cb \rangle_{\{P_\ell\}}} \odot \cw_n & = \mathcal{O}_{Q_1 \times \cdots \times Q_M}\,, 
\end{align}
where $Q^{\star}_m = \cbr{Q_1 \ld Q_{m-1},Q_{m+1} \ld Q_M}$. The matrices $\mathbf{0}_{P_\ell \times R} \in \mathbb{R}^{P_\ell \times R}$ and $\mathbf{0}_{Q_m \times R} \in \mathbb{R}^{Q_m \times R}$  are zero matrices, and $\mathcal{O}_{Q_1 \times \cdots \times Q_M} \in \mathbb{R}^{Q_1 \times \cdots \times Q_M}$ is a zero tensor. The matrix $\mathbf{T}^{(-\ell)}_{\bU} \in \mathbb{R}^{R \times R}$ is given by
\begin{equation} \label{eq:TU}
\mathbf{T}^{(-\ell)}_{\bU} := \rbr{\bU_1^T\bU_1 \odot \cdots \odot \bU_{\ell-1}^T\bU_{\ell-1}\odot \bU_{\ell+1}^T\bU_{\ell+1} \odot \cdots \odot \bV_M^T\bV_M},
\end{equation}
and the matrix $\mathbf{T}^{(-m)}_{\bV} \in \mathbb{R}^{R \times R}$  by
\begin{equation} \label{eq:TV}
\mathbf{T}^{(-m)}_{\bV} := \rbr{\bU_1^T\bU_1 \odot \cdots \odot \bV_{m-1}^T\bV_{m-1}\odot \bV_{m+1}^T\bV_{m+1} \odot \cdots \odot \bV_M^T\bV_M}.
\end{equation}
The elements of the tensor $\cc_n \in \mathbb{R}^{P_\ell \times R \times Q_1 \times \cdots \times Q_M}$ 
are given by
\begin{equation} \label{eq:Cl}
c_{n, p_\ell r q_1 \ldots q_M}^\ell = \sumpl \rbr{\ximp u_{p_1r}^{(1)} \cdots u_{p_{\ell-1}r}^{(\ell-1)}u_{p_{\ell+1}r}^{(\ell+1)}\cdots u_{p_Lr}^{(L)}  v_{q_1r}^{(1)} \cdots v_{q_Mr}^{(M)}},
\end{equation}
where $L^{\star}_{\ell} = \{1 \ld \ell-1, \ell +1 \ld L\}$. The elements of the tensor $\cd_n \in \mathbb{R}^{Q_1 \times \cdots \times Q_{m-1} \times R \times Q_{m+1} \times \cdots \times Q_M}$
are
\begin{equation} \label{eq:Dm}
d_{n, q_1 \ldots q_{m-1} \, r \, q_{m+1} \ldots q_M}^m = \sump  \rbr{\ximp u_{p_1r}^{(1)}\cdots u_{p_Lr}^{(L)}v_{q_{1}r}^{(1)} \cdots v_{q_{m - 1}r}^{(m - 1)}  v_{q_{m + 1}r}^{(m + 1)} \cdots v_{q_Mr}^{(M)}}.
\end{equation}
Using the notation $\psi_j=\rho_j'$ and defining the weight function $w_j(z) = \psi_j(z) / z$ for $j = 1, 2$, the weight tensor $\cw_n = \sbr{w_{n, q_1 \cdots q_M}} \in \mathbb{R}^{Q_1 \times \ldots \times Q_M}$ is defined as
\begin{equation}
\label{eq:Wm}
\cw_n = \Woutl_n \odot \Wcase_n \odot \Wcell_n \odot \co_n,
\end{equation}
where the entries of $\Wcell_n \in \mathbb{R}^{Q_1 \times \ldots \times Q_M}$ are given by $\wcell = w_1\left( \rs / \sgcell \right)$, while $\Wcase_n \in \mathbb{R}^{Q_1 \times \ldots \times Q_M}$ and $\Woutl_n \in \mathbb{R}^{Q_1 \times \ldots \times Q_M}$ contain the same value in all their entries, given by $\wcase = w_2\left( t_n / \sgcase \right)$ and $\woutl$ respectively. By convention, $w_j(0) = 1$ so that cells or cases with zero residual receive full weight.

As the system of equations~\eqref{eq:condi1}, \eqref{eq:condi2}, and~\eqref{eq:condi3} is non-linear because the weight tensor depends on the estimates and the estimates also depend on the weights, it can be solved by alternating least squares \citep{Gabriel:ALS}. 
The IRLS algorithm starts with the initial estimate $\parini$ as explained in Section~\ref{sec:setup} below, and associated weight tensors $\cbr{\cw^{0}_n}$ obtained from~\eqref{eq:Wm} by using the corresponding residuals~\eqref{eq:cellres} and~\eqref{eq:caseres}. The set of matrices $\{\mathbf{T}_\bU^{(-\ell),0}\}$, $\{\mathbf{T}_\bV^{(-m),0}\}$ and tensors $\{\mathcal{C}_n^{\ell,0}\}$ and $\{\mathcal{D}_n^{m,0}\}$ are initialized following~\eqref{eq:TU}, \eqref{eq:TV}, \eqref{eq:Cl} and \eqref{eq:Dm}. 
Then, at each iteration $k = 1,2, \ldots$, updated estimates $\parnext$ and corresponding weight tensors $\cbr{\cw_n^{k +1} =\left[w_{n, q_1 \ldots q_M}^{k+1}\right]}$, as well as updated $\{\mathbf{T}_\bU^{(-\ell),k+1}\}$, $\{\mathbf{T}_\bV^{(-m),k+1}\}$, $\{\mathcal{C}_n^{\ell,k+1}\}$ and $\{\mathcal{D}_n^{m,k+1}\}$  are obtained from the current estimates $\parcurr$, $\cbr{\cw_n^{k}}$, $\{\mathbf{T}_\bU^{(-\ell),k}\}$, $\{\mathbf{T}_\bV^{(-m),k}\}$, $\{\mathcal{C}_n^{\ell,k}\}$ and $\{\mathcal{D}_n^{m,k}\}$  by the following procedure, which is derived in Section~\ref{app:algo} of the Supplementary Material.

\begin{itemize}
\item[\textbf{(a)}]
The objective function~\eqref{eq:robustWLS} is minimized with respect to $\bU_\ell$ by solving \eqref{eq:condi1} for $\ell=1,\ldots,L$, which yields 
\begin{align}
\begin{split}
&\vecop \rbr{\bU_\ell^{k+1}}  = \left[\sum_{n=1}^N\rbr{{\mathbf{C}_n^{\ell, k}}}^T {\mathbf{W}_n^k}^* {\mathbf{C}_n^{\ell, k}} + \mathbf{P}^{\ell,k}\right]^{\dagger}  \sum_{n=1}^N \rbr{\mathbf{C}_n^{\ell, k}}^T {\mathbf{W}_n^k}^* \vecop\rbr{\cy_n - \cb_0^k},
\label{eq:solU}
\end{split}
\end{align}
where  $^{\dagger}$ denotes the Moore-Penrose generalized inverse. The matrix ${\mathbf{C}_n^{\ell, k}} \in \mathbb{R}^{{Q} \times RP_\ell}$, with ${Q} = \prod_{m=1}^M Q_m$, is the matricization of $\mathcal{C}^{\ell,k}_n$. The weight diagonal matrix ${\mathbf{W}_n^k}^*$ is defined such that every element of its diagonal is composed of the vectorization of $\cw^{k}_n$. The penalization term is given by
 $   \mathbf{P}^{\ell,k} = 4\lambda o \rbr{\mathbf{T}^{(-\ell), k}_{\bU} \otimes \mathbf{I}_{P_\ell}}$,
where $\otimes$ denotes the Kronecker product and $\mathbf{I}_{P_\ell}$ the identity matrix of size $P_\ell \times P_\ell$.

\item[\textbf{(b)}] 
Using the updated estimates $\cbr{\bU_\ell^{k+1}}$, \eqref{eq:robustWLS} is then minimized with respect to $\bV_m$ by solving~\eqref{eq:condi2} for $m=1,\ldots,M$ which yields the updated estimates
\begin{align}
\begin{split}
&\vecop \rbr{\bV_m^{k+1}} = \rbr{\sum_{n=1}^N\rbr{{(\mathbf{D}_n^{m,k}})^T \otimes \mathbf{I}_{Q_m}} \widetilde{\mathbf{W}}_{n, m}^{k} \rbr{\mathbf{D}_n^{m,k} \otimes \mathbf{I}_{Q_m}}  + \mathbf{P}^{m,k}}^{\dagger} \\&\hspace{6cm} \sum_{n=1}^N \vecop \rbr{\rbr{\rbr{\mathbf{Y}_{n, m} - \mathbf{B}_{0,m}^{k}} \odot \mathbf{W}_{n, m}^{k}} \mathbf{D}_n^{m,k}},
\end{split}
\label{eq:solV}
\end{align}
where  ${Q}^\star = \prod_{i \in M^{\star}_m} Q_i^\star$ and $M^\star_m = \{1 \ld m-1,m+1, \ld M\}$, and $\mathbf{Y}_{n, m} \in \mathbb{R}^{Q_m \times {Q}^\star}$ and $\mathbf{B}_{0,m}^{k} \in \mathbb{R}^{Q_m \times {Q}^\star}$ is the matricization of $\cy_n$ and $\cb_0$ along the mode corresponding to $Q_m$. The matrix $\mathbf{D}_n^{m,k} \in \mathbb{R}^{{Q}^\star \times R}$ is the matricization of $\mathcal{D}^{m,k}_n$. The weight diagonal matrix $\widetilde{\mathbf{W}}_{n, m}^{k}$ is defined such that every element of its diagonal is composed of the vectorization of $\mathbf{W}_{n,m}^k \in \mathbb{R}^{Q_m \times {Q}^\star}$, which is the matricization of $\cw_n^k$ along the mode corresponding to $Q_m$. The penalization term is given by
 $   \mathbf{P}^{m, k} = 4\lambda o \rbr{\mathbf{T}^{(-m), k}_{\bV} \otimes \mathbf{I}_{Q_m}}$.

\item[\textbf{(c)}] 
Using the updated estimates $\cbr{\bU_\ell^{k+1}}$ and $\cbr{\bV_m^{k+1}}$, the intercept $\cb_0$ is then updated by solving \eqref{eq:condi3}, which yields
\begin{equation*}
\cb_0^{k+1} = \rbr{\sum_{n=1}^N \rbr{\cy_n - \left\langle \cx_n^{\text{\tiny imp}}, \cb^{k+1} \right\rangle_{\{P_\ell\}}} \odot \cw_n^k} \odot \ch^k,
\end{equation*}
where $\ch^k \in \mathbb{R}^{Q_1 \times \cdots \times Q_M}$ has elements $h_{q_1 \ldots q_M}^k = 1/ \sum_{n=1}^N w_{n,q_1\ldots q_M}^k$ and $\sum_{n=1}^N w_{n,q_1\ldots q_M}^k > 0$. 

\item[\textbf{(d)}] 
Finally, the updated estimates $\cbr{\bU_\ell^{k+1}}$, $\cbr{\bV_m^{k+1}}$, and $\cb_0^{k+1}$ are used to update the residuals in \eqref{eq:cellres} and \eqref{eq:caseres}, and the weight tensor $\cbr{\cw_n^{k +1}}$ in \eqref{eq:Wm}. Moreover, $\{\mathbf{T}_\bU^{(-\ell),k+1}\}$, $\{\mathbf{T}_\bV^{(-m),k+1}\}$, $\{\mathcal{C}_n^{\ell,k+1}\}$ and $\{\mathcal{D}_n^{m,k+1}\}$ are obtained from~\eqref{eq:TU}, \eqref{eq:TV}, \eqref{eq:Cl} and \eqref{eq:Dm}.
\end{itemize}
The algorithm iterates these steps until
\begin{equation}
\label{eq:ConvergencCrit}
\frac{ \mathcal{L}\rbr{\{\cx_n\},\{\cy_n\},\lbrace \bU_\ell^{k+1} \rbrace, \lbrace \bV_m^{k+1} \rbrace, \cb^{k+1}_0} - 
\mathcal{L}\rbr{\{\cx_n\},\{\cy_n\},\lbrace \bU_\ell^{k} \rbrace, \lbrace \bV_m^{k} \rbrace, \cb^{k}_0} } 
{\mathcal{L}\rbr{\{\cx_n\},\{\cy_n\},\lbrace \bU_\ell^{k} \rbrace, \lbrace \bV_m^{k} \rbrace, \cb^{k}_0}} \leqslant 10^{-5},
\end{equation}
with the default maximum number of iterations set to 100. The following proposition shows that the IRLS algorithm to compute the ROTOT estimates converges.
\begin{proposition} \label{the_1P}
If $\rho_1$ and $\rho_2$ are valid $\rho$-functions, each iteration step of the algorithm decreases 
the objective function \eqref{eq:robustWLS}, 
so
$\mathcal{L}\rbr{\{\cx_n\},\{\cy_n\},\lbrace \bU_\ell^{k+1} \rbrace, \lbrace \bV_m^{k+1} \rbrace, \cb^{k+1}_0} \leqslant 
\mathcal{L}\rbr{\{\cx_n\},\{\cy_n\},\lbrace \bU_\ell^{k} \rbrace, \lbrace \bV_m^{k} \rbrace, \cb^{k}_0}$\,.
\end{proposition}
The proof is given in Section~\ref{app:proofConv} of the Supplementary Material. Since the objective function is decreasing  and it has a lower bound of zero, the algorithm must converge.

\subsection{Initialisation}
\label{sec:setup}
The objective function~\eqref{eq:robustWLS} is nonconvex. Consequently, the IRLS algorithm described in the previous section may converge to a local rather than a global minimum.
To avoid that the algorithm converges to an undesirable non-robust local minimum, it is important that the initialization of the algorithm is chosen appropriately,
{as shown in Section~\ref{app:simres} of the Supplementary Material.} 

 To obtain an initial fit, 
we vectorize each tensor $\cy_n$ and stack the vectors rowwise into a matrix $\mathbf{Y} = \sbr{\text{vec}(\cy_1)\ld \text{vec}(\cy_N)}^T$. The Detecting Deviating Cells (DDC) algorithm \citep{Rousseeuw:DDC} is then applied to $\mathbf{Y}$. DDC flags cellwise outliers and missing values and provides imputed values for them, yielding the corresponding imputed tensors $\cy^{\text{\tiny DDC}}_n$. It also yields an index set $I_y$ of potential casewise outliers in $\cbr{\cy_n}$. Moreover, define the index set $I_x$ for the casewise outliers in the predictors as the set corresponding to predictors with $\woutl \ne 1$, which is obtained from applying ROMPCA to the predictors. We then select  the $H=\lceil 0.75N\rceil$ cases 
whose indices $I_h$ are not contained in $\cbr{I_x, I_y}$ and with the fewest DDC-flagged cells in  $\cy_n$. 
If $\lbrace I_x, I_y \rbrace$ contains more than $25\%$ of the cases, $I_{h}$ is composed of all the cases not in  $\lbrace I_x, I_y \rbrace$.
Finally, classical tensor-on-tensor regression is applied to the pairs $\cbr{\cx^{\text{\tiny imp}}_h,\cy^{\text{\tiny DDC}}_h}_{h \in I_{h}}$, yielding initial estimates of the intercept $\cb_0^0$ and the slope $\cb = \dbr{\bU_1^0 \ld \bU_L^0, \bV_1^0 \ld \bV_M^0}$.

Next, the cellwise residuals are  computed using~\eqref{eq:cellres}. The corresponding cellwise M-scale estimates $\sgcell$ are then obtained from~\eqref{eq:Mscale}. Based on the standardized cellwise residuals, the casewise deviations are computed using~\eqref{eq:caseres}, together with their M-scale estimate $\sgcase$.

{Note that, even with a robust initialization, convergence to the global optimum is not guaranteed. Nevertheless, the proposed initialization provides a sufficiently good starting point to obtain a valid robust solution in practice. This type of strategy is standard in the robust statistics literature; see, for example, \cite{Centofanti:cellpca}. The simulation results presented in Section~\ref{sec:simu} show that this approach also works well in the tensor-on-tensor regression setting.}

\subsection{Tuning Parameter Selection}
\label{sec:tuning}
The parameters $\lambda$ and $R$ in \eqref{eq:robustWLS} and \eqref{eq:cp} are selected through $K$-fold cross-validation. 
Specifically, the data is partitioned into $K$ disjoint folds of equal size. Each of the $K$ folds is held out in turn as the validation set, while the remaining observations form the training set.
Cross-validation is performed using the imputed tensors $\cbr{\cx^{\text{\tiny imp}}_n}$ corresponding to cases with $\woutl=1$, in order to account for casewise and cellwise outliers in the predictor tensors of the validation set.
On each training set and for every candidate pair $(\lambda, R)$, the intercept and slope tensor are estimated via ROTOT. 
Since the predictors and their corresponding weights used during cross-validation are fixed to the imputed tensors $\cbr{\cx^{\text{\tiny imp}}_n}$ and the weights $\cbr{w^x_n}$, respectively, they do not need to be recomputed.
The residuals are then computed on the corresponding validation set, yielding a casewise scale $\hat{\sigma}_{2,k}$. The scales $\hat{\sigma}_{2,k}$ are averaged across the $K$ folds, and the pair $(\lambda, R)$ with the smallest average casewise scale is selected over a finite grid of candidate parameters. In our implementation, the default number of folds for cross-validation is set to $K = 5$. 

{The ROMPCA ranks $(K_1,\ldots,K_L)$ are selected based on the proportion of explained variation, following the procedure described in \cite{Hirari:ROMPCA}. Note that, analogous to principal component regression, the ranks that explain the largest proportion of variation are not necessarily optimal for prediction \citep{Jolliffe:PCA}. In principle, these ranks could also be incorporated into the cross-validation procedure described above.
However, this would substantially increase the computational burden. 
 The simulation results presented in Section~\ref{sec:simu} indicate that the proposed strategy yields satisfactory performance.}

\section{Simulation Study}
\label{sec:simu}
In this section, the performance of the ROTOT method is evaluated through a Monte Carlo simulation study.
The predictors are generated similarly to \cite{Hirari:ROMPCA}. 
The $L$th-order predictor tensors $\cx_n \in \mathbb{R}^{P_1 \times \cdots \times P_L}$, are generated according to the model  $\cx_n = \dbr{\core_n^x; \bV_1^x \ld \bV_L^x} + \ce_{n}^x$. Here, $\core_n^x = \sbr{u_{n, p_1 \ldots p_L}^x} \in \mathbb{R}^{P_1 \times \cdots \times P_L}$ are the core tensors, $\bV_{\ell}^x = \sbr{v_{p_\ell k_\ell}^{x(\ell)}} \in \mathbb{R}^{P_\ell\times K_\ell}$ are the $\ell$-mode projection matrices with rank $K_\ell$, and $\ce_{n}^x = \sbr{\epsilon_{n,p_1 \dots p_L}^x} \in \mathbb{R}^{P_1 \times \cdots \times P_L}$ are the noise tensors. The $u_{n, p_1 \ldots p_L}^x$ are generated  from $N(0,1)$, and then multiplied by $[(P_1, \ld P_L)/(\prod_{\ell=1}^L p_\ell )]^{0.9}$. 
The columns of the matrices $\bV_{\ell}^x$ are the first $K_\ell$ eigenvectors of the covariance matrix $\mathbf{\Sigma}$ where each entry $(i_1,i_2)$ is given by $\rho_{i_1, i_2} = (-0.9)^{|i_1 - i_2|}$. The entries of the noise tensor $\ce_{n}^x$ are sampled from $N(0, 0.1)$ (zero mean Gaussian distribution with variance $0.1$).
The response $\cy_n$ is generated as in \eqref{eq:TOTmodel}
where $\cb_0 = \mathcal{O}_{Q_1 \times \cdots \times Q_M}$ and $\cb = c\dbr{\bU_1 \ld \bU_L,\bV_1 \ld \bV_M}$, 
with $\bU_\ell = \sbr{u_{p_\ell r}^{(\ell)}} \in \mathbb{R}^{P_\ell \times R}$ for $\ell = 1 \ld L$ and $\bV_m = \sbr{v_{q_m r}^{(m)}} \in \mathbb{R}^{Q_m \times R}$ for $m = 1 \ld M$ whose elements  are generated as $N(0,1)$.
The elements of the error $\ce_n \in \mathbb{R}^{Q_1 \times \cdots \times Q_M}$ are generated from $N(0,1)$. The scalar $c$ is obtained such that
$\sum_{n=1}^N \abr{\langle \cx_n, \cb \rangle_{\{P_\ell\}}}^2_F/\sum_{n=1}^N \abr{\ce_n}^2_F = \text{SNR}$,
where $\text{SNR}$ stands for signal-to-noise ratio.

We evaluate the robustness of the methods by varying the degree of casewise and cellwise outliers in $\cbr{\cy_n}$ and in $\cbr{\cx_n}$. 
To generate cellwise outliers in the response tensors, a percentage $\varepsilon_{\text{\tiny cell}}$ of the total $NQ$ entries $\y$ are replaced by $\gamma_{\text{\tiny cell}} s_{y,q_1 \ldots q_M}$, where $s_{y,q_1 \ldots q_M}$ is the standard deviation of $\cbr{\y}_{n=1}^N$, and $\gamma_{\text{\tiny cell}}= \gamma \cdot c^{\text{\tiny cell}}$ controls the contamination magnitude. Cellwise outliers in $\cbr{\cx_n}$ are generated similarly.
Casewise outliers in $\cbr{\cy_n}$ are generated by shifting a percentage $\varepsilon_{\text{\tiny case}}$ of the $N$ tensors as $\cy^{\star}_n = \cy_n + \ce_n^{\star}$, where the elements of $\ce_n^{\star}$ are drawn from $N(\gamma_{\text{\tiny case}}, 2)$, with $\gamma_{\text{\tiny case}} = \gamma \cdot c^{\text{\tiny case}}$.
 To generate casewise outliers in $\cbr{\cx_n}$, $\varepsilon_{\text{\tiny case}}$ of the $N$ predictors are replaced by tensors generated as $\mathcal{X}^{\star}_n = \dbr{\core^{x,\star}_n; \bV_1^x, \ldots, \bV_L^x} + \ce_n^x$. The elements of $\core_n^{x,\star} \in \mathbb{R}^{P_1 \times \cdots \times P_L}$ are drawn from $N(\gamma_{\text{\tiny case}}, 1)$ only at index positions where, for each mode $\ell$, the index lies in either $\cbr{1, 2}$ or $\cbr{K_\ell + 1, K_\ell + 2}$ and all other entries are set to zero. 

We consider two simulation scenarios. 
In the first scenario, the predictors $\cbr{\cx_n}$ are contaminated under a severe contamination level with $\gamma_{\text{\tiny cell}} = 30$, $\gamma_{\text{\tiny case}} = 10$, and  $\varepsilon_{\text{\tiny cell}} = \varepsilon_{\text{\tiny case}} = 5\%$, while the contamination in the responses $\cbr{\cy_n}$ varies.
This includes cellwise contamination only, where $\varepsilon_{\text{\tiny cell}} = 10\%$ of outlying values in the response tensors are introduced with $c^{\text{\tiny cell}} = 4.5$, and  casewise contamination only, where $\varepsilon_{\text{\tiny case}} = 10\%$ of outlying response tensors are introduced with $c^{\text{\tiny case}} = 0.5$. In the last setting, the responses are contaminated with both $\varepsilon_{\text{\tiny cell}} = 10\%$ of cellwise and $\varepsilon_{\text{\tiny case}} = 10\%$ of casewise outliers, using  $c^{\text{\tiny cell}} = 4.5$ and $c^{\text{\tiny case}} = 0.5$ as before. We consider two signal-to-noise ratios: $\text{SNR} = 1$ and $\text{SNR} = 5$. The parameter $\gamma$ ranges from 0 to 8, data are uncontaminated when $\gamma = 0$.

In the second scenario, the responses $\cbr{\cy_n}$ are contaminated with $\gamma_{\text{\tiny cell}} = 20$ and $\gamma_{\text{\tiny case}} = 3.5$ for $\text{SNR} = 1$ and  $\gamma_{\text{\tiny case}} = 4$ for $\text{SNR} = 5$, and $\varepsilon_{\text{\tiny cell}} = \varepsilon_{\text{\tiny case}} = 10\%$, while the contamination in the predictors $\cbr{\cx_n}$ varies. This includes cellwise contamination only, where $\varepsilon_{\text{\tiny cell}} = 10\%$ of outliers are introduced with $c^{\text{\tiny cell}} = 1.5$, and  casewise contamination only, where $\varepsilon_{\text{\tiny case}} = 10\%$ of outliers are introduced with $c^{\text{\tiny case}} = 1$. In the last setting, the predictors are contaminated with $\varepsilon_{\text{\tiny cell}} = 5\%$ cellwise and $\varepsilon_{\text{\tiny case}} = 5\%$ casewise outliers, using  $c^{\text{\tiny cell}} = 1.5$ and $c^{\text{\tiny case}} = 1$ as before.

Note that, cellwise outliers are always generated such that the observations already contaminated by casewise outliers remain untouched. 
Moreover, casewise outliers are introduced in a way that ensures the indices of outliers in ${\cx_n}$ do not overlap with those in ${\cy_n}$.

For each contamination scenario, $N = 200$ samples are generated, where the dimension of $\cb$ is set to $\rbr{P_1 = 15, P_2 = 20, Q_1 = 5, Q_2 = 10}$. The core tensors $\core^x_n$ of the predictors have rank $K_1=4$ and $K_2=6$
and $\cb$ is assumed to have a low rank representation with $R = 2$. 

To measure the performance of tensor-on-tensor regression methods, we consider the Relative Prediction Error (RPE) on an uncontaminated validation set $\{(\cx_v^{\text{\tiny val}},\cy_v^{\text{\tiny val}})\}_{v=1}^{N_v}$ of size $N_v=100$ defined as
\begin{equation}
\label{eq:RPE}
\operatorname{RPE} := \frac{ \sum_{v=1}^{N_v}\abr{\cy_v^{\text{\tiny val}} - \widehat{\cb}_0 - \langle \cx_v^{\text{\tiny val}}, \widehat{\cb} \rangle_{\{P_\ell\}}}_F}{ \sum_{v=1}^{N_v}\abr{\cy_v^{\text{\tiny val}} - \bar{\cy}_v^{\text{\tiny val}}}_F}
\end{equation}
where $\bar{\cy}_v^{\text{\tiny val}}$ is the mean response tensor. 
The simulations are replicated 100 times for each of the scenarios to acquire the median RPE.

We evaluate ROTOT with other competitors. The first benchmark is the tensor-on-tensor regression method proposed by \cite{lock:TOT} denoted as TOT. 
{The second benchmark is the robust tensor-on-tensor regression method proposed by \cite{Lee:RTOT}, denoted by RTOT. They introduce a robust regularized tensor-on-tensor regression approach proposed to handle cellwise contamination in the response tensor. In their work, the authors explicitly state that outliers in $\cbr{\cx_n}$ should be handled through preprocessing. We therefore consider an additional competitor, denoted by RTOT-Ximp, in which RTOT is fitted using the imputed predictor tensors $\cbr{\cx_n^{\text{\tiny imp}}}$ to account for cellwise contamination in the predictors.
We also compare ROTOT with robust multivariate regression approaches where the predictor  $\cbr{\cx_n}$ and response  $\cbr{\cy_n}$ tensors are vectorized, and the resulting vectors are stacked to form predictor and response matrices.  The first approach, named DDC-ridge, 
 applies DDC to both matrices to detect and impute cellwise outliers followed by the multivariate ridge regression method of \citet{Brown:MRR}. 
In addition, we consider a casewise robust multivariate regression approach, denoted RMCD,
 that first applies the Minimum Regularized Covariance Determinant estimator \citep{Boudt:RMCD} to the
 concatenated predictor and response matrices.
 Then,  the robust regression method described in \cite{Rousseeuw:RMCD} is used to construct the regression coefficient matrix, which is subsequently reshaped into a coefficient tensor.}
The optimal value of $\lambda$ for each method is selected as the one that yields the best performance on the uncontaminated data.
{Additional simulation results are provided in Section~\ref{app:simres} of the supplementary material, where the median RPE is reported over a grid of ranks $R$ and regularization parameters $\lambda$ to assess the sensitivity of ROTOT to the choice of $\lambda$ and $R$.}
All methods are implemented in \texttt{R}~\citep{RTeam:R}. For TOT regression, we use the implementation in \texttt{R} package \texttt{MultiwayRegression}~\citep{Lock:R_MultiwayRegression}.

\begin{figure}[!ht]
\centering
\begin{tabular}{cccc}
& \textbf{Cellwise in $\brc{\cy_n}$}  & \textbf{Casewise in $\brc{\cy_n}$} &{\textbf{Casewise \& Cellwise in $\brc{\cy_n}$}}\\
 \rotatebox{90}{\normalsize {\parbox{5cm}{\centering $\text{SNR} = 1$ }}} & 
\includegraphics[width=.30\textwidth]{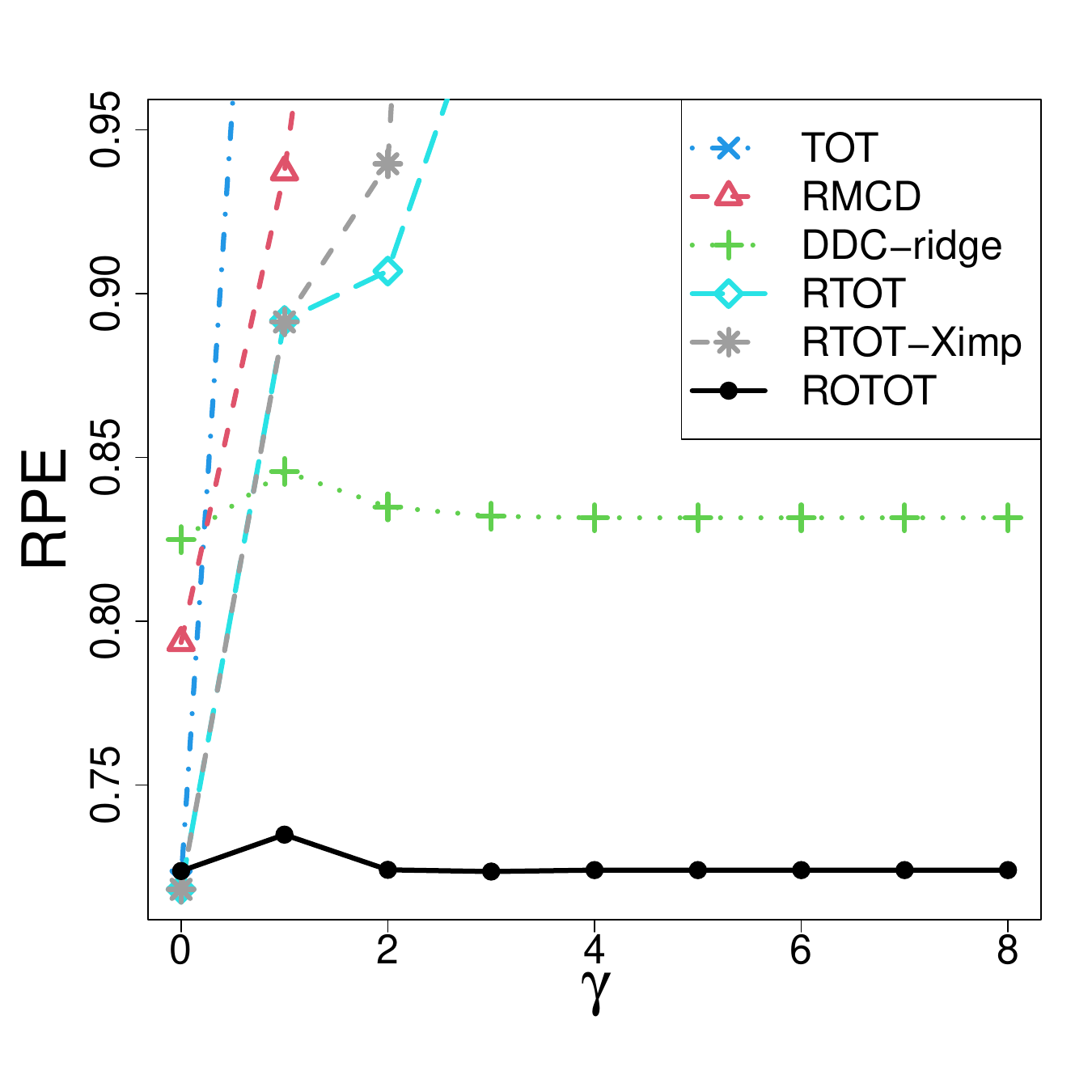} &%
\includegraphics[width=.30\textwidth]{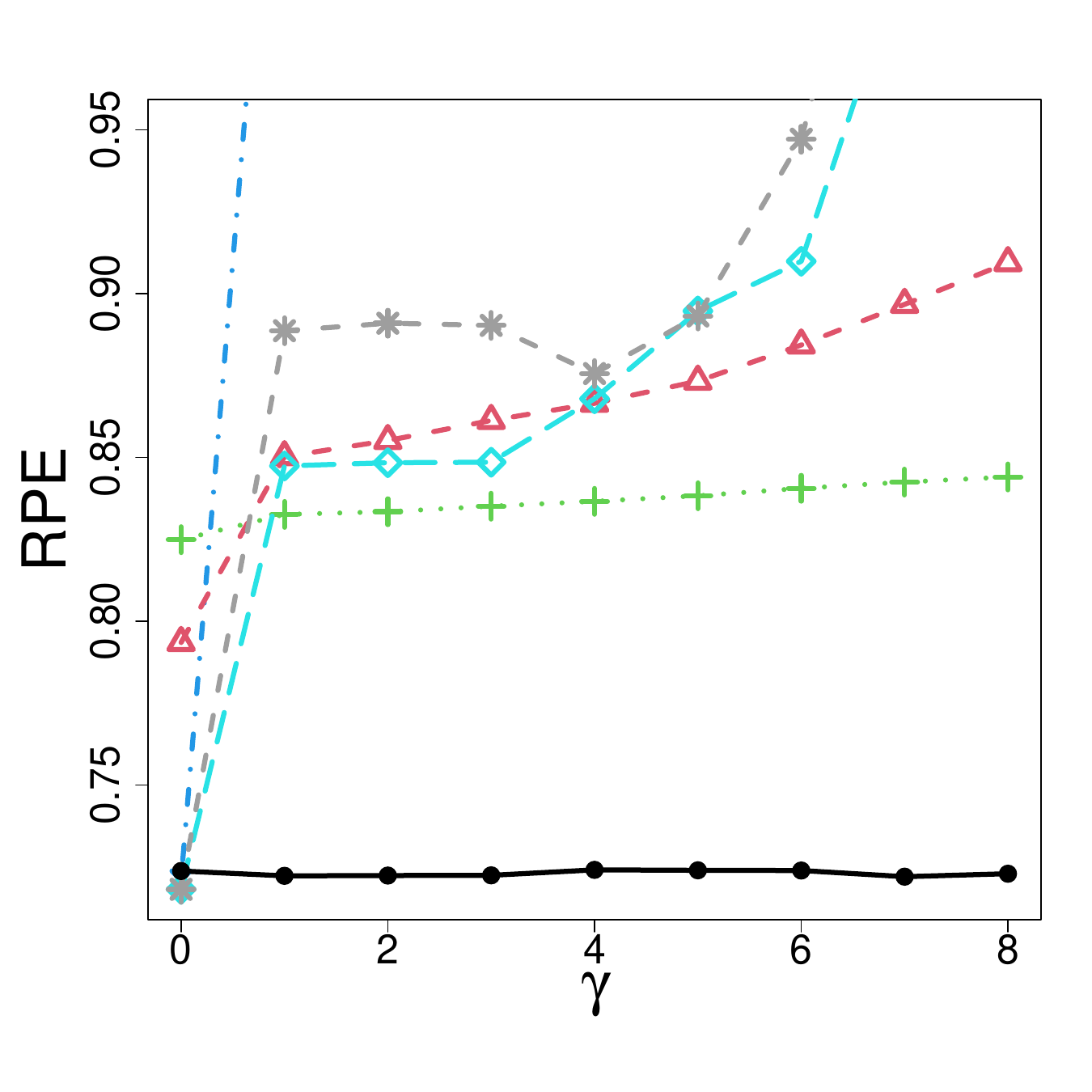}& %
\includegraphics[width=.30\textwidth]{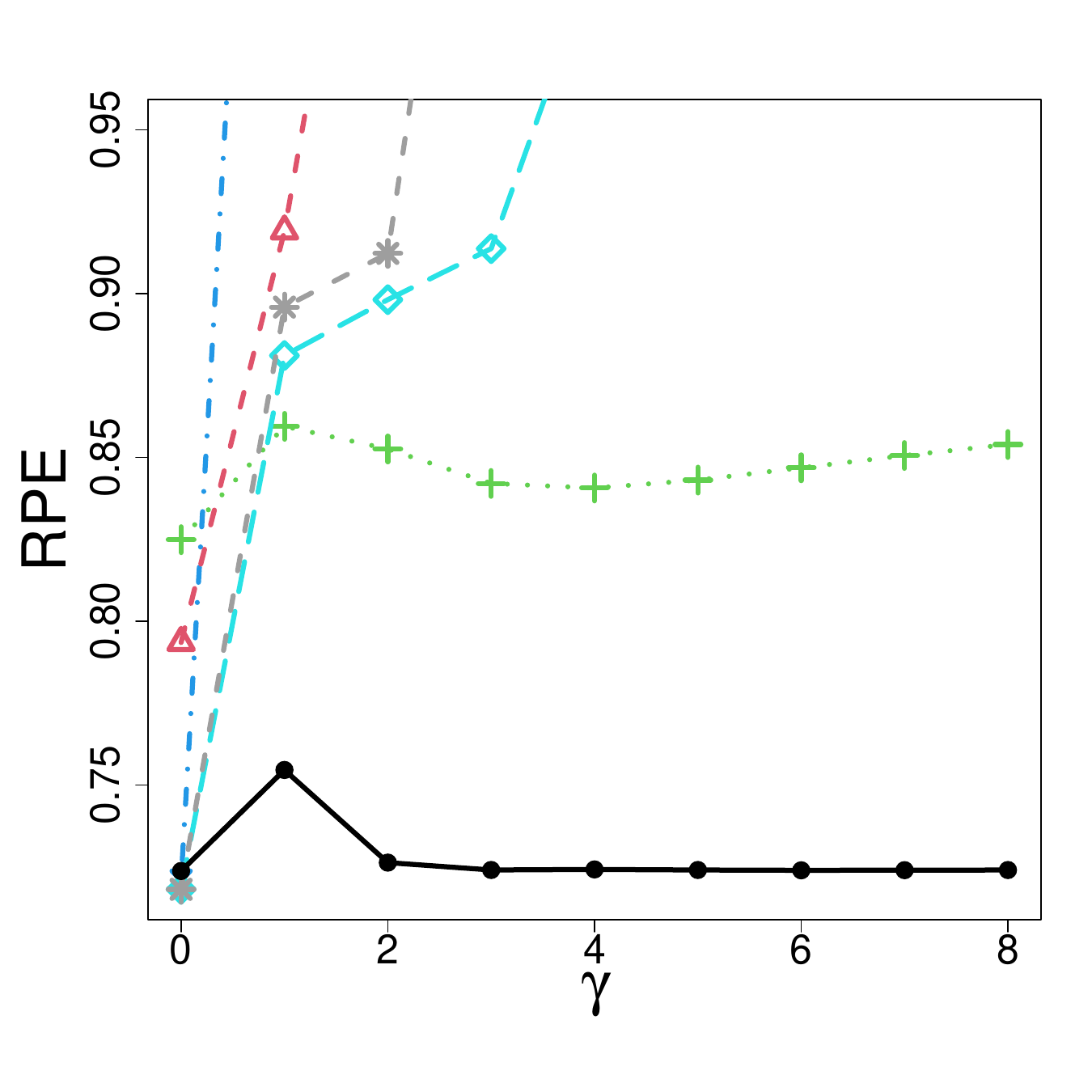}\\ 
\rotatebox{90}{\normalsize  {\parbox{5cm}{\centering $\text{SNR} = 5$}}} &  
\includegraphics[width=.30\textwidth]{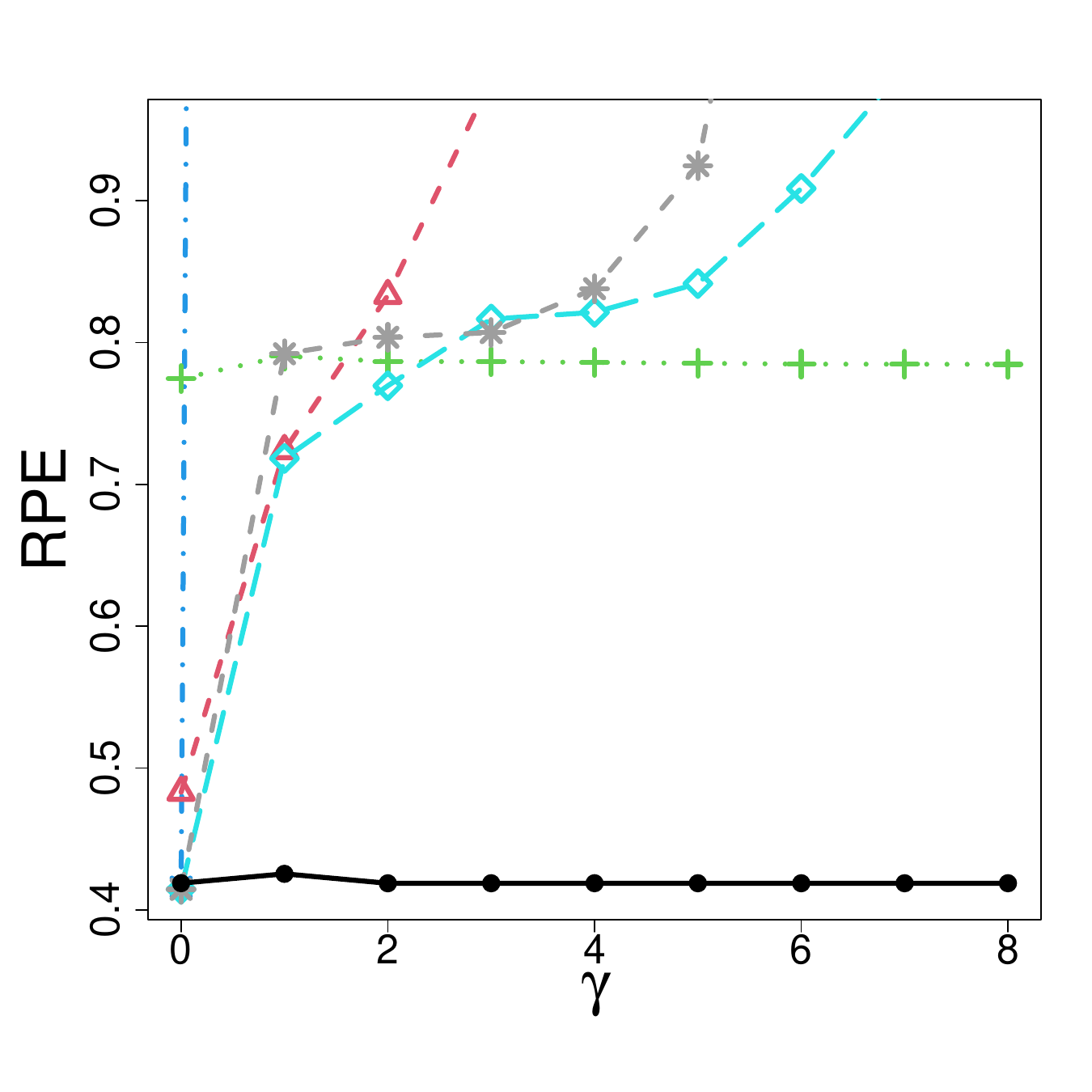} &%
\includegraphics[width=.30\textwidth]{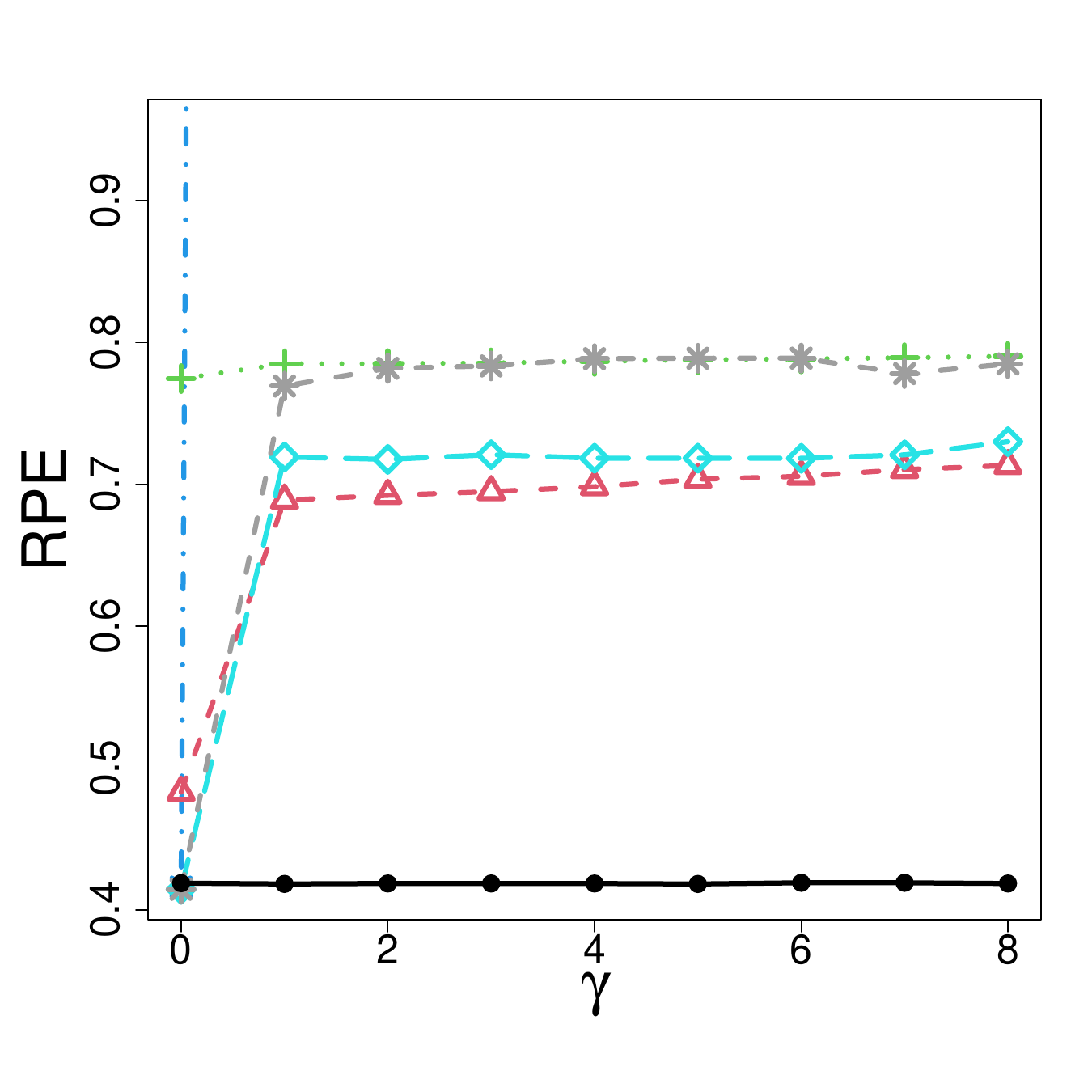} &%
\includegraphics[width=.30\textwidth]{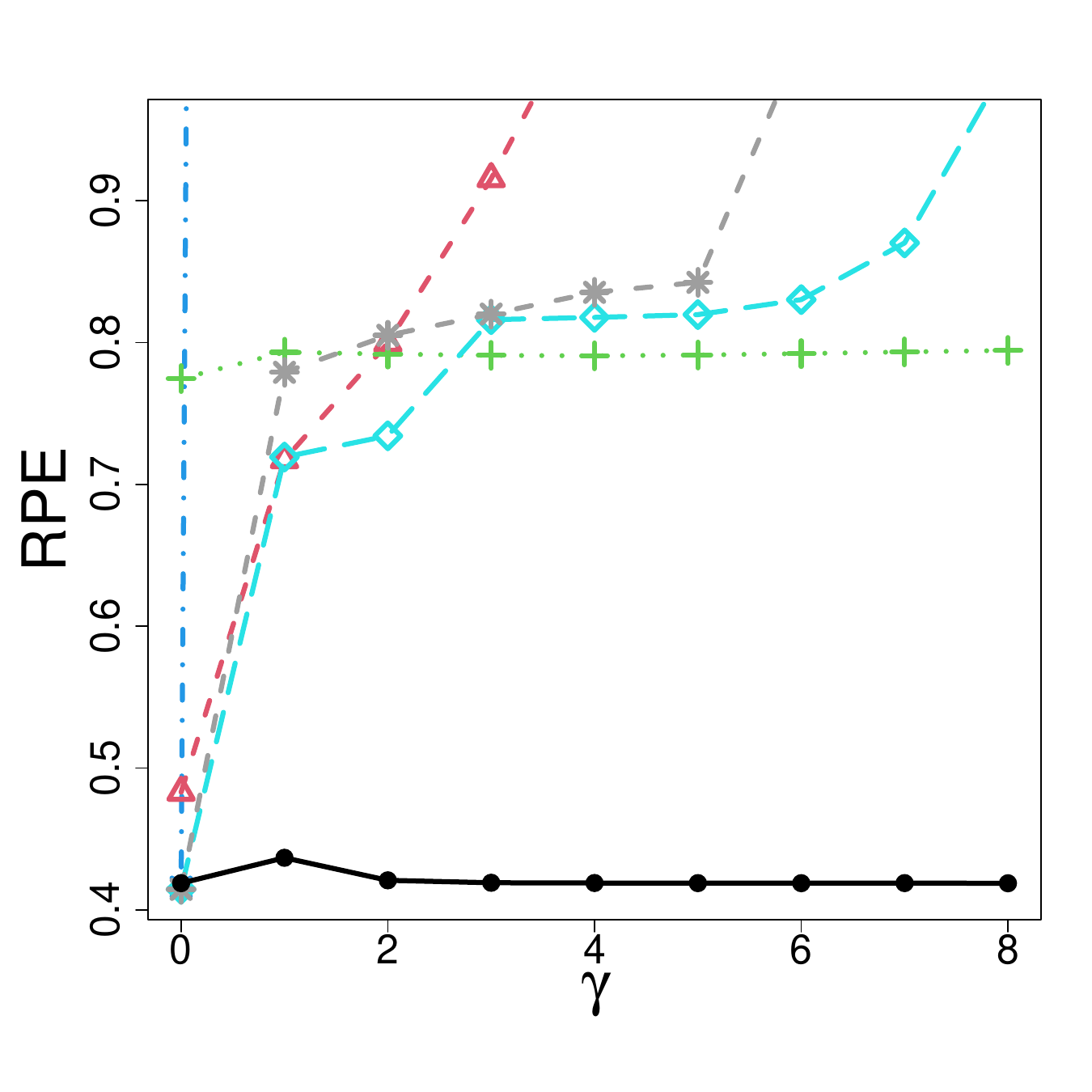} 
\end{tabular}
\caption{{Median RPE attained by TOT, RMCD, DDC-ridge, RTOT, RTOT-Ximp and ROTOT across the different contamination settings, with fixed contamination in the predictors and varying contamination magnitudes in the responses, for $\text{SNR}\in \{1,5\}$.}}%
 \label{fig:RPEvarY}%
\end{figure}

     {Figure~\ref{fig:RPEvarY} shows the median RPE of all methods across the different contamination settings and contamination magnitudes in the \textit{responses}, for $\text{SNR}\in\cbr{1,5}$ under the first contamination scenario. 
ROTOT achieves the best performance across the different contamination settings and magnitudes. The differences in performance among the competing methods can be explained by the types of contamination they are designed to handle. TOT is strongly affected by all types of contamination. RTOT also shows limited robustness across these settings, since it does not account for contamination in the predictor tensors and is sensitive to casewise outliers. RTOT-Ximp improves the handling of cellwise contamination in the predictors through imputation, but remains sensitive to casewise outliers in the response and predictor tensors. 
{A setting in which RTOT-Ximp performs well is presented in Section~\ref{app:simres} of the Supplementary Material, where only cellwise contamination is introduced in both the response and predictor tensors.} 
RMCD is designed to provide robustness against casewise outliers, but cannot handle cellwise contamination. Moreover, both RMCD and DDC-ridge operate on vectorized representations of the data and do not exploit the underlying tensor structure. Although DDC-ridge is robust against cellwise contamination, both methods exhibit higher RPE than the tensor-based methods even in the uncontaminated setting. This highlights the advantage of preserving and exploiting the natural tensor structure of the data.
This conclusion remains unchanged when the SNR is varied.
}

\begin{figure}[!ht]
\centering
\begin{tabular}{cccc}
 & \textbf{Cellwise in $\brc{\cx_n}$}  & \textbf{Casewise in $\brc{\cx_n}$} &{\textbf{Casewise \& Cellwise in $\brc{\cx_n}$}}\\
 \rotatebox{90}{\normalsize {\parbox{5cm}{\centering $\text{SNR} = 1$ }}} & 
\includegraphics[width=.30\textwidth]{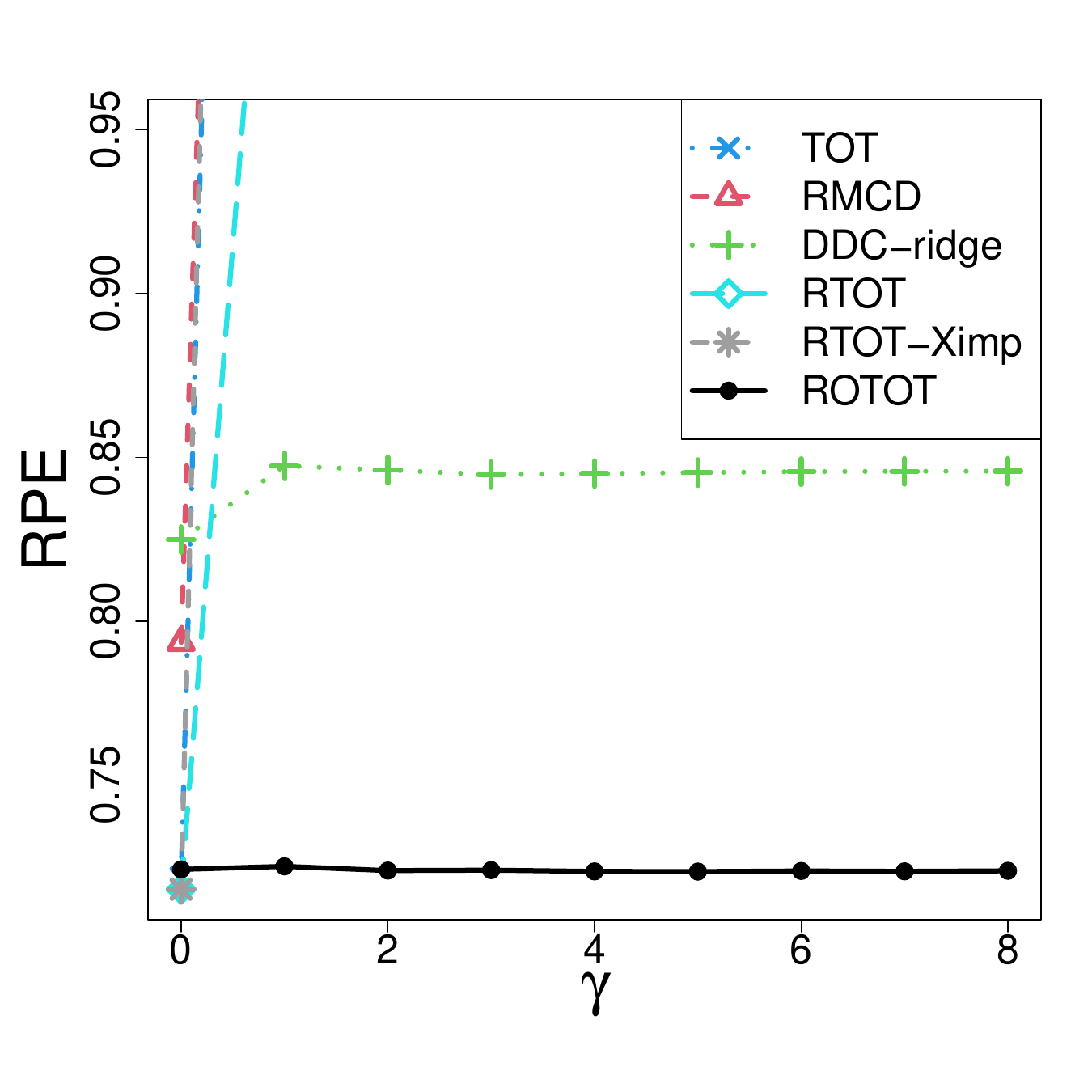} &%
\includegraphics[width=.30\textwidth]{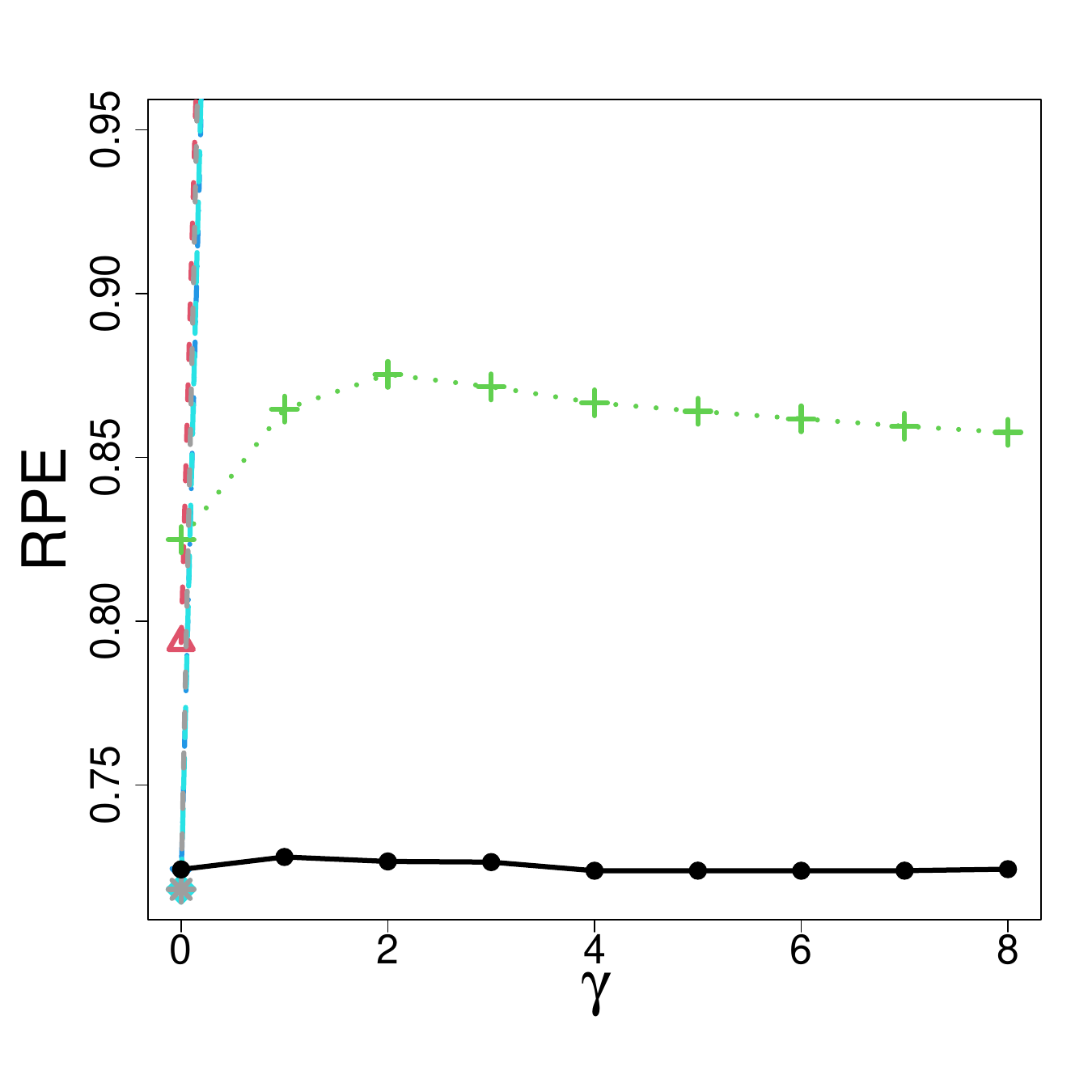}& %
\includegraphics[width=.30\textwidth]{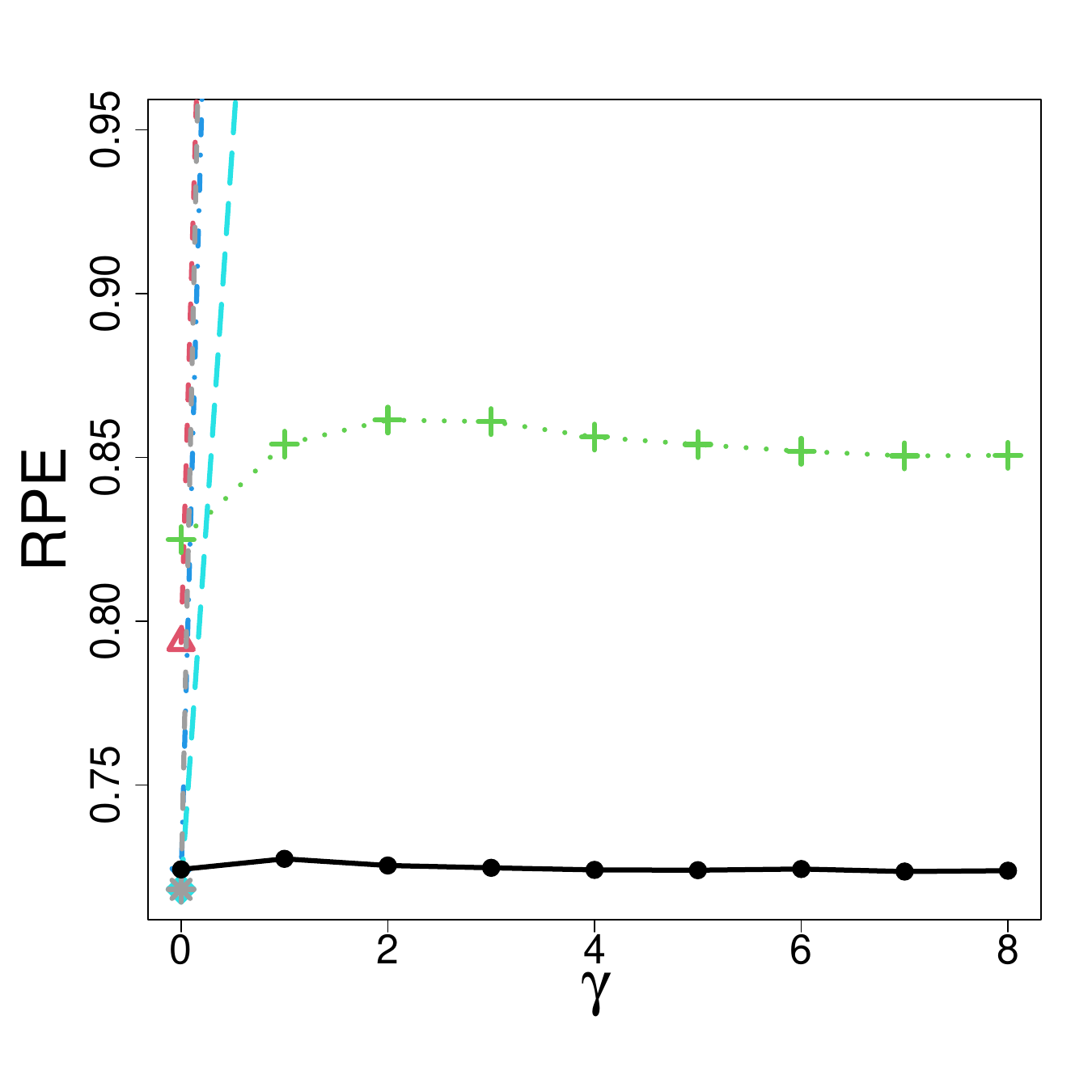}\\ 
\rotatebox{90}{\normalsize  {\parbox{5cm}{\centering $\text{SNR} = 5$}}} &  
\includegraphics[width=.30\textwidth]{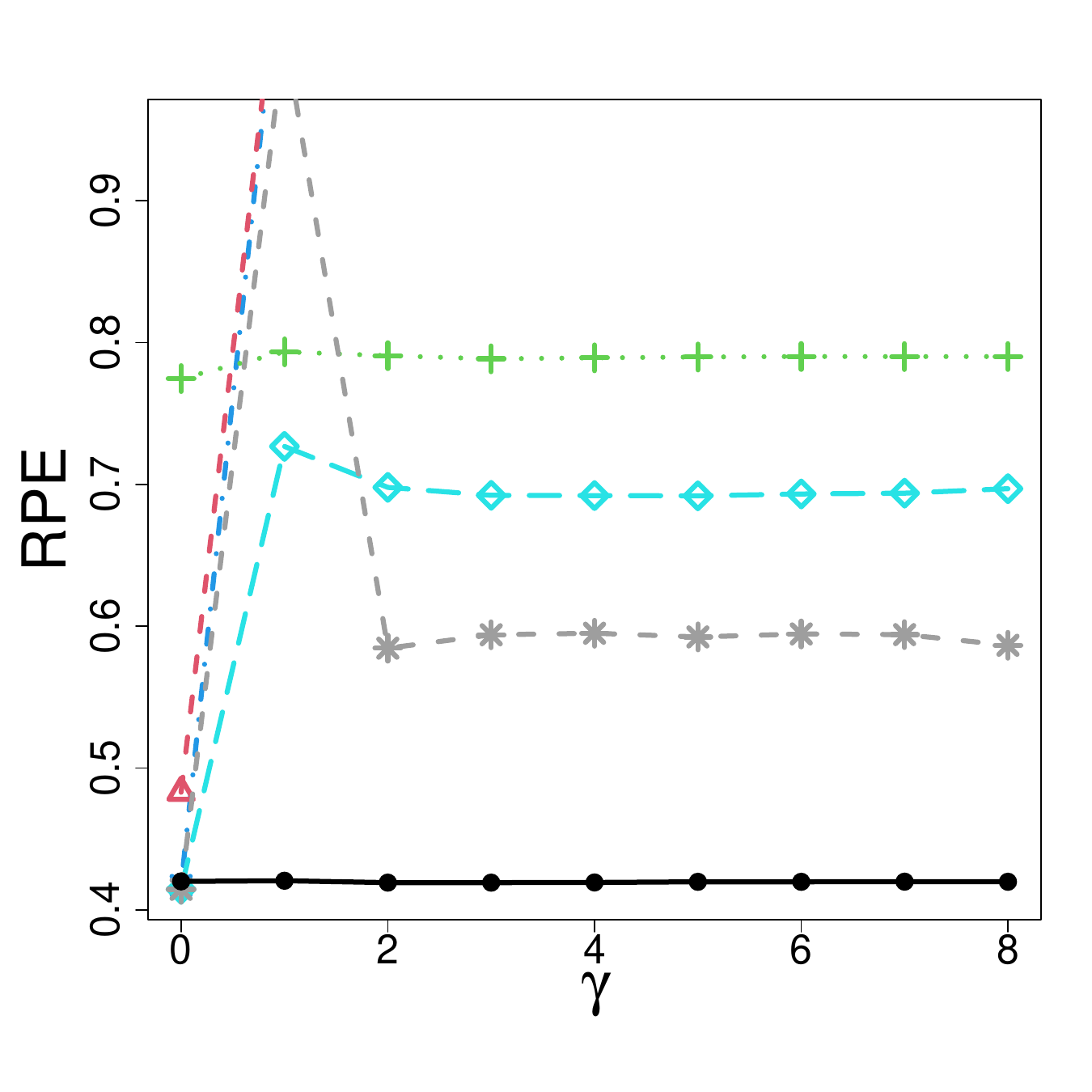} &%
\includegraphics[width=.30\textwidth]{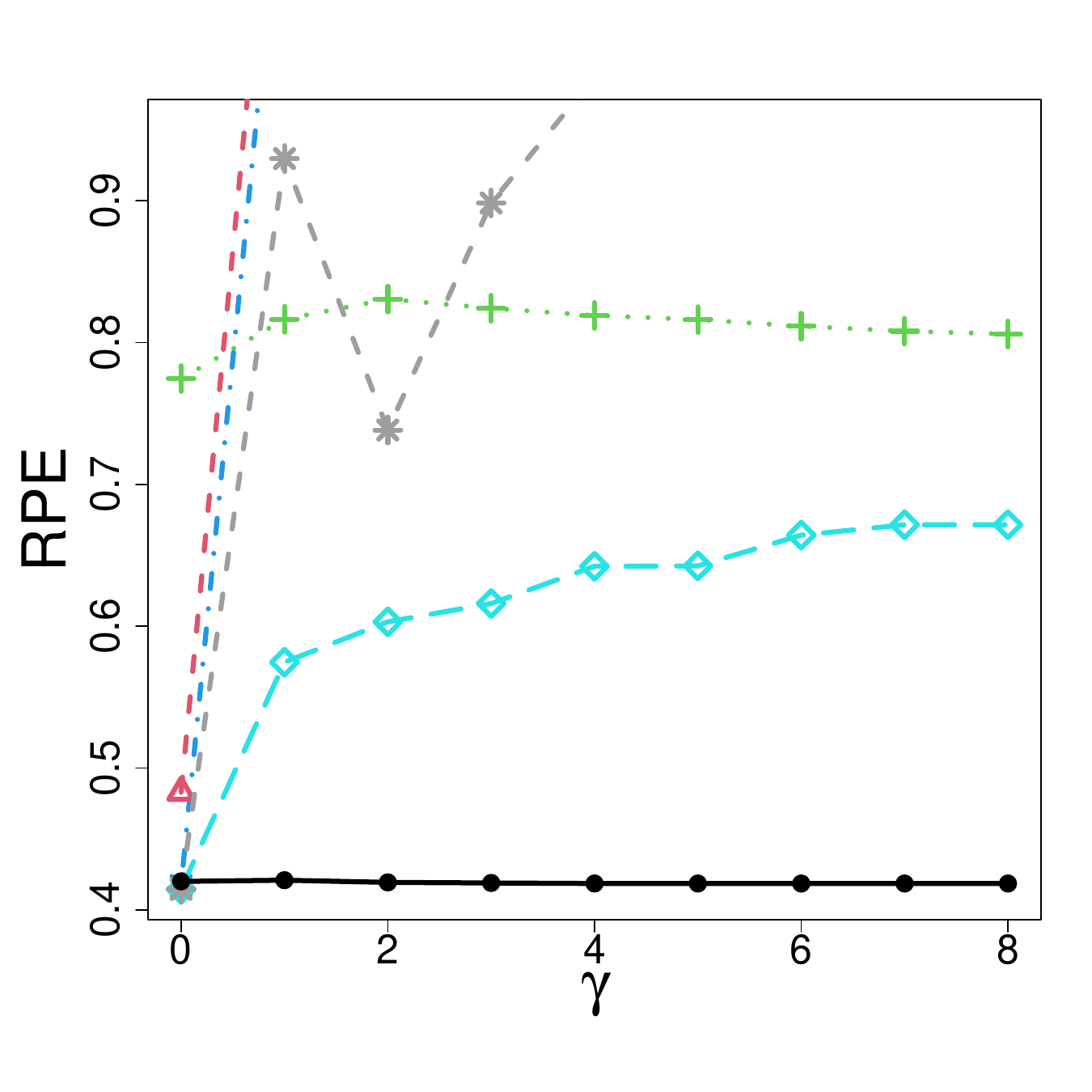} &%
\includegraphics[width=.30\textwidth]{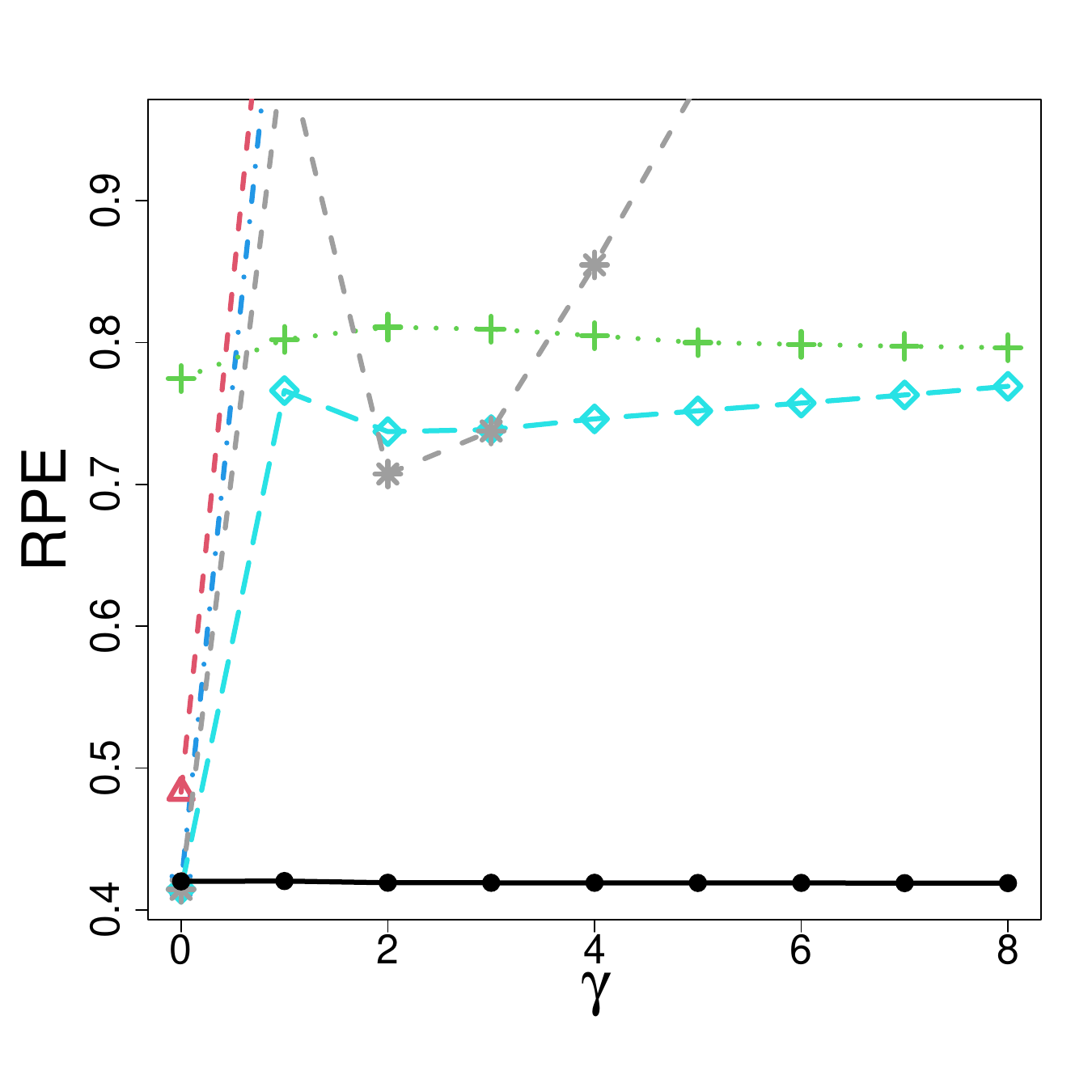} 
\end{tabular}
\caption{{Median RPE attained by TOT, RMCD, DDC-ridge, RTOT, RTOT-Ximp and ROTOT across the different contamination settings, with fixed contamination in the responses and varying contamination magnitudes in the predictors, for $\text{SNR}\in \{1,5\}$.}}%
 \label{fig:RPEvarX}%
\end{figure}

   {Figure~\ref{fig:RPEvarX} shows the median RPE for all methods across the different contamination settings and contamination magnitudes in the \textit{predictors}, with $\text{SNR} = \{1,5\}$ under the second contamination scenario}. 
As expected, ROTOT consistently outperforms all other methods. {Across all contamination settings, TOT and RMCD are affected by the outliers. RTOT is sensitive to contamination in the predictor tensor, since it is designed to be robust against cellwise contamination in the response. 
{Also RTOT-Ximp performs poorly in these settings because they all include casewise outliers, which are not addressed by imputation.}
  DDC-ridge remains relatively robust across the different contamination settings, however, its performance is modest because it does not exploit the inherent tensor structure of the data.}
{In contrast, ROTOT maintains good performance across the various contamination settings, indicating that the combination of imputation of $\cbr{\cx_n}$ and the downweighting of casewise outliers in $\cbr{\cx_n}$ is effective in handling contamination in the predictor tensors.}

{The same conclusions hold in the presence of missing values, as shown by the additional results in Section~\ref{app:simres} of the Supplementary Material. In that section we also compare ROTOT with OnlyCase-TOT and OnlyCell-TOT, which correspond to ROTOT with $\rho_1(z)=z^2$ and $\rho_2(z)=z^2$, respectively, and therefore address only casewise or only cellwise outliers.}

{To assess the efficiency of ROTOT, we compare its predictive performance with that of TOT by computing the ratio of their average RPEs on uncontaminated data ($\gamma=0$). 
We obtain a ratio that is close to one for both $\mathrm{SNR}=1$ and $\mathrm{SNR}=5$, indicating only a small loss in predictive performance relative to TOT on uncontaminated data.}

\section{Outlier Detection}
\label{sec:outdetect}
Based on the ROTOT estimates, we can construct numerical and graphical diagnostics to gain more insight into the outlying cells and cases in the response and predictor tensors. First, we compute the ROTOT residual tensors $\mathcal{R}_n = [\rs]=\cy_n - \widehat{\cy}_n$ following~\eqref{eq:fitted}. Then, we compute the M-scales of the $\{\rs\}_{n=1}^N$, yielding the scales $\tsgcell$ and the final standardized residual tensors $\widetilde{\mathcal{R}}_n = [\trs] = [ \rs/\tsgcell]$. Their elements can be displayed in a 
\textit{residual cellmap} which is a heatmap that visualizes the cellwise outlyingness of entries in the data \citep{Rousseeuw:DDC}. 
{Standardized cellwise residuals with absolute values below the threshold $c_{\text{\tiny cell}}=\sqrt{\chi^2_{1,0.998}}=3.09$ are considered regular and colored yellow, while missing values are shown in white. For outlying cells, the color indicates both the direction and magnitude of the standardized residual. Positive residuals exceeding $c_{\text{\tiny cell}}$ are colored from light orange to red, whereas negative residuals below $-c_{\text{\tiny cell}}$ are colored from purple to dark blue. Darker colors indicate a greater degree of outlyingness.}

\begin{figure}[h]
\begin{center}
  \includegraphics[width=1\textwidth]{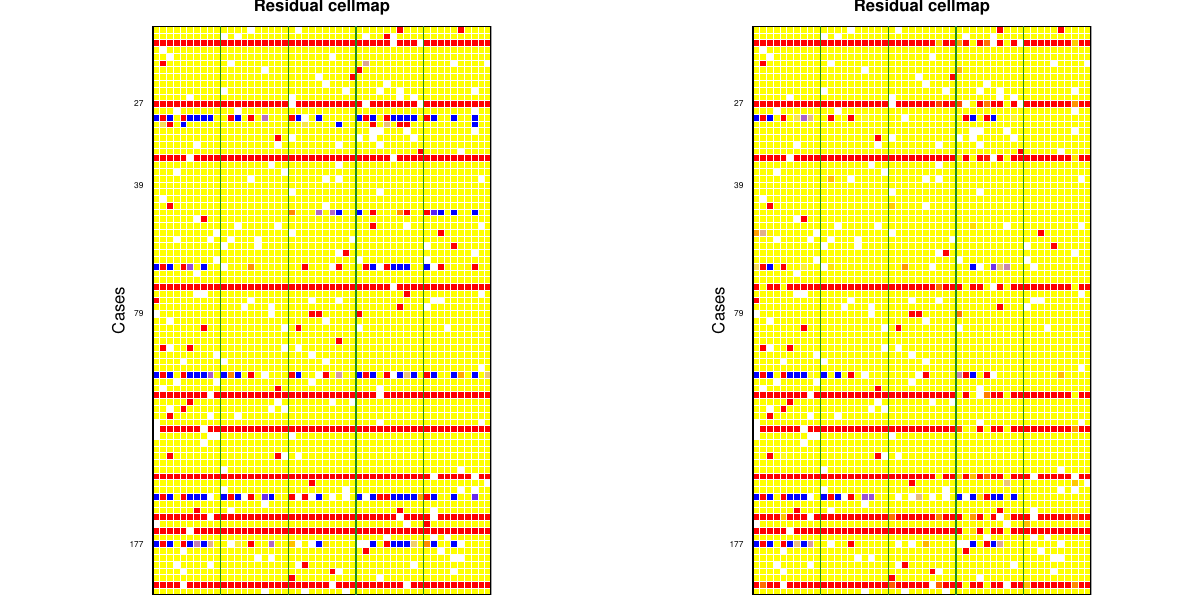}
 \end{center}
\vspace{-0.5cm}
\caption{{Residual cellmaps obtained from ROTOT (left) and by using the true coefficient tensor $\cb$ (right) of a simulated dataset with cellwise and casewise contamination and missing values in the response and predictor tensors.}}
\label{fig:Simulation_UnfResMap}
\end{figure}

We can display a residual cellmap of all individual data cells by vectorizing each residual tensor $\widetilde{\mathcal{R}}_n$ and stacking these vectors rowwise into a matrix (as in Section~\ref{sec:setup}). 
As an illustration, the left panel of Figure~\ref{fig:Simulation_UnfResMap} displays the residual cellmaps for a simulated dataset 
obtained from the data generating process described in Section~\ref{sec:simu}, with $(P_1,P_2)=(15,20)$, $(Q_1,Q_2)=(5,10)$, and $\text{SNR} = 5$.
It is contaminated with both casewise and cellwise outliers and  missing values in both the response and the predictor tensors. 
{
As a benchmark for assessing the reliability of the residual cellmap produced by ROTOT, the right panel of Figure~\ref{fig:Simulation_UnfResMap} presents the residual cellmap obtained using the true coefficient tensor $\cb$. The two cellmaps are highly similar, supporting the validity of the residual patterns identified by ROTOT.}
The green vertical lines separate the $Q_1=5$ slices resulting from the vectorisation of each residual tensor. 

From this residual cellmap, we can clearly observe the difference between cases that contain many outlying entries in their residual tensor and those that contain only a few. 
However, the plot does not allow us to determine whether the large absolute residuals of an outlying case are caused by a poorly fitting response or by an outlying predictor.

\begin{figure}[!ht]
\begin{center}
\includegraphics[width=0.55\textwidth]{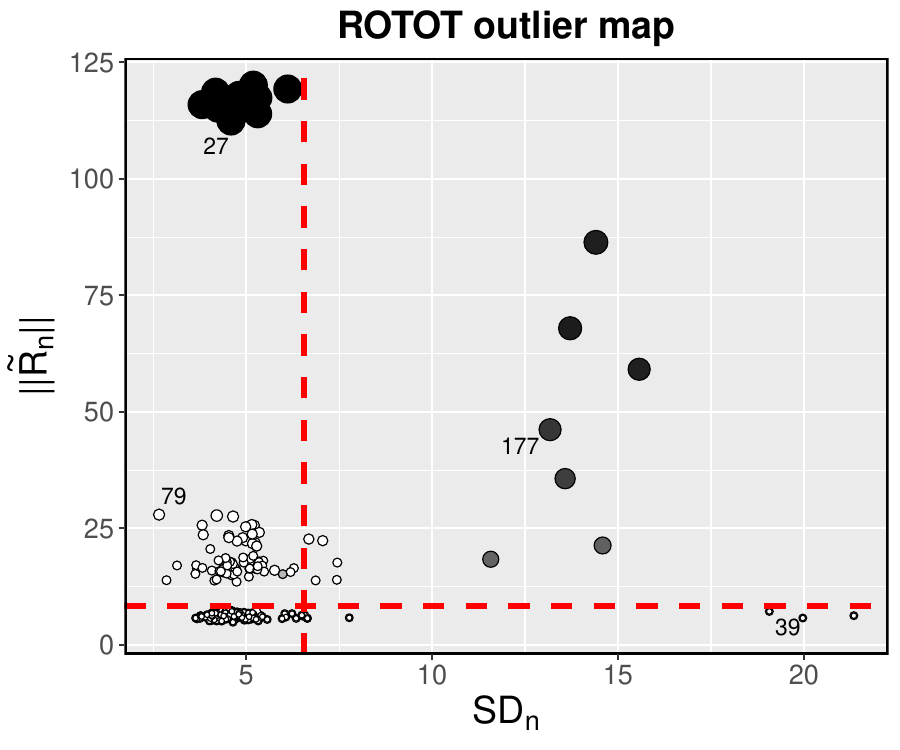}
\end{center}
\vspace{-0.5cm}
\caption{Outlier map of ROTOT applied to 
a simulated dataset with different types of outliers.}
\label{fig:Simulation_outmap}
\end{figure}

The residual cellmap focuses solely on cellwise deviations in the residual tensors. To also identify anomalies in the predictors, 
we propose a new outlier map based on the ROTOT and ROMPCA outputs. This tool is inspired by the regression outlier map as described in \cite{Hubert:ReviewHighBreakdown} and the enhanced PCA outlier map from \cite{Centofanti:cellpca}. Figure~\ref{fig:Simulation_outmap} shows the outlier map from the simulated dataset of Figure~\ref{fig:Simulation_UnfResMap}. It displays the residual distance of each observation, defined as $\|\widetilde{\mathcal{R}}_n\|_F$, versus its score distance 
$\text{SD}_n$. This score distance measures the outlyingness of the predictor 
using its vectorized core tensor as $\text{SD}_n = \sqrt{\text{vec}(\widehat{\core}_n^x)^{T}\widehat{\mathbf{\Sigma}}_u^{-1}\text{vec}(\widehat{\core}_n^x)}$,
where $\widehat{\mathbf{\Sigma}}_u$ is the  Minimum Regularized Covariance Determinant \citep{Boudt:RMCD} estimate of the covariance of the $\text{vec}(\widehat{\core}_n^x)$. The size of each point is proportional to the Percentage of Outlying Cells (POC) in the residual tensor, which is defined as 
\begin{equation*}
    \text{POC}_n = \frac{1}{Q} \sumq I \left(|\trs| > c_{\text{\tiny cell}} \right),
\end{equation*}
where $I$ is the indicator function \citep{Hubert:MacroPARAFAC} and $Q=\prod_{m=1}^{M} Q_m$. Larger points thus correspond to observations with many outlying cells in the residual tensor. Finally, points are colored based on their standardized casewise deviation $\tilde{t}_n$. First we compute each $t_n$ from \eqref{eq:caseres} based on $\widetilde{\mathcal{R}}_n$ and all the $\tsgcell$. Then we compute their M-scale $\tilde{\sigma}_2$ which results in $\tilde{t}_n = t_n/\tilde{\sigma}_2$. Observations below a cutoff $c_{\text{\tiny case}}$ are colored white and  those above are shaded from light gray to black, with black indicating strong outliers. 
The cutoff $c_{\text{\tiny case}}$ is defined as the 0.99 quantile of the distribution of $\tilde{\boldsymbol{r}}$ for uncontaminated data. 
The red vertical line represents the cutoff $c_{\text{\tiny SD}} = \sqrt{\chi^2_{\prod_\ell K_\ell, 0.99}}$, and the red horizontal line indicates the cutoff $c_{\text{\tiny res}}$, defined as the 0.99 quantile of the distribution of the $\|\widetilde{\mathcal{R}}_n\|_F$, obtained via simulation. 

On this outlier map we spot different groups of outlying cases. Contamination can occur in the response only. These so-called \textit{vertical outliers} can be divided into two types. Casewise vertical outliers, such as case 27, have a large residual distance, a large standardized casewise deviation, a large number of outlying cells but no outlying score distance. Hence they stand out as black big circles in the left upper corner. Cases with only cellwise outliers  in the response (e.g.\ case 79) show up as the group of white circles with enlarged size and large residual distance but small SD. When the predictor is contaminated, the residual distance can still be small, giving rise to \textit{good leverage points} such as case 39.  \textit{Bad leverage points} (like case 177) on the other hand have an outlying predictor and do not fit the regression model well. 

\section{Real Data Example}
\label{sec:realdata}
The Labeled Faces in the Wild (LFW) dataset \citep{Learned:LFW} contains over 13,000 publicly available images collected from the internet, each depicting the face of an individual.
The images are unposed and show wide variation in lighting, image quality, angle, and other conditions. For any image, a corresponding set of 72 describable attributes measured on a continuous scale is available. 
These attributes are quantitative variables derived from characteristics of the individual, e.g., gender, ethnicity, age,  their expression (e.g., smiling, frowning, eyes open), and their accessories (e.g., glasses, make-up, jewelry). 
 
{Although more flexible nonlinear approaches, such as neural networks, may be better suited to capture the complexity of facial image features, we use the tensor-on-tensor regression model in~\eqref{eq:TOTmodel} as an illustrative application of ROTOT to predict attributes from facial images.}
 {To train the predictive model, each image is converted to grayscale and downsampled via image decimation to a resolution of $30 \times 30$ pixels.
We consider random subsets of $N =\lbrace 400, 1000, 2000\rbrace$ observations  from unique individuals. Note that $N=1000$ matches the sample size used in \cite{lock:TOT}.}
  The ROTOT method is compared with the TOT regression method. The parameters $\lambda$ and $R$ are chosen by $5$-fold cross-validation as described in Section \ref{sec:tuning}.
To evaluate estimation performance while accounting for potential outliers in the data, we use the trimmed  Mean Squared Prediction Error referred to as robMSPE. Consider a set of $N_v$ response tensors $\cy_v$, the corresponding predictions  $\widehat{\cy}_v$, and the residual tensors $\mathcal{R}_v =\cy_v-\widehat{\cy}_v$.
Then, the robMSPE is defined as
\begin{equation*}
\text{robMSPE} = \frac{1}{HQ} \sum_{h = 1}^H \sumq (r^*_{h,q_1 \ldots q_M})^2,
\end{equation*}
where $|r^*_{1,q_1 \ldots q_M}| \leqslant \dots \leqslant |r^*_{H,q_1 \ldots q_M}|$ are the $H = \lceil 0.75N_v \rceil$ smallest absolute residuals among $r_{1,q_1 \ldots q_M},\dots,r_{v,q_1 \ldots q_M}$, and $Q=\prod_{m=1}^{M} Q_m$.

\begin{figure}[h]
\begin{center}
   \includegraphics[width=0.7\textwidth]{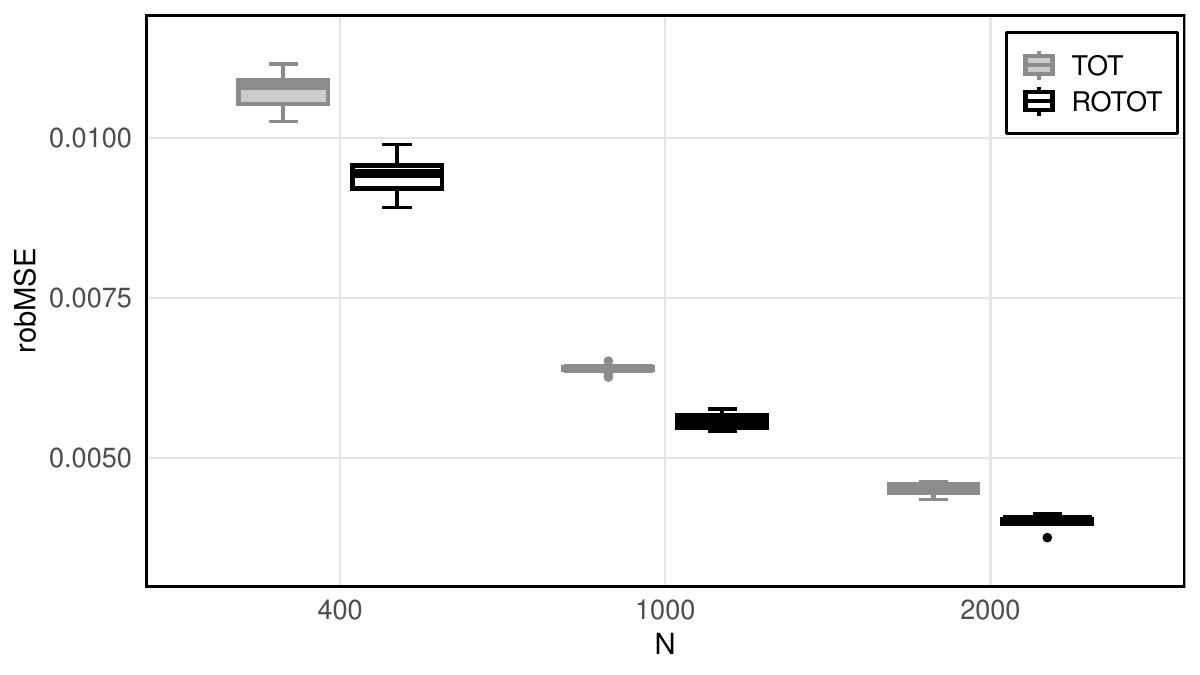} %
\end{center}
\vspace{-0.5cm}
\caption{{Boxplots of the robMSPE for the LFW dataset comparing TOT and ROTOT with
$N =\lbrace 400, 1000, 2000\rbrace$.}}
\label{fig:LFW_Boxplot}
\end{figure}

 {The robMSPE is computed using a 10-fold cross-validation procedure. For each considered sample size, the data are partitioned into 10 subsets, of which 9 are used for training and one for validation. For example, when $N=400$, this corresponds to $360$ training observations and $40$ validation observations. ROTOT and TOT are applied to each training set, and the robMSPE is evaluated on the corresponding validation set. This process is repeated over all folds, and the resulting 10 robMSPE values are summarized separately for each sample size in Figure~\ref{fig:LFW_Boxplot}. The results indicate that ROTOT not only achieves a lower median robMSPE than TOT, but also consistently outperforms it across all folds. As expected, the robMSPE also decreases as the sample size increases.}

\begin{figure}[h]
\begin{center}
\includegraphics[width=0.4\textwidth]{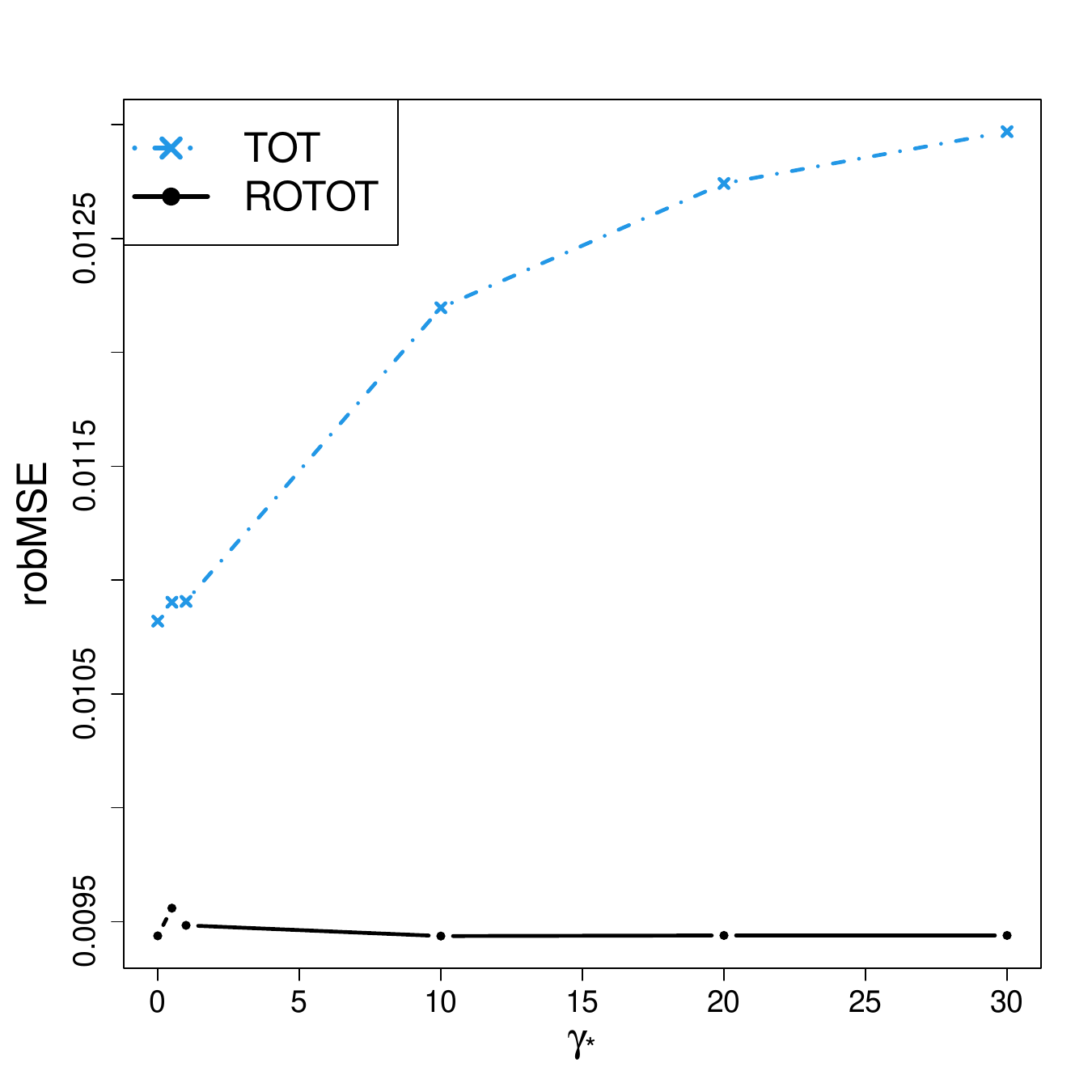}
\end{center}
\vspace{-0.5cm}
\caption{robMSPE over different values of $\gamma^*$ of the Labeled Faces in the Wild Data.}
\label{fig:LFW_Simulation}
\end{figure}

 {To further evaluate the performance of our method, artificial outliers are introduced into the dataset with $N=400$, providing a more challenging setting in which the strength of ROTOT can be more clearly assessed.} 
Specifically, cellwise outliers are introduced by replacing 5\% of the cells in $\cbr{\cy_n}$ with fixed values $\gamma^* \in \cbr{0.5,\ 1,\ 10,\ 20,\ 30}$. The  median robMSPE for different values of $\gamma^*$ for both TOT and ROTOT are displayed in Figure~\ref{fig:LFW_Simulation}. For $\gamma^*=0$, we report the median robMSPE on the actual dataset. As expected, the performance of ROTOT remains consistently good across all levels of contamination, whereas the accuracy of TOT decreases as the magnitude of contamination increases.

\begin{figure}[h]
\begin{center}
\includegraphics[width=0.8\textwidth]{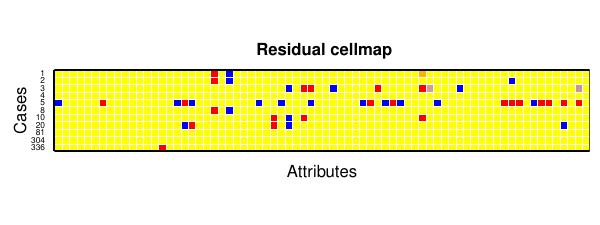}
\end{center}
\vspace{-1.5cm}
\caption{ROTOT residual cellmap for a subset of the observations in the LFW dataset.}
\label{fig:LFW_UnfResMap}
\end{figure}

\begin{figure}[h]
 \begin{center}
\includegraphics[width=0.5\textwidth]{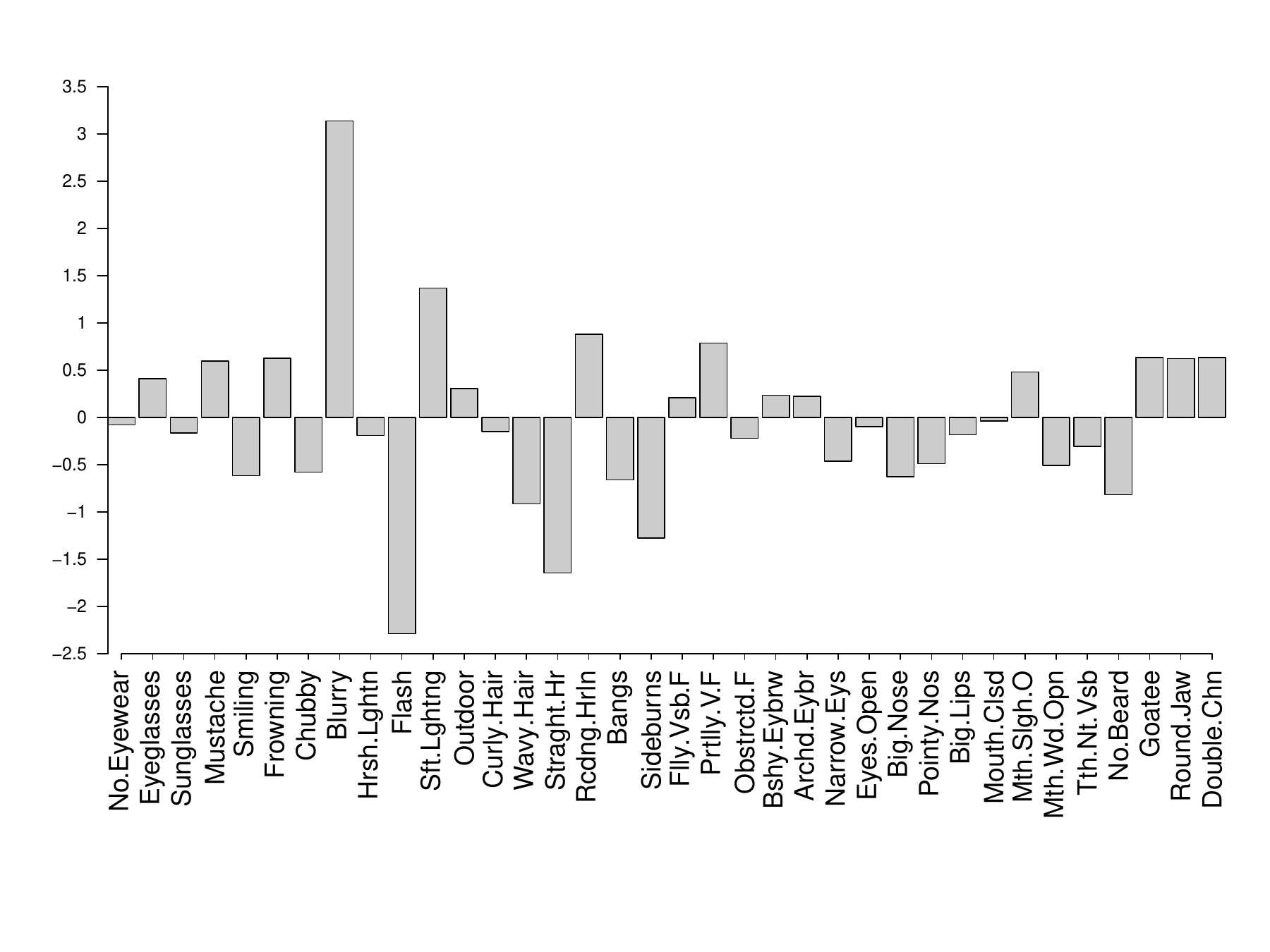}
\hspace{1cm}
\includegraphics[width=0.4\textwidth]{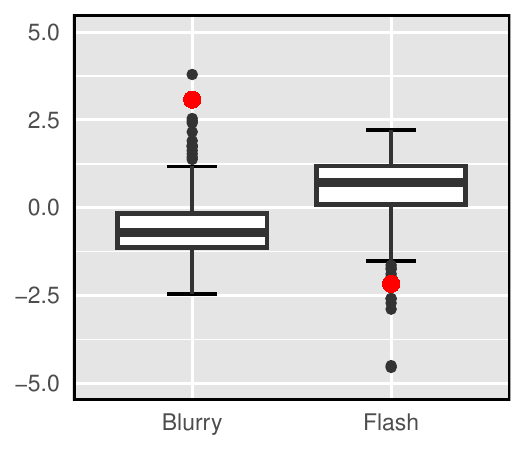}
\end{center}
\vspace{-0.5cm}
\caption{Attributes for observation 8 (left) and boxplots of the attributes ``Blurry'' and ``Flash'' (right) in the LFW dataset.}
\label{fig:LFW_celloutl_indv}
\end{figure}

Figure~\ref{fig:LFW_UnfResMap} presents the residual cellmap produced by ROTOT for a subset of the observations. In this plot, we observe the presence of several cellwise outliers. In particular, in case 8, ``Blurry'' and ``Flash'' stand out as outlying attributes, with ``Blurry'' showing a large positive residual and ``Flash'' a large negative one. 
The left panel of Figure~\ref{fig:LFW_celloutl_indv} displays the standardized response for observation 8.
The bar plot reveals high marginal values for the attributes ``Blurry'' and ``Flash'', with a large positive value for ``Blurry'' and a large negative value for ``Flash''.
However, determining whether these values correspond to cellwise outliers requires comparison with the distribution of the attributes across all observations.
The right panel of Figure~\ref{fig:LFW_celloutl_indv} presents boxplots of the attributes ``Blurry'' and ``Flash''. The red point indicates the value for observation 8. It can be seen that these values lie in the extreme tails of the distributions, confirming that they correspond to unusually large marginal values for these attributes.


\begin{figure}[h]
\begin{center}
\includegraphics[width=0.65\textwidth]{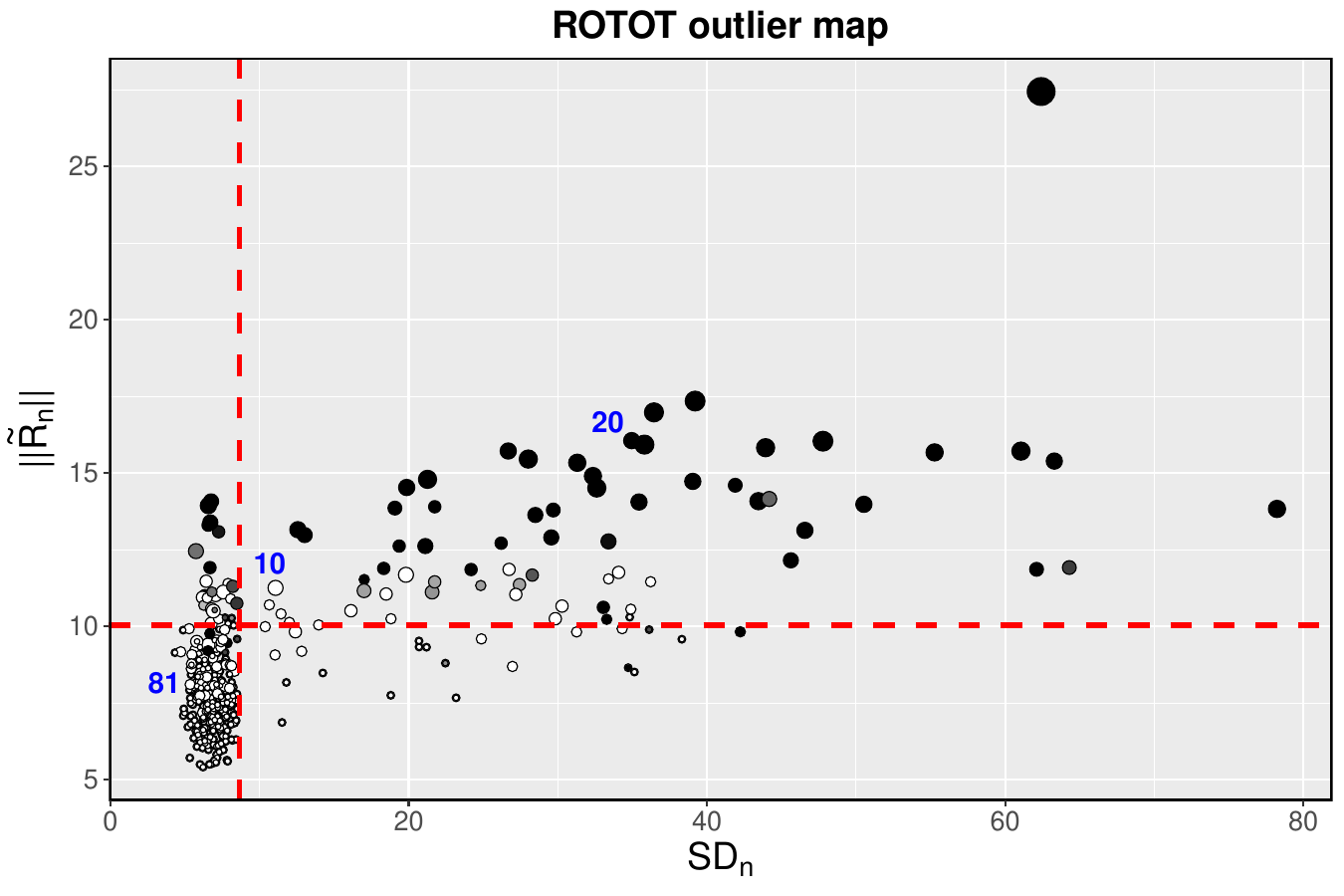}
\end{center}
\vspace{-0.5cm}
\caption{ROTOT outlier map for the LFW dataset.}
\label{fig:LFW_outmap}
\end{figure}

 {The outlier map in Figure~\ref{fig:LFW_outmap} provides further insight into the nature of the outlyingness in the predictor and response tensors. We focus on the three observations whose face images are shown in Figure~\ref{fig:LFW_obs_comp}. The left image corresponds with observation~81. It has a small score distance and a regular residual distance as it exhibits no outlying cells in the residual cellmap in Figure~\ref{fig:LFW_UnfResMap}. In contrast, observation 10 (the middle image) has a mildly deviating score and residual distance, whereas observation 20 (the right image) is highly outlying. Moreover, it is displayed in black, indicating a casewise outlyingness.}

\begin{figure}[h]
\begin{center}
\includegraphics[width=0.85\textwidth]{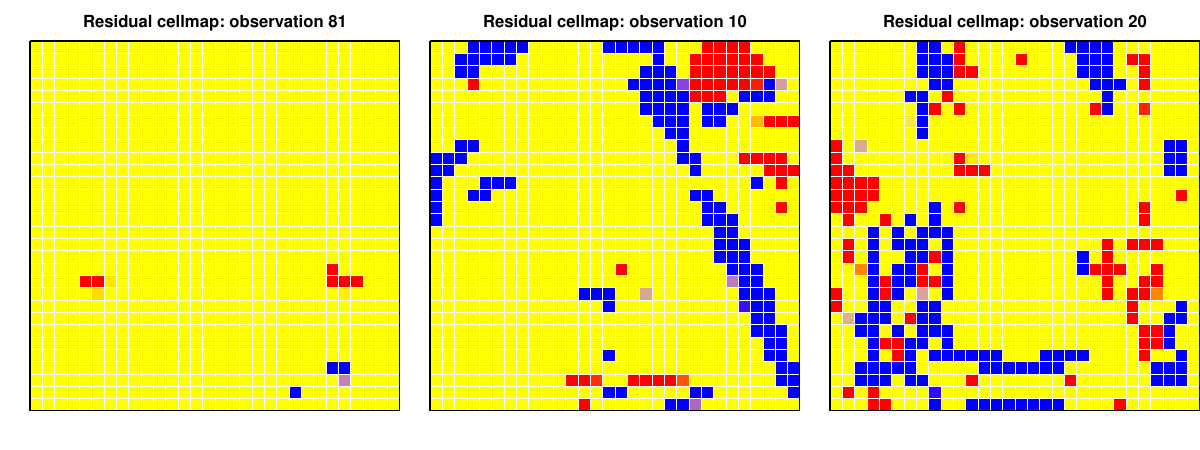}
\end{center}
\vspace{-0.7cm}
\caption{ROMPCA residual cellmap for observations 81 (left), 10 (middle), and 20 (right).}
\label{fig:LFW_ROMPCA_resmap}
\end{figure}

{To further investigate the source of outlyingness in the images $\cx_n$, Figure~\ref{fig:LFW_ROMPCA_resmap} presents the ROMPCA residual cellmaps for the selected observations. These maps visualise the deviations $\cx_n - \widehat{\cx}_n$ as described in \cite{Hirari:ROMPCA}. The colors have an interpretation similar to that of the ROTOT residual cellmap. Image 81 has a largely clean residual cellmap, with almost no cells flagged as outlying. For image 10, the golf club appears to be the main source of outlyingness, as the largest residuals coincide with the region obscured by the club. Similarly, for image 20, its large score distance can be attributed to the hair covering large parts of the face.} 

\section{Conclusion}
\label{sec:disc}
We introduced a novel robust tensor-on-tensor (ROTOT) regression method that is the first approach capable of simultaneously addressing casewise outliers, cellwise outliers, and missing data in both the response and predictor tensors. Its objective function combines two robust loss functions to reduce the influence of cellwise and casewise outliers in the response tensor. The imputed tensor and the weights obtained from the ROMPCA method make it possible to address cellwise and casewise contamination in the predictor tensor. 
ROTOT estimates are obtained via an iteratively reweighted least squares algorithm, which yields standardized cellwise residuals used to construct the residual cellmap for detecting cellwise outliers in the residual tensor. In addition, the combination of these deviations with the score distances from ROMPCA produces an outlier map that visualizes both casewise and cellwise outliers in the response, as well as good and bad leverage points. Also robust predictions for possibly contaminated new predictors are obtained. The favorable performance of ROTOT is demonstrated through extensive simulations and an application to predict people's attributes from facial images using the Labeled Faces in the Wild dataset.
 
{ROTOT can simultaneously handle casewise and cellwise outliers in both the predictor and response tensors. However, in particular applications contamination may exhibit a structure that lies between these two extremes. We refer to such intermediate forms as structured outliers, an example is discussed in \cite{Mou:StrData}. Section~\ref{app:stroutl} of the Supplementary Material evaluates the performance of ROTOT under this contamination setting. Although ROTOT performs well in the presence of structured outliers, methods specifically designed to address this type of contamination may yield further improvements. Developing such methods is an interesting direction for future research.}

\vspace{3mm}
\noindent{\bf Software availability.} The R code
that reproduces the example is
available at \url{https://wis.kuleuven.be/statdatascience/robust/software}.

\vspace{3mm}
\noindent{\bf Data Availability.} 
The Labeled Faces in the Wild dataset used in Section~\ref{sec:realdata} is provided in the Supplementary Materials and can be downloaded 
 from \url{https://doi.org/10.6084/m9.figshare.31819423}.

\vspace{3mm}
\noindent{\bf Supplementary Materials.} These consist 
of a text with additional material, as well as R code 
for the proposed method and a script that reproduces 
the example. 

\vspace{3mm}
\noindent{\bf Disclosure Statement.} The authors 
report there are no competing interests to declare.

\bibliographystyle{chicago}
\setlength{\bibsep}{5pt plus 0.2ex}
{\small
\spacingset{1}
\bibliography{mybibliography}
}

\end{document}


\def\spacingset#1{\renewcommand{\baselinestretch}
{#1}\small\normalsize} \spacingset{1}

\newcommand{\blind}{0}

\if0\blind
{
\title{\bf Supplementary Materials to: \\ {Casewise and Cellwise Robust\\ Tensor-on-Tensor Regression}}
\author[1]{Mehdi Hirari}
\author[1]{Fabio Centofanti}
\author[1]{Mia Hubert}
\author[1]{Stefan Van Aelst}
\affil[1]{Section of Statistics and Data Science, Department 
          of Mathematics, KU Leuven, Belgium}
\setcounter{Maxaffil}{0}
\renewcommand\Affilfont{\itshape\small}
\date{}         
  \maketitle
} \fi

\if1\blind
{
  \bigskip
  \bigskip
  \bigskip
  \begin{center}
    {\LARGE\bf Supplementary Materials to: \\ {Casewise and Cellwise Robust\\ \vspace{0.35cm}Tensor-on-Tensor Regression}}
\end{center}
  \medskip
} \fi

\spacingset{1.7}

\renewcommand{\thesection}{\Alph{section}}
\renewcommand\thefigure{S.\arabic{figure}}   
\renewcommand{\theequation}{S.\arabic{equation}} 
\setcounter{figure}{0}    
\setcounter{equation}{0}

\section{Robust Multilinear Principal Component
Analysis}
\label{app:Def_ROMPCA}
This section describes the Robust MPCA (ROMPCA) method used to address outliers in the predictors. For each given $\mathcal{X}_n$, the tensor $\mathcal{O}_n^x=[o_{n,p_1 \ldots p_L}^x] \in \mathbb{R}^{P_1 \times \cdots \times P_L}$ has value $o_{n,p_1 \ldots p_L}^x=1$ if $x_{n,p_1 \ldots p_L}$ is observed, and 0 otherwise. The scalar $o_{n}^x = \sump o_{n,p_1 \ldots p_L}^x$ denotes the number of observed cells in $\mathcal{X}_n$ and $o^x = \sum_{n=1}^N o_n^x$ the total number of observed cells in the set $\{\mathcal{X}_n\}$. ROMPCA estimates a set of core tensors $\{\core_n^x = [u_{n,k_1\ldots k_L}^x] \in \mathbb{R}^{K_1 \times \cdots \times K_L}\}_{n=1}^N$, projection matrices $\{\bV_\ell^x = [v_{p_\ell k_\ell}^{x(\ell)}] \in \mathbb{R}^{P_\ell \times K_\ell}\}_{\ell=1}^L$, and a robust center $\mathcal{C}^x = [c_{p_1 \ldots p_L}^x] \in \mathbb{R}^{P_1 \times \cdots \times P_L}$ by minimizing 
\begin{equation}
\label{eq:obj}
\frac{\sgcasexs}{o^x} \sum_{n=1}^N \ o_n^x \rho_2\left(\frac{1}{\sgcasex}\sqrt{\frac{1}{o_n^x} \sump  o_{n,p_1 \ldots p_L}^x \sgcellxs \, \rho_1\left(\frac{r_{n,p_1 \ldots p_L}^x}{\sgcellx}\right)}\ \right),\\
\end{equation}
where the cellwise residuals are given by
\begin{equation*} 
r_{n,p_1 \ldots p_L}^x:= {x}_{n,p_1 \ldots p_L}- c_{n,p_1 \ldots p_L}^x -  \sum_{k_1 \ldots k_L}^{K_1 \ldots K_L} {u}_{n,k_1\ldots k_L}^x v_{p_1 k_1}^{x(1)} \cdots v_{p_L k_L}^{x(L)}\,.
\end{equation*}
Both $\rho_1$ and $\rho_2$ are chosen to be the hyperbolic tangent, see Section~\ref{app:rho_func} of the Supplementary Material.
Each of the scales $\sgcellx$ standardizes the corresponding cellwise residuals $r_{n,p_1 \ldots p_L}^x$ while  the scale $\sgcasex$  standardizes the casewise deviations
\begin{equation*} 
t_{n}^x := \sqrt{\frac{1}{o_n^x} \sump o_{n,p_1 \ldots p_L}^x \sgcellxs \rho_1\left(\frac{r_{n,p_1 \ldots p_L}^x}{\sgcellx}\right)}.
\end{equation*}
The objective in~\eqref{eq:obj} is minimized by an iteratively reweighted least squares algorithm, and the scales are computed as M-scales, as detailed in \cite{Hirari:ROMPCA}. The resulting estimates are denoted as $\widehat{\core}_n^x$, $\widehat{\bV}_\ell^x$ and $\widehat{\mathcal{C}}^x$.

For each tensor $\cx_n$, ROMPCA outputs a casewise weight $\wcasex$ and a tensor with cellwise weights $\Wcellx_n = \sbr{\wcellx}$, all taking values between 0 (outlying) and 1 (regular). 
The casewise weight reflects how large the standardized casewise deviation is, but it does not measure the degree of outlyingness of the core tensor $\widehat{\core}_n$.  
For this, we compute 
the score distance 
as
\begin{equation*}
\text{SD}_n = \sqrt{\text{vec}(\widehat{\core}_n^x)^{T}\widehat{\mathbf{\Sigma}}_u^{-1}\text{vec}(\widehat{\core}_n^x)},
\end{equation*}
where $\widehat{\mathbf{\Sigma}}_u$ is the Minimum Regularized Covariance Determinant \citep{Boudt:RMCD}  estimate of the covariance of the $\text{vec}(\widehat{\core}_n^x)$. 
 Following \cite{Simpson:One-GM}, we define the casewise weight based on the score distance as $w_n^u = \min \sbr{1, \rbr{{b}\big /{\text{SD}_n^2}}^{1/2}}$, where $b$ is chosen as the $(1-\gamma)$-quantile of a chi-squared distribution with $\prod_\ell K_\ell$ degrees of freedom, with $\gamma=0.01$. The overall casewise weight is then given by $w_n^x = \wcasex w_n^u$.
 
The imputed tensor $\cx^{\text{\tiny imp}}_n$ is obtained as
\begin{equation*}
\cx_n^{\text{\tiny imp}} := \widehat{\cx}_n + \widetilde{\cw}_n^{x} \odot \rbr{\cx_n - \widehat{\cx}_n},
\end{equation*}
where $\widehat{\cx}_n = \widehat{\mathcal{C}}^x + \dbr{\widehat{\core}_n^x; \widehat{\mathbf{V}}_1^x, \ldots, \widehat{\mathbf{V}}_L^x}$ and $\widetilde{\cw}_n^{x} = \Wcellx_n \odot \co_n^x$.

Given a new tensor $\cx_*$, its 
estimated core tensor $\widehat{\core}_*^x $ is obtained by minimizing the inner part of the objective \eqref{eq:obj} with respect to $\core_x= \sbr{{u}_{k_1 \ldots k_L}^x}$, given the estimated center and projection matrices, that is
\begin{equation*}
    \sump o_{*,p_1 \ldots p_L}^x \sgcellxs \rho_1 \rbr{\frac{x_{*,p_1 \ldots p_L} - \hat{c}_{p_1 \ldots p_L}^x - \sum_{k_1 \ldots k_L}^{K_1 \ldots K_L} {u}_{k_1 \ldots k_L}^x \hat{v}_{p_1 k_1}^{x(1)} \cdots \hat{v}_{p_L k_L}^{x(L)}}{\sgcellx}},
\end{equation*}
where $o_{*,p_1 \ldots p_L}^x$ are the observed indicators of $\cx_*$. The tensor ${\widetilde{\mathcal{W}}_*^x} = \sbr{w_{*, p_1 \ldots p_L}^{x}}$ stores the cellwise weights resulting from the minimization multiplied by the observed indicator tensor $\co_*$.
The
imputed version of $\cx_*$ is then constructed as
$\cx_*^{ \text{\tiny imp}} := \widehat{\cx}_* + \widetilde{\mathcal{W}}_*^x \odot \rbr{\cx_* - \widehat{\cx}_*}$
where $\widehat{\cx}_* = \widehat{\mathcal{C}}^x + \dbr{\widehat{\core}_*^x; \widehat{\mathbf{V}}_1^x, \ldots, \widehat{\mathbf{V}}_L^x}$. 

\section{Selection of Robust Loss Functions}
\label{app:rho_func}
\begin{figure}[H]
    \centering
    \includegraphics[width=0.49\textwidth]{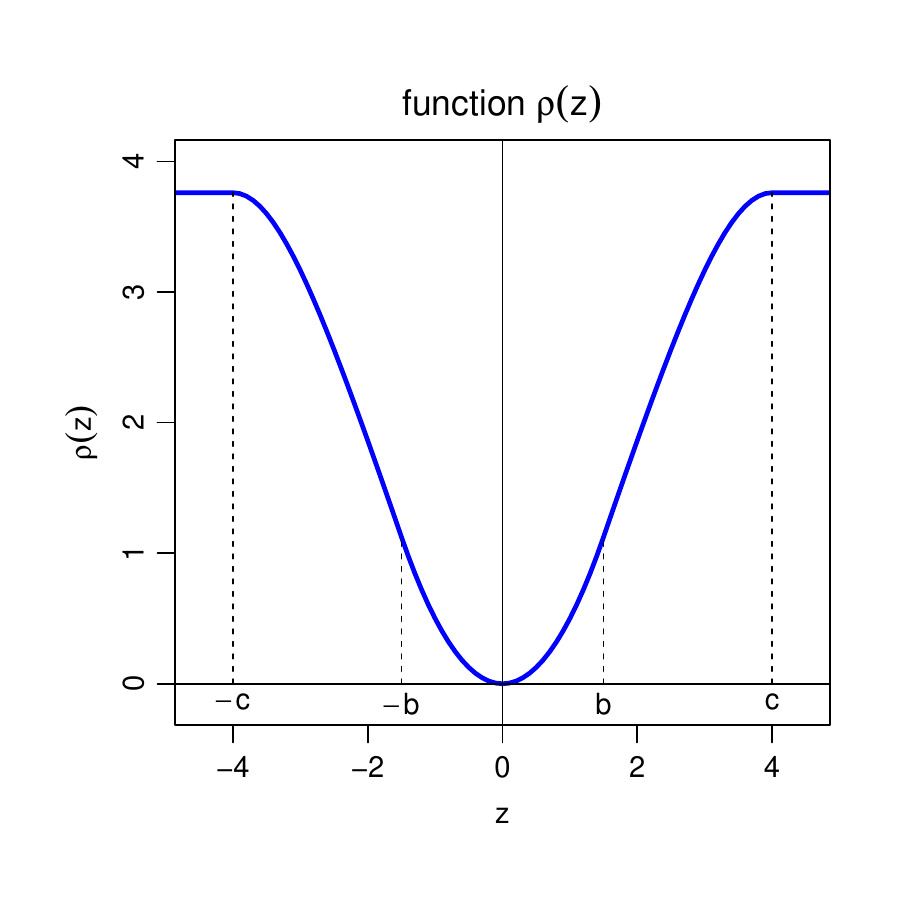} %
    \includegraphics[width=0.49\textwidth]{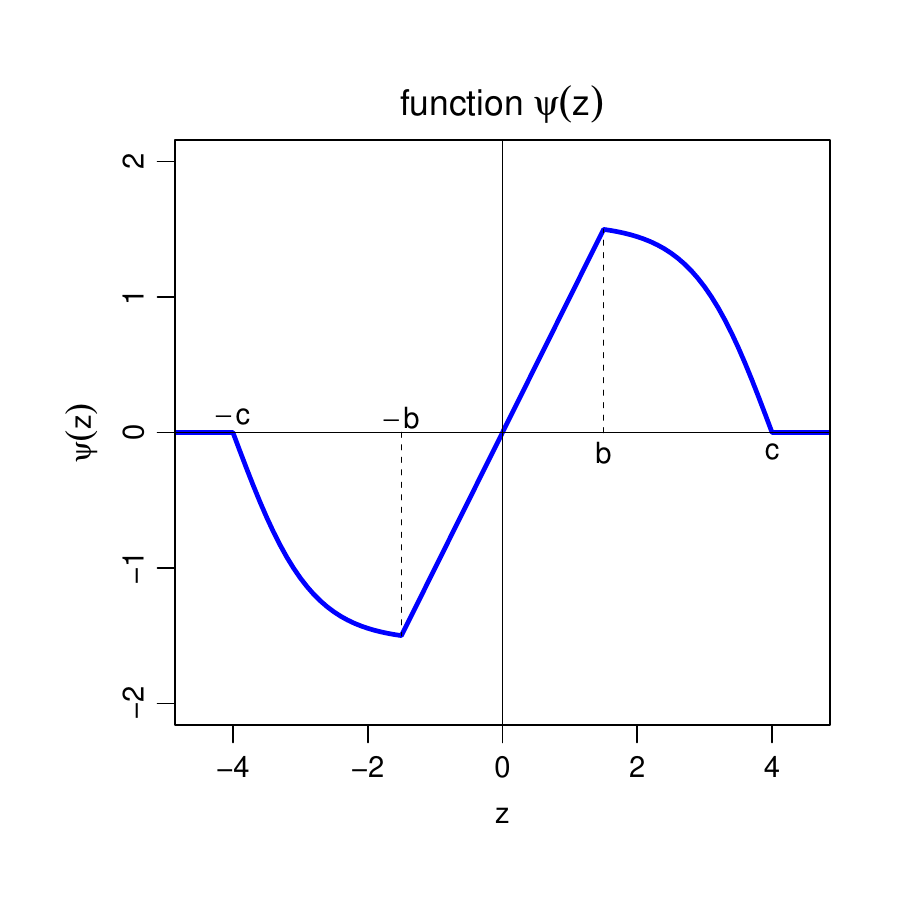} %
    \vspace{-0.5cm}
    \caption{The function $\rho_{b,c}$ with $b = 1.5$ and $c = 4$ (left) and its derivative $\psi_{b,c}$ (right).}%
    \label{fig:HypTan}%
\end{figure}

In this paper both $\rho$-functions in~\eqref{eq:robustWLS} and in~\eqref{eq:obj} are chosen to be the hyperbolic tangent $(tanh)$ \citep{Hampel:CVC}, defined by
\begin{equation*}
\rho_{b,c}(z) = 
\begin{cases} 
    z^2/2 & \text{if } 0 \leqslant |z| \leqslant b, \\
    d - \left(q_1/q_2\right) \ln\left(\cosh\left(q_2\left(c - |z|\right)\right)\right) & \text{if } b \leqslant |z| \leqslant c, \\
    d & \text{if } c \leqslant |z|,
\end{cases}
\end{equation*}
where $d = \left( {b^2}/{2} \right) + \left( q_1/q_2 \right) \ln\left(\cosh\left(q_2 (c - b)\right)\right)$. Its derivative, denoted by $\psi_{b,c} = \rho'_{b,c}$, is
\begin{equation*}
\psi_{b,c}(z) = 
\begin{cases} 
    z & \text{if } 0 \leqslant |z| \leqslant b, \\
    q_1 \tanh\left(q_2\left(c - |z|\right)\right) \, \text{sign}(x) & \text{if } b \leqslant |z| \leqslant c, \\
    0 & \text{if } c \leqslant |z|.
\end{cases}
\end{equation*}
The hyperbolic tangent $\rho$-function and its derivative are shown in Figure~\ref{fig:HypTan} for ${b=1.5}$, ${c=4}$, ${q_1=1.540793}$, and ${q_2=0.8622731}$. These choices offer a good balance between efficiency and robustness, as discussed by~\cite{Raymaekers:FastCorr} and empirically studied in Section~H of the Supplementary Material of \cite{Hirari:ROMPCA}.
This corresponds approximately to an efficiency of  95\% and a gross-error sensitivity of $ 1.78$, where it measures the local robustness of the estimator. The smaller its value, the less sensitive the estimator is to infinitesimal contamination.
Since $\psi_{1.5,4}(z) = 0$ for $|z| \geqslant 4$, large outlying cells and cases will be completely downweighted in the estimation procedure.
Note that while any redescending $\rho$-function could serve as a valid choice and would yield comparable performance, the hyperbolic tangent is preferred due to its theoretical optimality among redescending M-estimators. It is specifically designed to maximize efficiency for a given level of robustness, according to the change-of-variance curve criterion introduced by \cite{Hampel:CVC}.

 {The robustness of $\widehat{\sigma}$, given by~\eqref{eq:Mscale}, depends on the value of $\delta$. The \emph{explosion breakdown value} of $\widehat{\sigma}$ is the smallest percentage of outliers in the dataset that can make $\widehat{\sigma}$ arbitrarily large, and equals $\delta/\max(\rho)$. On the other hand, the \emph{implosion breakdown value} of $\widehat{\sigma}$ is the smallest percentage of outliers that can bring $\widehat{\sigma}$ arbitrarily close to zero, and equals $(\max(\rho)-\delta)/\max(\rho)$. Choosing $\delta=\max(\rho)/2$ balances these two breakdown values and maximizes their minimum, yielding a breakdown value of $50\%$.}

 {The constant $a$ in \eqref{eq:Mscale} is introduced to obtain consistency of the estimator at the Gaussian distribution. It is determined such that $E[\rho(Z/a)]=\delta$ for $Z\sim N(0,1)$. In the ROTOT algorithm, we use the M-scale based on the function $\rho_{1.5,4}$, shown in the top left panel of Figure~\ref{fig:HypTan}. Since $\max(\rho)=3.7622$, we set $\delta=1.8811$, which gives $a=0.3431$.}
 {Note that $a$ can be calibrated for other reference distributions. However, such calibration is not straightforward, since the underlying true distribution is generally unknown in advance. 
Section~\ref{app:simres} further investigates the impact of  $\delta$ and $a$ on the ROTOT performance.}

\clearpage
\section{First-Order Conditions}
\label{app:firstorder}
In the following, the derivation of the first-order necessary equations is presented. Following~\eqref{eq:robustWLS} the objective function $\cl \rbr{\cbr{\cx_n}, \cbr{\cy_n}, \bUbV, \cb_0}$ of ROTOT is given by
\begin{equation*}
\frac{\sgcase^2}{o} \sum_{n=1}^N o_n \woutl \rho_2\rbr{\frac{1}{\sgcase}\sqrt{\frac{1}{o_n}\sumq \ocell \sgcell^2 \rho_1\rbr{\frac{\rs}{\sgcell}}}} + \lambda \abr{\cb}^2_F.
\end{equation*}
The derivative of $\cl$ with respect to $u^{(\ell)}_{p_\ell r}$ is:
\begin{align*}
\frac{\partial \cl}{\partial u^{(\ell)}_{p_\ell r}}
&= \frac{\sgcase^2}{o}  \sum_{n=1}^N o_n \woutl \rho'_2\left( t_n/\sgcase \right) \frac{1}{\sgcase}\frac{\partial}{\partial u^{(\ell)}_{p_\ell r}} \sqrt{\frac{1}{o_n} \sumq \ocell \sgcell^2\, \rho_1\left(\frac{\rs}{\sgcell}\right)} \\&\hspace{7cm} + \lambda \frac{\partial}{\partial u^{(\ell)}_{p_\ell r}} \sum_{p_1 \cdots q_M}^{P_1 \cdots Q_M} \rbr{\sum_{r}^{R} u_{p_1r}^{(1)} \cdots v_{q_Mr}^{(M)}}^2 \nonumber \\ 
&=\frac{1}{2o} \sum_{n=1}^{N} \woutl \frac{\psi_2\left( t_n/\sgcase\right)}{t_n/\sgcase} \frac{\partial}{\partial u_{p_\ell r}^{(\ell)}} \left(\sumq \ocell \sgcell^2 \rho_{1}\left(\frac{\rs}{\sgcell} \right)\right) \\&\hspace{7cm} + \lambda \sum_{p_1 \cdots q_M}^{P_1 \cdots Q_M} \frac{\partial}{\partial u^{(\ell)}_{p_\ell r}} \rbr{\sum_{r=1}^{R} u_{p_1r}^{(1)} \cdots v_{q_Mr}^{(M)}}^2 \nonumber \\ 
&= \frac{1}{2o} \sum_{n=1}^{N} \woutl \wcase \,  \sumq \sgcell^2\, \psi_1(\rs/\sgcell) \frac{\ocell}{\sgcell} \frac{\partial \rs}{\partial u_{p_\ell r}^{(\ell)}} \\&\hspace{4cm} + 2\lambda \sumpl \sumq \rbr{\sum_{r=1}^{R} 
u_{p_1r}^{(1)} \cdots v_{q_Mr}^{(M)}} \beta_{p_1 \ldots p_{\ell-1} \, r \, p_{\ell+1}\ldots p_L q_1 \ldots q_M}^{(-\ell)} 
\nonumber\\
&= -\frac{1}{2o}\sum_{n=1}^{N}  \sumq \Big[\woutl \wcase  \wcell \ocell \rs\\&\hspace{4cm} \frac{\partial}{\partial u_{p_\ell r}^{(\ell)}} \sum_{r=1}^{R} \sum_{p_{\ell=1}}^{P_{\ell}} \rbr{\sumpl\left( \ximp \beta_{p_1 \ldots p_{\ell-1} \, r \, p_{\ell+1} \ldots p_L q_1 \ldots q_M}^{(-\ell)} \right)}u_{p_\ell r}^{(\ell)}\Big] \\&\hspace{4cm} + 2\lambda \sumpl \sumq \rbr{\sum_{r=1}^{R} 
u_{p_1r}^{(1)} \cdots v_{q_Mr}^{(M)}} 
\beta_{p_1 \ldots p_{\ell-1} \, r \, p_{\ell+1}\ldots p_L q_1 \ldots q_M}^{(-\ell)} 
\nonumber \\
 &= -\frac{1}{2o}\sum_{n=1}^{N}  \sumq \Big[\woutl \wcase  \wcell \ocell \rs \sumpl \left( \ximp b_{p_1 \ldots p_{\ell-1} \, r \, p_{\ell+1} \ldots p_L q_1 \ldots q_M}^{(-\ell)} \right)\Big] \\&\hspace{4cm} + 2\lambda \sumpl \sumq \rbr{\sum_{r=1}^{R} u_{p_1r}^{(1)} \cdots v_{q_Mr}^{(M)}}\beta_{p_1 \ldots p_{\ell-1} \, r \, p_{\ell+1} \ldots p_L q_1 \ldots q_M}^{(-\ell)}, \nonumber\\
\end{align*}
where $\beta_{p_1 \ldots p_{\ell-1} \, r \, p_{\ell+1} \ldots p_L q_1 \ldots q_M}^{(-\ell)}=u_{p_1r}^{(1)} \cdots u_{p_{\ell-1}r}^{(\ell-1)}u_{p_{\ell+1}r}^{(\ell+1)}\cdots u_{p_Lr}^{(L)}  v_{q_1r}^{(1)} \cdots v_{q_Mr}^{(M)}$ and
 $L^{\star}_{\ell} = \{1 \ld \ell-1, \ell +1 \ld L\}$. This leads to~\eqref{eq:condi1}:
\begin{equation*}
\frac{\partial \cl}{\partial \bU_\ell} = \sum_{n=1}^N \left\langle \rbr{\cy_n - \cb_0 - \langle \cx_n^{\text{\tiny imp}}, \cb \rangle_{\{P_\ell\}}} \odot \cw_n, \cc_n \right\rangle_{\{Q_m\}} - (4\lambda m) \bU_\ell \, \mathbf{T}^{(-\ell)}_{\bU} = \mathbf{0}_{P_\ell \times R}.
\end{equation*}

The derivative of $\cl$ with respect to $v^{(m)}_{q_m r}$ is:
\begin{align*}
\frac{\partial \cl}{\partial v^{(m)}_{q_m r}}
&= \frac{\sgcase^2}{o}  \sum_{n=1}^N o_n \woutl \rho'_2\left( t_n/\sgcase \right) \frac{1}{\sgcase}\frac{\partial}{\partial v^{(m)}_{q_m r}} \sqrt{ \frac{1}{m_n }\sumq  \ocell \sgcell^2\, \rho_1\left(\frac{\rs}{\sgcell}\right)} \\&\hspace{7cm} + \lambda \frac{\partial}{\partial v^{(m)}_{q_m r}} \sum_{p_1 \cdots q_M}^{P_1 \cdots Q_M} \rbr{\sum_{r=1}^{R} u_{p_1r}^{(1)} \cdots v_{q_Mr}^{(M)}}^2 \nonumber \\ 
&=\frac{1}{2o} \sum_{n=1}^{N} \woutl \frac{\psi_2\left( t_n/\sgcase\right)}{t_n/\sgcase} \frac{\partial}{\partial v^{(m)}_{q_m r}} \left(\sumq \ocell \sgcell^2 \rho_{1}\left(\frac{\rs}{\sgcell} \right)\right) \\&\hspace{7cm} + \lambda \sump \sumqm \frac{\partial}{\partial v^{(m)}_{q_m r}} \rbr{\sum_{r=1}^{R} u_{p_1r}^{(1)} \cdots v_{q_Mr}^{(M)}}^2 \nonumber \\ 
&=\frac{1}{2o} \sum_{n=1}^{N} \woutl \wcase \, \sumqm \sgcell^2\, \psi_1(\rs/\sgcell) \frac{\ocell}{\sgcell} \frac{\partial \rs}{\partial v^{(m)}_{q_m r}} \\&\hspace{4cm} + 2\lambda \sump \sumqm \rbr{\sum_{r=1}^{R} u_{p_1r}^{(1)} \cdots v_{q_Mr}^{(M)}} 
\beta_{p_1 \ldots p_L q_1 \ldots q_{m-1} \, r \, q_{m+1} \ldots q_M}^{(-m)}
\nonumber\\
&= -\frac{1}{2o}\sum_{n=1}^{N} \sumqm \Big[\woutl \wcase  \wcell \ocell \rs\\&\hspace{4cm} \frac{\partial}{\partial v^{(m)}_{q_m r}} \sum_{r=1}^{R} \rbr{\sump  \ximp \beta_{p_1 \ldots p_L q_1 \ldots q_{m-1} \, r \, q_{m+1} \ldots q_M}^{(-m)} }v^{(m)}_{q_m r}\Big] \\&\hspace{4cm} + 2\lambda \sum_{p_1 \cdots p_L}^{P_1 \cdots P_L} \sumqm \rbr{\sum_{r=1}^{R} u_{p_1r}^{(1)} \cdots v_{q_Mr}^{(M)}} 
\beta_{p_1 \ldots p_L q_1 \ldots q_{m-1} \, r \, q_{m+1} \ldots q_M}^{(-m)}
\nonumber \\
&= -\frac{1}{2o}\sum_{n=1}^{N} \sumqm \Big[\woutl \wcase  \wcell \ocell \rs \sump  \ximp \beta_{p_1 \ldots p_L q_1 \ldots q_{m-1} \, r \, q_{m+1} \ldots q_M}^{(-m)} \Big] \\&\hspace{4cm} + 2\lambda \sum_{p_1 \cdots p_L}^{P_1 \cdots P_L} \sumqm \rbr{\sum_{r=1}^{R} u_{p_1r}^{(1)} \cdots v_{q_Mr}^{(M)}} \beta_{p_1 \ldots p_L q_1 \ldots q_{m-1} \, r \, q_{m+1} \ldots q_M}^{(-m)}, \nonumber
\end{align*}
where $\beta_{p_1 \ldots p_L q_1 \ldots q_{m-1} \, r \, q_{m+1} \ldots q_M}^{(-m)}=u_{p_1r}^{(1)}\cdots u_{p_Lr}^{(L)}v_{q_{1}r}^{(1)} \cdots v_{q_{m - 1}r}^{(m - 1)}  v_{q_{m + 1}r}^{(m + 1)} \cdots v_{q_Mr}^{(M)}$ and $M^\star_m = \{1 \ld m-1, m+1 \ld M\}$.
This leads to~\eqref{eq:condi2}:
\begin{equation*}
\frac{\partial \cl}{\partial \bV_{m}} = \sum_{n=1}^N \left\langle \rbr{\cy_n - {\cb}_0 - \langle \cx_n^{\text{\tiny imp}}, \cb \rangle_{\{P_\ell\}}} \odot \cw_n, \cd_n \right\rangle_{Q^{\star}_m} - (4\lambda m) \bV_m \, \mathbf{T}^{(-m)}_{\bV} = \mathbf{0}_{Q_m \times R}.
\end{equation*}
The derivative of $\cl$ with respect to $\bint$ is:
\begin{align*}
\frac{\partial \cl}{\partial \bint}
&= \frac{\sgcase^2}{o}  \sum_{n=1}^N m_n \woutl \rho'_2\left( t_n/\sgcase \right) \frac{1}{\sgcase}\frac{\partial}{\partial \bint} \sqrt{\frac{1}{o_n} \sumq  \ocell \sgcell^2\, \rho_1\left(\frac{\rs}{\sgcell}\right)} \\ 
&=\frac{1}{2o} \sum_{n}^{N} \woutl \frac{\psi_2\left( t_n/\sgcase\right)}{t_n/\sgcase} \frac{\partial}{\partial \bint} \left(\sumq \ocell \sgcell^2 \rho_{1}\left(\frac{\rs}{\sgcell} \right)\right) \\ 
&= \frac{1}{2o} \sum_{n}^{N} \woutl \wcase \,   \sgcell^2\, \psi_1(\rs/\sgcell) \frac{\ocell}{\sgcell} \frac{\partial \rs}{\partial \bint} \nonumber\\
&= -\frac{1}{2o}\sum_{n}^{N} \woutl \wcase  \wcell \ocell \rs, \nonumber
\end{align*}
which leads to~\eqref{eq:condi3}:
\begin{equation*}
\frac{\partial \cl}{\partial \cb_0} = \sum_{n=1}^N \rbr{\cy_n - \cb_0 - \langle \cx_n^{\text{\tiny imp}}, \cb \rangle_{\{P_\ell\}}} \odot \cw_n = \mathcal{O}_{Q_1 \times \cdots \times Q_M}.
\end{equation*}

In what follows, it is shown that the first-order conditions of the ROTOT objective function \eqref{eq:robustWLS} are identical to those of the following objective function 
\begin{equation*}
    \tcl\rbr{\cbr{\cx_n}, \cbr{\cy_n}, \bUbV, \cb_0} = \frac{1}{4m}\sum^N_{n=1}  \abr{\left(\cw_{n}\right)^{\frac{1}{2}} \odot \left(\cy_n - \cb_0 - \langle \cx_n^{\text{\tiny imp}}, \cb \rangle_{\{P_\ell\}} \right)}_F^2 + \lambda\abr{\cb}_F^2.
\end{equation*}

The derivative of $\tcl$ with respect to $u^{(\ell)}_{p_\ell r}$ is:
\begin{align*}
\frac{\partial \tcl}{\partial u^{(\ell)}_{p_\ell r}}
&=  \frac{1}{4m}\sum_{n=1}^N \sumq w_{n, q_1 \cdots q_M} \frac{\partial}{\partial u^{(\ell)}_{p_\ell r}} \left(\y - \bint - \sump \ximp \bs \right)^2 \\&\hspace{8cm} + \lambda \frac{\partial}{\partial u^{(\ell)}_{p_\ell r}} \sum_{p_1 \cdots q_M}^{P_1 \cdots Q_M} \rbr{\sum_{r=1}^{R} u_{p_1r}^{(1)} \cdots v_{q_Mr}^{(M)}}^2 \nonumber \\ 
&=  \frac{1}{4m}2\sum_{n=1}^N \sumq w_{n, q_1 \cdots q_M} \rs \frac{\partial}{\partial u^{(\ell)}_{p_\ell r}} \left(\y - \bint - \sump \ximp \bs \right)\\&\hspace{4cm} + 2\lambda \sumpl \sumq\rbr{\sum_{r=1}^{R} u_{p_1r}^{(1)} \cdots v_{q_Mr}^{(M)}} 
\beta_{p_1 \ldots p_{\ell-1} \, r \, p_{\ell+1} \ldots p_L q_1 \ldots q_M}^{(-\ell)}
\nonumber \\ 
&= - \frac{1}{2m}\sum_{n=1}^{N}  \sumq \Big[w_{n, q_1 \cdots q_M} \rs \sumpl  \ximp \beta_{p_1 \ldots p_{\ell-1} \, r \, p_{\ell+1} \ldots p_L q_1 \ldots q_M}^{(-\ell)}\Big] \\&\hspace{4cm} + 2\lambda \sumpl \sumq \rbr{\sum_{r=1}^{R} u_{p_1r}^{(1)} \cdots v_{q_Mr}^{(M)}} 
\beta_{p_1 \ldots p_{\ell-1} \, r \, p_{\ell+1} \ldots p_L q_1 \ldots q_M}^{(-\ell)}.
\nonumber
\end{align*}
This leads to: 
\begin{equation*}
\frac{\partial \tcl}{\partial \bU_\ell} = \sum_{n=1}^N \left\langle \rbr{\cy_n - \cb_0 - \langle \cx_n^{\text{\tiny imp}}, \cb \rangle_{\{P_\ell\}}} \odot \cw_n, \cc_n \right\rangle_{\{Q_m\}} - (4\lambda m) \bU_\ell \, \mathbf{T}^{(-\ell)}_{\bU} = \mathbf{0}_{P_\ell \times R},
\end{equation*}
that is identical to \eqref{eq:condi1}. The derivative of $\tcl$ with respect to $v^{(m)}_{q_m r}$:
\begin{align*}
\frac{\partial \tcl}{\partial v^{(m)}_{q_m r}}
&=  \frac{1}{4m}\sum_{n=1}^N \sumqm w_{n, q_1 \cdots q_M} \frac{\partial}{\partial v^{(m)}_{q_m r}} \left(\y - \bint - \sump \ximp \bs \right)^2  \\&\hspace{7cm} + \lambda \frac{\partial}{\partial v^{(m)}_{q_m r}} \sum_{p_1 \cdots q_M}^{P_1 \cdots Q_M} \rbr{\sum_{r=1}^{R} u_{p_1r}^{(1)} \cdots v_{q_Mr}^{(M)}}^2 \nonumber \\ 
&=  \frac{1}{4m}2\sum_{n=1}^N \sumqm w_{n, q_1 \cdots q_M} \rs \frac{\partial}{\partial v^{(m)}_{q_m r}} \left(\y - \bint - \sump \ximp \bs \right)  \\&\hspace{4cm} + 2\lambda \sump \sumqm \rbr{\sum_{r=1}^{R} 
u_{p_1r}^{(1)} \cdots v_{q_Mr}^{(M)}} 
\beta_{p_1 \ldots p_L q_1 \ldots q_{m-1} \, r \, q_{m+1} \ldots q_M}^{(-m)} 
\nonumber\\
&= - \frac{1}{2m}\sum_{n=1}^N \sumqm \Big[w_{n, q_1 \cdots q_M} \rs \sump  \ximp \beta_{p_1 \ldots p_L q_1 \ldots q_{m-1} \, r \, q_{m+1} \ldots q_M}^{(-m)} \Big]   \\&\hspace{4cm} + 2\lambda \sump \sumqm \rbr{\sum_{r=1}^{R} u_{p_1r}^{(1)} \cdots v_{q_Mr}^{(M)}}
\beta_{p_1 \ldots p_L q_1 \ldots q_{m-1} \, r \, q_{m+1} \ldots q_M}^{(-m)} 
\nonumber
\end{align*}
This leads to:
\begin{equation*}
\frac{\partial \tcl}{\partial \bV_{m}} = \sum_{n=1}^N \left\langle \rbr{\cy_n - \cb_0 - \langle \cx_n^{\text{\tiny imp}}, \cb \rangle_{\{P_\ell\}}} \odot \cw_n, \cd_n \right\rangle_{Q^{\star}_m} - (4\lambda m) \bV_m \, \mathbf{T}^{(-m)}_{\bV} = \mathbf{0}_{Q_m \times R},
\end{equation*}
that is identical to \eqref{eq:condi2}.

The derivative of $\tcl$ with respect to $\bint$ is
\begin{align*}
\frac{\partial \tcl}{\partial \bint}
&=   \frac{1}{4m}\sum^N_{n=1} w_{n, q_1 \cdots q_M} \frac{\partial}{\partial \bint} \left(\y - \bint - \sump \ximp \bs \right)^2 \\ 
&=   \frac{1}{4m}2\sum^N_{n=1} w_{n, q_1 \cdots q_M} \rs \frac{\partial}{\partial \bint} \rbr{\y - \bint - \sump \ximp \bs} \\ 
&=  - \frac{1}{2m}\sum^N_{n=1} w_{n, q_1 \cdots q_M} \rs,
\end{align*}
which results in
\begin{equation*}
\frac{\partial \tcl}{\partial \cb_0} = \sum_{n=1}^N \rbr{\cy_n - \cb_0 - \langle \cx_n^{\text{\tiny imp}}, \cb \rangle_{\{P_\ell\}}} \odot \cw_n = \mathcal{O}_{Q_1 \times \cdots \times Q_M}
\end{equation*}
that is identical to \eqref{eq:condi3}.

\section{Additional Notation and Tensor Operation}
\label{app:addNot}
In this section, we present results from tensor algebra that will be used in Section~\ref{app:algo} of the Supplementary Material to simplify the first-order conditions derived in Section~\ref{app:firstorder}.

Consider a tensor $\cx = \sbr{x_{p_1\ldots p_L}} \in \mathbb{R}^{P_1 \times \cdots \times P_L}$  and its matricization $\mathbf{X}_{(I_r \times I_c)} = \sbr{x_{j k}} \in \mathbb{R}^{J \times K}$ defined as in Section~\ref{sec:notations}. It is assumed that $I_r \cup I_c = \{1,\ldots,L\}$,  which together index all modes of the tensor, and $I_r \cap I_c = \varnothing$.
The entry $x_{p_1 \ldots p_L}$ maps to the $(j,k)$th element of the matrix $\mathbf{X}_{(I_r \times I_c)}$, 
that is, $x_{p_1\ldots p_L} = \rbr{\mathbf{X}_{(I_r \times I_c)}}_{jk}= x_{jk}$, 
where $j = 1 + \sum_{\ell=1}^{|I_r|} \sbr{(p_{r_\ell} - 1) \prod_{\ell' = 1}^{\ell-1} P_{r_{\ell'}}}$ and $k = 1 + \sum_{m=1}^{|I_c|} \sbr{(p_{c_m} - 1) \prod_{m' = 1}^{m-1} P_{c_{m'}}}$. The mapping from the tensor index tuple $(p_{r_1},\ldots,p_{r_{|I_r|}})$ to the row index $j$ defined above is bijective. In particular, each row index $j \in \{1,\ldots,J\}$ corresponds to exactly one tensor index tuple $(p_{r_1},\ldots,p_{r_{|I_r|}})$, and vice versa. The tensor indices associated with the row modes can be recovered from $j$ as
\begin{equation*}
p_{r_\ell} = 1+ \rbr{\Big\lfloor \frac{j-1}{\prod_{h=1}^{\ell-1}P_{r_h}}\Big\rfloor \bmod P_{r_\ell}} , \qquad \ell=1,\ldots,|I_r|,
\end{equation*}
with the convention that an empty product is equal to 1.

An analogous bijection holds between the tensor index tuple $(p_{c_1},\ldots,p_{c_{|I_c|}})$ and the column index $k$. Consequently, each matrix index pair $(j,k)$ corresponds uniquely to one tensor index tuple $(p_1,\ldots,p_L)$ and vice versa.

Using the definition of matricization, the tensor contraction can also be expressed in matrix form. Denote the tensor $\ca = \sbr{a_{i_1 \ldots i_K q_1 \ldots q_M}} \in \mathbb{R}^{I_1 \times \cdots \times I_K \times Q_1 \times \cdots \times Q_M}$ and the tensor $\cb = \sbr{b_{q_1 \ldots q_M p_1 \ldots p_L}} \in \mathbb{R}^{Q_1 \times \cdots \times Q_M \times P_1 \times \cdots \times P_L}$. It can be shown that the tensor contracted product $\langle \ca, \cb \rangle_{\cbr{Q_m}} \in \mathbb{R}^{I_1 \times \cdots \times I_K \times P_1 \times \cdots \times P_L}$ can be reformulated into matrix form 
$\langle \mathbf{A}, \mathbf{B} \rangle_{\cbr{Q}}$,
where $\mathbf{A} = \sbr{a_{iq}} \in \mathbb{R}^{I \times Q}$, with $I = \prod_{k=1}^K I_k$ and $Q = \prod_{m=1}^M Q_m$, denotes the matricization of $\ca$ obtained by stacking the first $K$ modes  into rows and the last $M$ modes into columns. The matrix $\mathbf{B} = \sbr{b_{qp}} \in \mathbb{R}^{Q \times P}$, with $P = \prod_{\ell=1}^L P_\ell$, denotes the matricization of $\cb$ obtained by stacking the first $M$ modes into rows and the last $L$ modes into columns. The elements of $\langle \ca, \cb \rangle_{\cbr{Q_m}}$ are given by $\sum_{q_1 \ldots q_M}^{Q_1 \ldots Q_M} a_{i_1 \ldots i_K q_1 \ldots q_M} b_{q_1 \ldots q_M p_1 \ldots p_L}$. By definition of matricization, each element $a_{i_1 \ldots i_K q_1 \ldots q_M}$ of the tensor $\ca$ corresponds uniquely to an entry $a_{iq}$ of its matricization, where the indices $i$ and $q$ follow the definition of matricization. Similarly, each element $b_{q_1 \ldots q_M p_1 \ldots p_L}$ of the tensor $\cb$ corresponds uniquely to an entry $b_{qp}$ of its matricization, where the indices $q$ and $p$ follows the definition of matricization. 
Thus, the elements of $\langle \ca, \cb \rangle_{\cbr{Q_m}}$ can be equivalently expressed as $\sum_{q = 1}^{Q} a_{iq} b_{q p}$, which corresponds to the elements of $\langle \mathbf{A}, \mathbf{B} \rangle_{\{Q\}}$. From Section \ref{sec:notations}, it directly follows that $\langle \mathbf{A}, \mathbf{B} \rangle_{\cbr{Q}} = \mathbf{A} \mathbf{B}$.

Using the definition of matricization, it can be shown that
\begin{equation}
\label{eq:VecExpC}
\vecop \rbr{\langle \cx_n^{\text{\tiny imp}}, \cb \rangle_{\{P_{\ell}\}}} 
= \mathbf{C}_n^{\ell}\vecop \rbr{\bU_\ell},
\end{equation}
where $\mathbf{C}_n^{\ell} = [c_{n, jk}] \in \mathbb{R}^{Q \times P_\ell R}$ denotes the matricization of $\cc_n$ (defined in~\eqref{eq:Cl}) obtained by stacking the last $M$ modes into rows and the first two modes into columns. From the elements of $\widehat{\cy}_n - \cb_0 = \langle \cx_n^{\text{\tiny imp}}, \cb \rangle_{\{P_{\ell}\}}$, we may write
\begin{align*}
\widehat{y}_{n, q_1 \ldots q_M} - \bint
&=\sump \ximp\bs \\
&= \sum_{p_\ell=1}^{P_\ell}\sum_{r=1}^{R}
\Bigg(
\sumpl \ximp
\beta^{(-\ell)}_{p_1 \ldots p_{\ell - 1} \, r \, p_{\ell + 1} \ldots p_L q_1 \ldots q_M}
\Bigg)
u_{p_{\ell}r}^{(\ell)} \\
&= \sum_{p_\ell=1}^{P_\ell}\sum_{r=1}^{R}
c_{n, p_\ell r q_1 \ldots q_M}^\ell \,
u_{p_{\ell}r}^{(\ell)},
\end{align*}
where $\beta^{(-\ell)}_{p_1 \ldots p_{\ell - 1} \, r \, p_{\ell + 1} \ldots p_L q_1 \ldots q_M} = u_{p_1r}^{(1)} \cdots u_{p_{\ell-1}r}^{(\ell-1)} u_{p_{\ell+1}r}^{(\ell+1)} \cdots u_{p_Lr}^{(L)} v_{q_1r}^{(1)} \cdots v_{q_Mr}^{(M)}$.
Each element $c_{n, p_\ell r q_1 \ldots q_M}^\ell$ of the tensor $\cc_n$ corresponds uniquely to an entry $c_{n, jk}$, and each element $(\widehat{y}_{n, q_1 \ldots q_M} - \bint)$ of $\vecop \rbr{\widehat{\cy}_n - \cb_0}$ corresponds uniquely to an entry $(\widehat{y}_{n,j} - \beta_{0, j})$, where the indices $j$ and $k$ follow the definition of matricization. Thus, the elements of $\widehat{\cy}_n - \cb_0$ can be equivalently expressed as
\begin{equation*}
\widehat{y}_{n, j} - \beta_{0, j}
=
c_{n, j 1} \, u_{11}^{(\ell)}
+ \cdots +
c_{n, j (P_\ell R)} \, u_{P_\ell R}^{(\ell)}.
\end{equation*}
This expression corresponds exactly to the elements of the following equation
\begin{equation*}
\vecop \rbr{\widehat{\cy}_n - \cb_0} = \vecop \rbr{\langle \cx_n^{\text{\tiny imp}}, \cb \rangle_{\{P_{\ell}\}}} = \mathbf{C}_n^\ell \vecop\!\big(\bU_\ell\big).
\end{equation*}

Moreover, it can be shown that
\begin{equation}
\label{eq:MatExpV}
\rbr{\langle \cx_n^{\text{\tiny imp}}, \cb \rangle}_{m} = \widehat{\mathbf{Y}}_{n,m} - \mathbf{B}_{0, m} = \bV_m{\mathbf{D}_n^{m}}^T,
\end{equation}
where $\rbr{\langle \cx_n^{\text{\tiny imp}}, \cb \rangle}_{m} \in \mathbb{R}^{Q_m \times Q^\star}$ and  $(\widehat{\mathbf{Y}}_{n,m} - \mathbf{B}_{0,m}) \in \mathbb{R}^{Q_m \times Q^\star}$, with $Q^\star = \prod_{i \in M_m^{\star}} Q_i^\star$, denotes the matricization of $\langle \cx_n^{\text{\tiny imp}}, \cb \rangle_{\{P_{\ell}\}}$ and $\widehat{\cy}_n$, respectively, obtained by stacking the mode $q_m$ into rows and the modes $(q_1,\ldots,q_{m-1},q_{m+1},\ldots,q_M)$ into columns.
The matrix $\mathbf{D}_n^{m} \in \mathbb{R}^{Q^\star \times R}$ is the matricization of $\cd_n$ obtained by stacking the modes $(q_1,\ldots,q_{m-1},q_{m+1},\ldots,q_M)$ into rows and the mode $q_r$ into columns. From the elements of $\widehat{\cy}_n - \cb_0 = \langle \cx_n^{\text{\tiny imp}}, \cb \rangle_{\{P_{\ell}\}}$, we may write
\begin{align*}
\widehat{y}_{n,q_1 \ldots q_M} - \bint
= \sump \ximp\bs
&= \sum_{r=1}^{R}
\Bigg(
\sump \ximp
\beta^{(-m)}_{p_1 \ldots p_L q_1 \ldots q_{m-1} \, r \, q_{m+1} \ldots q_M}
\Bigg)
v_{q_{m}r}^{(m)} \\
&= \sum_{r=1}^{R}
d_{n, q_1 \ldots q_{m-1} \, r \, q_{m+1} \ldots q_M}^m \,
v_{q_{m}r}^{(m)},
\end{align*}
where $\beta^{(-m)}_{p_1 \ldots p_L q_1 \ldots q_{m-1} \, r \, q_{m+1} \ldots q_M}
=
u_{p_1r}^{(1)}\cdots u_{p_Lr}^{(L)}v_{q_{1}r}^{(1)} 
\cdots v_{q_{m - 1}r}^{(m - 1)}  v_{q_{m + 1}r}^{(m + 1)} 
\cdots v_{q_Mr}^{(M)}$.
Each element $d_{n, q_1 \ldots q_{m-1} \, r \, q_{m+1} \ldots q_M}^m$ of the tensor $\cd_n$ corresponds uniquely to an entry $d_{n, jr}$, and each element $(\widehat{y}_{n, q_1 \ldots q_M} - \bint)$ of $(\widehat{\mathbf{Y}}_{n,m} - \mathbf{B}_{0, m})$ corresponds uniquely to an entry $(\widehat{y}_{n,q_m j} - \beta_{0, q_m j})$, where the index $j$ follows the definition of matricization. Thus, the elements of $(\widehat{\cy}_n - \cb_0)$ can be equivalently expressed as
\begin{equation*}
\widehat{y}_{n, q_m j} - \beta_{0, q_m j} = \sum_{r = 1}^R v_{q_m r}^{(m)} d_{n, jr},
\end{equation*}
which corresponds to the elements of $\rbr{\langle \cx_n^{\text{\tiny imp}}, \cb \rangle}_{m}= \widehat{\mathbf{Y}}_{n,m} - \mathbf{B}_{0, m} = \bV_m{\mathbf{D}_n^{m}}^T$.

\section{Description of the Algorithm}
\label{app:algo}
The IRLS starts with the initial estimate $\parini$ defined in Section~\ref{sec:setup}, and the corresponding weight $\cbr{\cw_n^0}$ obtained from \eqref{eq:Wm}. The set of matrices $\{\mathbf{T}_\bU^{(-\ell),0}\}$, $\{\mathbf{T}_\bV^{(-m),0}\}$ and tensors $\{\mathcal{C}_n^{\ell,0}\}$ and $\{\mathcal{D}_n^{m,0}\}$ are initialized following~\eqref{eq:TU}, \eqref{eq:TV}, \eqref{eq:Cl} and \eqref{eq:Dm}. Then for each $k = 1, 2 \ld$ $\parnext$ are obtained from $\parcurr$ and $\cbr{\cw_n^{k}}$ by the following procedure:
\begin{itemize}
\item[\textbf{(a)}] 
Minimize \eqref{eq:robustWLS} with respect to $\bU_\ell$ by substituting $\cbr{\bV_\ell^k}$, $\cb_0^k$, and $\cbr{\cw^k_n}$ in~\eqref{eq:condi1}. That is, for ${\ell=1,\ldots,L}$
\begin{align*}
&\sum_{n=1}^N \left\langle \rbr{\cy_n - \cb_0^{k}} \odot\cw_n^{k} , \mathcal{C}_n^{\ell,k} \right\rangle_{\cbr{Q_m}} = \sum_{n=1}^N \left\langle \langle \cx_n^{\text{\tiny imp}}, \cb^{k} \rangle_{\{P_{\ell}\}} \odot \cw_n^{k}, \mathcal{C}_n^{\ell,k} \right\rangle_{\cbr{Q_m}}   + (4\lambda m) \bU_\ell\, \mathbf{T}^{(-\ell),k}_{\bU}.
\end{align*}
Let $\mathbf{C}_n^{\ell, k} \in \mathbb{R}^{Q \times RP_\ell}$, with $Q = \prod_{m=1}^M Q_m$, be the matricization of $\mathcal{C}_n^{\ell,k}$. By reformulating the previous equation in matrix form as shown in Section~\ref{app:addNot} of the Supplementary Material, we have
\begin{align*}
\sum_{n=1}^N \left\langle \vecop\rbr{\rbr{\cy_n - \cb_0^k} \odot \cw_n^k}^{T}, \mathbf{C}_n^{\ell, k} \right\rangle_{\cbr{Q}} &= \sum_{n=1}^N\left\langle \vecop \rbr{\langle \cx_n^{\text{\tiny imp}}, \cb^k \rangle_{\{P_{\ell}\}} \odot \cw_n^k}^T, \mathbf{C}_n^{\ell, k} \right\rangle_{\cbr{Q}} \\[1.2ex]& \hspace{34mm} + (4\lambda m) \text{vec}\rbr{\bU_\ell \, \mathbf{T}^{(-\ell),k}_{\bU}}^T;\\[1.5ex]
\sum_{n=1}^N \left\langle \vecop\rbr{\rbr{\cy_n - \cb_0^k} \odot \cw_n^k}^{T}, \mathbf{C}_n^{\ell, k} \right\rangle_{\cbr{Q}} &= \sum_{n=1}^N \left\langle \rbr{\vecop \rbr{\langle \cx_n^{\text{\tiny imp}}, \cb^{k} \rangle_{\{P_{\ell}\}}} \odot \vecop\rbr{\cw_n^k}}^T, \mathbf{C}_n^{\ell, k} \right\rangle_{\cbr{Q}} \\[1.2ex]& \hspace{34mm} + (4\lambda m) \text{vec}\rbr{\bU_\ell \, \mathbf{T}^{(-\ell),k}_{\bU}}^T.
\end{align*}
From~\eqref{eq:VecExpC} we have,
\begin{align*}
& \sum_{n=1}^N \vecop\rbr{\rbr{\cy_n - \cb_0^k} \odot \cw_n^k}^{T} \mathbf{C}_n^{\ell, k} = \sum_{n=1}^N \rbr{\rbr{\mathbf{C}_n^{\ell, k} \vecop \rbr{\bU_\ell}} \odot \vecop\rbr{\cw_n^k}}^T \mathbf{C}_n^{\ell, k} \\[1.2ex]&\hspace{10.7cm} + (4\lambda m) \text{vec}\rbr{\bU_\ell \, \mathbf{T}^{(-\ell),k}_{\bU}}^T;\\[1.2ex]
&\sum_{n=1}^N {\mathbf{C}_n^{\ell, k}}^T \vecop\rbr{\rbr{\cy_n - \cb_0^k} \odot \cw_n^k} = \sum_{n=1}^N {\mathbf{C}_n^{\ell, k}}^T\rbr{\rbr{\mathbf{C}_n^{\ell, k} \vecop \rbr{\bU_\ell}} \odot \vecop\rbr{\cw_n^k}}\\[1.2ex]&\hspace{10.7cm}  + 4\lambda m \rbr{\mathbf{T}^{(-\ell), k}_{\bU} \otimes \mathbf{I}_{P_\ell}} \text{vec}\rbr{\bU_\ell};\\[1.2ex]
&\sum_{n=1}^N {\mathbf{C}_n^{\ell, k}}^T {\mathbf{W}_n^k}^* \vecop\rbr{\rbr{\cy_n - \cb_0^k}} =\sum_{n=1}^N {\mathbf{C}_n^{\ell, k}}^T {\mathbf{W}_n^k}^* {\mathbf{C}_n^{\ell, k}} \vecop \rbr{\bU_\ell} + \mathbf{P}^{\ell, k}\text{vec}\rbr{\bU_\ell},
\end{align*}
where the diagonal matrix ${\mathbf{W}_n^k}^*$ is defined such that every element of its diagonal is composed of the vectorization of $\cw^{k}_n$. Thus,
\begin{equation*}
    \vecop \rbr{\bU_\ell}  = \rbr{\sum_{n=1}^N {\mathbf{C}_n^{\ell, k}}^T \mathbf{W}_n^* \mathbf{C}_n^{\ell, k} + \mathbf{P}^{\ell, k}}^{\dagger} \sum_{n=1}^N {\mathbf{C}_n^{\ell, k}}^T {\mathbf{W}_n^k}^* \vecop\rbr{\rbr{\cy_n - \cb_0^k}},
\end{equation*}
where $^{\dagger}$ denotes the Moore-Penrose generalized inverse.  
\item[\textbf{(b)}] Minimize \eqref{eq:robustWLS} with respect to $\bV_m$ by substituting $\lbrace \bU_\ell^{k+1} \rbrace$, $\cb_0^k$, and $\cbr{\cw^k_n}$ in~\eqref{eq:condi2}. That is, for ${m=1,\ldots,M}$
\begin{align*}
&\sum_{n=1}^N \left\langle \rbr{\cy_n - \cb_0^k} \odot \cw_n^k , \mathcal{D}_n^{m, k} \right\rangle_{Q^{\star}_m} = \sum_{n=1}^N \left\langle \langle \cx_n^{\text{\tiny imp}}, \cb^k \rangle_{\{P_{\ell}\}} \odot \cw_n^k, \mathcal{D}_n^{m, k} \right\rangle_{Q^{\star}_m} + (4\lambda m) \bV_m \, \mathbf{T}^{(-m),k}_{\bV};
\end{align*}
By appropriately modifying the dimensions, the contracted tensor can be matricized as $\mathbf{D}_n^m  \in \mathbb{R}^{{Q}^\star \times R}$, where ${Q}^\star = \prod_{i \in M^{\star}_m} Q_i^\star$,
and $\mathbf{Y}_{n,m} \in \mathbb{R}^{Q_m \times {Q}^\star}$. Then, from the results in Section~\ref{app:addNot} of the Supplementary Material, we obtain
\begin{align*}
&\sum_{n=1}^N \left\langle \rbr{\mathbf{Y}_{n, m} - \mathbf{B}_{0,m}^k} \odot \mathbf{W}_{n,m}^k, \mathbf{D}_n^{m, k} \right\rangle_{\cbr{{Q}^*}} = \sum_{n=1}^N \left\langle \rbr{\bV_m{\mathbf{D}_n^{m, k}}^T}\odot \mathbf{W}_{n,m}^k, \mathbf{D}_n^{m, k} \right\rangle_{\cbr{{Q}^*}} \\[1.2ex]&\hspace{10.7cm} + (4\lambda m) \bV_m \, \mathbf{T}^{(-m),k}_{\bV};\\[1.2ex]
&\sum_{n=1}^N \rbr{\rbr{\mathbf{Y}_{n, m} - \mathbf{B}_{0,m}^k} \odot \mathbf{W}_{n, m}^k} \mathbf{D}_n^{m, k} = \sum_{n=1}^N\rbr{\rbr{\bV_m{\mathbf{D}_n^{m, k}}^T}\odot \mathbf{W}_{n,m}^k} \mathbf{D}_n^{m, k} \\[1.2ex]&\hspace{10.7cm} + (4\lambda m) \bV_m\, \mathbf{T}^{(-m),k}_{\bV}.
\end{align*}
By vectorizing this last equation, we have that
\begin{align*}
&\sum_{n=1}^N \vecop \rbr{\rbr{\rbr{\mathbf{Y}_{n, m} - \mathbf{B}_{0,m}^k} \odot \mathbf{W}_{n, m}^k} \mathbf{D}_n^{m,k}}  = \sum_{n=1}^N \vecop \rbr{\rbr{\rbr{\bV_m{\mathbf{D}_n^{m,k}}^T}\odot \mathbf{W}_{n, m}^k} \mathbf{D}_n^{m,k}} \\[1.2ex]&\hspace{11.5cm} + (4\lambda m) \vecop \rbr{\bV_m\, \mathbf{T}^{(-m),k}_{\bV}};\\[1.2ex]
&\sum_{n=1}^N \vecop \rbr{\rbr{\rbr{\mathbf{Y}_{n, m} - \mathbf{B}_{0,m}^k} \odot \mathbf{W}_{n, m}^k} \mathbf{D}_n^{m,k}}  = \\[1.2ex]&\hspace{4cm}\sum_{n=1}^N \rbr{{\mathbf{D}_n^{m,k}}^T \otimes \mathbf{I}_{Q_m}} \widetilde{\mathbf{W}}_{n, m}^k \rbr{\mathbf{D}_n^{m,k} \otimes \mathbf{I}_{Q_m}} \vecop \rbr{\bV_m} + \mathbf{P}^{m,k} \vecop \rbr{\bV_m};\\[1.2ex]
&\sum_{n=1}^N \vecop \rbr{\rbr{\rbr{\mathbf{Y}_{n, m} - \mathbf{B}_{0,m}^k} \odot \mathbf{W}_{n, m}^k} \mathbf{D}_n^{m,k}}  = \\[1.2ex]&\hspace{4cm} \rbr{\sum_{n=1}^N \rbr{{\mathbf{D}_n^{m, k}}^T \otimes \mathbf{I}_{Q_m}} \widetilde{\mathbf{W}}_{n, m}^k \rbr{\mathbf{D}_n^{m, k} \otimes \mathbf{I}_{Q_m}} + \mathbf{P}^{m, k}} \vecop \rbr{\bV_m}.
\end{align*}
Thus,
\begin{align*}
    &\vecop \rbr{\bV_m^{k+1}} = \rbr{\sum_{n=1}^N\rbr{{\mathbf{D}_n^{m, k}}^T \otimes \mathbf{I}_{Q_m}} \widetilde{\mathbf{W}}_{n, m}^k \rbr{\mathbf{D}_n^{m, k} \otimes \mathbf{I}_{Q_m}}  + \mathbf{P}^{m, k}}^{\dagger} \\[1.2ex]&\hspace{8cm} \sum_{n=1}^N \vecop \rbr{\rbr{\rbr{\mathbf{Y}_{n, m} - \mathbf{B}_{0,m}^k} \odot \mathbf{W}_{n, m}^k} \mathbf{D}_n^{m, k}}.
\end{align*}
\item[\textbf{(c)}] Minimize \eqref{eq:robustWLS} with respect to $\cb_0$ by substituting $\lbrace \bU_\ell^{k+1} \rbrace$, $\lbrace \bV_\ell^{k+1} \rbrace$ and $\cbr{\cw^k_n}$ in \eqref{eq:condi3}. That is
\begin{align*}
&\sum_{n=1}^N \rbr{\cy_n - \cb_0 - \langle \cx_n^{\text{\tiny imp}}, \cb^{k+1} \rangle_{\{P_{\ell}\}}} \odot \cw_n^k = 
\mathcal{O}_{Q_1 \times \cdots \times Q_M};\\
&\cb_0  \odot \rbr{\sum_{n=1}^N \cw_n^k} = \sum_{n=1}^N \rbr{\cy_n - \langle \cx_n^{\text{\tiny imp}}, \cb^{k+1} \rangle_{\{P_{\ell}\}}} \odot \cw_n^k.
\end{align*}
Thus,
\begin{equation*}
\cb_0^{k+1} = \rbr{\sum_{n=1}^N \rbr{\cy_n - \langle \cx_n^{\text{\tiny imp}}, \cb^{k+1} \rangle_{\{P_{\ell}\}}} \odot \cw_n^k} \odot \ch^k.
\end{equation*}
\item[\textbf{(d)}] Update $\cbr{\cw_n}$ using \eqref{eq:Wm} with $\parnext$.
\end{itemize}

\section{Proof of Algorithm Descent}
\label{app:proofConv}
 {In this section Proposition \ref{the_1P} is proved, which ensures that each step of the algorithm decreases the objective function \eqref{eq:robustWLS}. Since the ROTOT objective function is nonconvex, the proposed algorithm is only guaranteed to converge to a stationary point and not necessarily to a global optimum. 
Convergence results for IRLS algorithms are well established in the classical regression literature. However, these results do not directly carry over to the present tensor-on-tensor regression setting. Proposition~\ref{the_1P} therefore verifies that the successive reweighting and parameter-update steps of ROTOT preserve the descent of the proposed objective function. 
The proof relies on the following three lemmas.}

 Let us define the regularized weighted tensor-on-tensor regression objective function as
\begin{equation}
 \frac{1}{4m}\sum^N_{n=1}  \abr{\left(\cw_{n}\right)^{\frac{1}{2}} \odot \rbr{\cy_n - \cb_0 - \langle \cx_n^{\text{\tiny imp}}, \cb \rangle_{\{P_{\ell}\}}}}_F^2 + \lambda\abr{\cb}_F^2,
 \label{eq:regWTOT}
\end{equation}

\begin{lemma} For a given weight tensor $\cbr{\cw^k_n}$, each of the update steps (a), (b), and (c) of the algorithm in Section \ref{app:algo} decreases the regularized weighted tensor-on-tensor regression objective function \eqref{eq:regWTOT}. 
\end{lemma}
\begin{proof} It is shown in Section \ref{app:firstorder} of the Supplementary Material that minimizing \eqref{eq:robustWLS}  is the same as minimizing \eqref{eq:regWTOT}, for
fixed $\cw_{n}$. 
We start from the parameters $\parcurr$ with objective
\begin{equation}
	 \frac{1}{4m}\sum_{n=1}^N \abr{{\left(\cw_{n}^k\right)}^{\frac{1}{2}} \odot \rbr{\cy_n - \cb_0^k - \langle \cx_n^{\text{\tiny imp}}, \cb^k \rangle_{\{P_{\ell}\}}}}_F^2 + \lambda\abr{\cb^k}_F^2.
    \label{eq:regWTOTk}
\end{equation}
Step (a) minimizes the squared norm \eqref{eq:regWTOTk} with respect to $\{\bU_{\ell}\}$, for $\ell = 1 \ld L$. The closed form solution is obtained as in \eqref{eq:solU} and the objective \eqref{eq:regWTOTk} becomes a function of the new triplet $\rbr {\cbr{\bU_{\ell}^{k+1}}, \cbr{\bV_{m}^{k}}, \cb_0^{k}}$.

Step (b) minimizes the squared norm \eqref{eq:regWTOTk} with respect to $\cbr{\bV_{m}}$, for $m = 1 \ld M$, which becomes a function of the new triplet $\rbr {\cbr{\bU_{\ell}^{k+1}}, \cbr{\bV_{m}^{k+1}}, \cb_0^{k}}$.

Step (c) minimizes the squared norm \eqref{eq:regWTOTk} with respect to $\cb_0$, which becomes a function of the new triplet $\rbr {\cbr{\bU_{\ell}^{k+1}}, \cbr{\bV_{m}^{k+1}}, \cb_0^{k+1}}$. 
\end{proof}
Denote the set of parameters to be estimated by $\Theta := \cbr{\cb_0, \cbr{\bU_\ell}, \cbr{\bV_m}}$ and introduce the notation $\boldsymbol{f}\rbr{\Theta}$ to indicate the elements
\begin{equation*}
  \lbrace \text{vec}(( \cy_n - \cb_0 - \langle \cx_n^{\text{\tiny imp}}, \cb \rangle_{\{P_{\ell}\}}) \odot (\cy_n - \cb_0 - \langle \cx_n^{\text{\tiny imp}}, \cb \rangle_{\{P_{\ell}\}}) \rbrace_{n = 1}^N = \lbrace \text{vec}(( \cy_n -\widehat{\cy}_n) \odot (\cy_n -\widehat{\cy}_n)) \rbrace_{n = 1}^N  ,
\end{equation*}
\sloppy stacked  into a column vector. The vector $\boldsymbol{f}\rbr{\Theta}$ has $p = N\prod_{m = 1}^M Q_m$ entries with values $\left(\y - \widehat{y}_{n, q_1 \ldots q_M} \right)^2$. We introduce an index function that maps an element of the tensor into the corresponding element of its vectorized form
\begin{equation*}
    d_{n, q_1 \ldots q_M} = q_1 + \sum_{m = 2}^M \left(\prod_{\ell = 1}^{n-1} Q_\ell \right) \left(q_m - 1  \right) + \left(n - 1 \right) \prod_{m = 1}^M Q_m.
\end{equation*}
Thus, an element of $\boldsymbol{f}\rbr{\Theta}$ is denoted by $\fd$. To write the tensor-on-tensor regression objective function \eqref{eq:robustWLS}, we denote the regularization term as $\mathcal{P}\rbr{\cbr{\bU_\ell}, \cbr{\bV_m}}$ such that
\begin{equation*}
    L \left(\boldsymbol{f} \rbr{\Theta} \right) + \mathcal{P}\rbr{\cbr{\bU_\ell}, \cbr{\bV_m}} := \mathcal{L} \rbr{\Theta}.
\end{equation*}

\begin{lemma} The function $\boldsymbol{f} \rightarrow L\left(\boldsymbol{f} \right)$ is concave.
\end{lemma}
\begin{proof} 
Let us define the functions $h_1: \mathbb R_{+} \rightarrow \mathbb R_{+}:z \mapsto 
 \rho_{1}(\sqrt{z})$ and $h_2: \mathbb R_{+} \rightarrow \mathbb R_{+}:
 z \mapsto \rho_{2}(\sqrt{z})$. As $\rho_{1}$ and $\rho_{2}$ are valid $\rho$-functions, $h_1$ and $h_2$ are concave.

By the definition of concavity of a multivariate function, we also prove that for any column vectors $\boldsymbol{f}$, $\boldsymbol{g}$ in $\mathbb{R}^{p}_{+}$ and any $\lambda$ in $\left(0,1 \right)$ it holds that $L \left(\lambda \boldsymbol{f} + \left(1 - \lambda \right){\boldsymbol{g}} \right) \geqslant  \lambda L\left(\boldsymbol{f} \right) + \left(1 - \lambda \right) L\left({\boldsymbol{g}} \right)$. This follows from:
\begin{align*}
    L \rbr{\lambda \boldsymbol{f} + (1 - \lambda)\boldsymbol{g}} &= 
    \frac{\sgcase^2}{m} \sum_{n=1}^N h_2 \rbr{\frac{1}{m_n \hat{\sigma}_{2}^2} \sumq \ocell \sgcell^2 h_1 \rbr{\frac{\lambda \fd + (1 - \lambda)\gd}{\sgcell^2}}} \\
    &\geqslant 
    \frac{\sgcase^2}{m} \sum_{n=1}^N h_2 \left( \frac{1}{m_n \sgcase^2} \sumq \ocell \sgcell^2 \bigg[ \lambda h_1 \left( \frac{\fd}{\sgcell^2} \right) \right. \\
    &\quad + \left. (1 - \lambda) h_1 \left( \frac{\gd}{\sgcell^2} \right) \bigg] \right) \\
    &= 
    \frac{\sgcase^2}{m} \sum_{n=1}^N h_2 \left( \lambda \frac{1}{m_n \sgcase^2} \sumq \ocell \sgcell^2 h_1 \left( \frac{\fd}{\sgcell^2} \right) \right. \\
    &\quad + \left. (1 - \lambda) \frac{1}{m_n \sgcase^2} \sumq \ocell \sgcell^2 h_1 \left( \frac{\gd}{\sgcell^2} \right) \right) \\
    &\geqslant 
    \frac{\sgcase^2}{m} \sum_{n=1}^N \left[ \lambda h_2 \left( \frac{1}{m_n \sgcase^2} \sumq \ocell \sgcell^2 h_1 \left( \frac{\fd}{\sgcell^2} \right) \right) \right. \\
    &\quad + \left. (1 - \lambda) h_2 \left( \frac{1}{m_n \sgcase^2} \sumq \ocell \sgcell^2 h_1 \left( \frac{\gd}{\sgcell^2} \right) \right) \right] \\
    &= \lambda L\left(\boldsymbol{f} \right) + \left(1 - \lambda \right) L\left(\boldsymbol{g} \right).
\end{align*}
The first inequality follows from the concavity of $h_1$ and the fact that $h_2$ is nondecreasing. The second inequality comes from the concavity of $h_2$ . Thus, $L$ is a concave function. 
\end{proof}

The unregularized weighted tensor-on-tensor regression objective from \eqref{eq:regWTOT} can be written as a function of $\boldsymbol{f}$. Denote it as $L_{\mathcal{W}} (\boldsymbol{f} \rbr{\Theta} ) :=  \frac{1}{4m}\text{vec} \left(\cw\right)^T \boldsymbol{f} \rbr{\Theta}$, where $\cw$ is turned into a vector in the same way as the column vector $\boldsymbol{f}$. Lemma 3 links the regularized weighted tensor-on-tensor regression objective and the original objective \eqref{eq:robustWLS}.

\begin{lemma}
If $\Theta^1, \Theta^2$ satisfy $L_{\mathcal{W}} \left(\boldsymbol{f} \rbr{\Theta^1}\right) + \mathcal{P}\rbr{\cbr{\bU_\ell^1}, \cbr{\bV_m^1}} \leqslant L_{\mathcal{W}}\left( \boldsymbol{f}\rbr{\Theta^2} \right) + \mathcal{P}\rbr{ \cbr{\bU_\ell^2}, \cbr{\bV_m^2}}$ then $L\left(\boldsymbol{f}\rbr{\Theta^1}\right) + \mathcal{P}\rbr{\cbr{\bU_\ell^1}, \cbr{\bV_m^1}} \leqslant L\left( \boldsymbol{f}\rbr{\Theta^2} \right) + \mathcal{P}\rbr{\cbr{\bU_\ell^2}, \cbr{\bV_m^2}}$.
\end{lemma}

\begin{proof} From Lemma 2 we know that $L(\boldsymbol{f})$ is concave as a function of $\boldsymbol{f}$, and it is also differentiable because $h_1$ and $h_2$ are. Therefore
\begin{equation*}
    L\left(\boldsymbol{f}\rbr{\Theta^1}\right) \leqslant L\rbr{\boldsymbol{f}\rbr{\Theta^2}} + \rbr{\nabla L\rbr{\boldsymbol{f}\rbr{\Theta^2}}}^{T} \rbr{\boldsymbol{f}\rbr{\Theta^1} - \boldsymbol{f}\rbr{\Theta^2}}, 
\end{equation*}
where the column vector $\nabla L\rbr{\boldsymbol{f}\rbr{\Theta^2}}$ is the gradient of $L$ in $\boldsymbol{f}\rbr{\Theta^2}$. We further write
\begin{align*}
    &L\left(\boldsymbol{f}\rbr{\Theta^1}\right) + \mathcal{P}\rbr{\cbr{\bU_\ell^1}, \cbr{\bV_m^1}} \leqslant L\rbr{\boldsymbol{f}\rbr{\Theta^2}}  + \rbr{\nabla L\rbr{\boldsymbol{f}\rbr{\Theta^1}}}^{T} \rbr{\boldsymbol{f}\rbr{\Theta^1}  - \boldsymbol{f}\rbr{\Theta^2}} \\&\hspace{5.5cm} + \mathcal{P}\rbr{\cbr{\bU_\ell^1}, \cbr{\bV_m^1}} - \mathcal{P}\rbr{\cbr{\bU_\ell^2}, \cbr{\bV_m^2}} + \mathcal{P}\rbr{\cbr{\bU_\ell^2}, \cbr{\bV_m^2}}.
\end{align*}
We know that $\nabla L\rbr{\boldsymbol{f}\rbr{\Theta^2}}$ is equal to $\nabla L_{\cw}\rbr{\boldsymbol{f}\rbr{\Theta^2}}$ which equals $ \frac{1}{4m}\text{vec}\rbr{\cw}$ by construction, so $\rbr{\nabla L\rbr{\boldsymbol{f}\rbr{\Theta^2}}}^{T} \rbr{\boldsymbol{f}\rbr{\Theta^1} - \boldsymbol{f}\rbr{\Theta^2}}$ is equal to $L_{\cw}\rbr{\boldsymbol{f}\rbr{\Theta^1}} - L_{\cw}\rbr{\boldsymbol{f}\rbr{\Theta^2}}$.
Thus,
\begin{align*}    
 L\rbr{\boldsymbol{f}\rbr{\Theta^1}} + \mathcal{P}\rbr{\cbr{\bU_\ell^1}, \cbr{\bV_m^1}} & \leqslant L\rbr{\boldsymbol{f}\rbr{\Theta^2}} + \mathcal{P}\rbr{\cbr{\bU_\ell^2}, \cbr{\bV_m^2}}  \\&\quad + \rbr{L_{\cw}\rbr{\boldsymbol{f}\rbr{\Theta^1}} + \mathcal{P}\rbr{\cbr{\bU_\ell^1}, \cbr{\bV_m^1}}} \\& \quad -\rbr{L_{\cw}\rbr{\boldsymbol{f}\rbr{\Theta^2}} + \mathcal{P}\rbr{\cbr{\bU_\ell^2}, \cbr{\bV_m^2}}}.
\end{align*}
Since it is given that 
$L_{\mathcal{W}} \left(\boldsymbol{f}\rbr{\Theta^1} \right) + \mathcal{P}\rbr{\cbr{\bU_\ell^1}, \cbr{\bV_m^1}} \leqslant L_{\mathcal{W}}\left( \boldsymbol{f}\rbr{\Theta^2} \right) + \mathcal{P}\rbr{\cbr{\bU_\ell^2}, \cbr{\bV_m^2}}$, 
we have that
\begin{equation*}
    L(\boldsymbol{f}\rbr{\Theta^1}) + \mathcal{P}\rbr{\cbr{\bU_\ell^1}, \cbr{\bV_m^1}} \leqslant L\rbr{\boldsymbol{f}\rbr{\Theta^2}} + \mathcal{P}\rbr{\cbr{\bU_\ell^2}, \cbr{\bV_m^2}}.
\end{equation*}
\end{proof}

\begin{proof}[Proof of Proposition \ref{the_1P}.] When the parameters are updated from $\Theta^{k}$ to $\Theta^{k+1}$, Lemma $1$ shows that 
 \begin{align*}
    &L_{\cw} \left(\boldsymbol{f}\rbr{\Theta^{k+1}}\right) + \mathcal{P}\rbr{ \cbr{\bU_\ell^{k+1}}, \cbr{\bV_m^{k+1}}} \leqslant L_{\cw} \left(\boldsymbol{f}\rbr{\Theta^{k}}\right) + \mathcal{P}\rbr{ \cbr{\bU_\ell^{k}}, \cbr{\bV_m^{k}}}, 
\end{align*}
so using Lemma $3$ it follows that 
\begin{align*}
    &L\left(\boldsymbol{f}\rbr{\Theta^{k+1}}\right) + \mathcal{P}\rbr{ \cbr{\bU_\ell^{k+1}}, \cbr{\bV_m^{k+1}}} \leqslant L\left(\boldsymbol{f}\rbr{\Theta^{k}}\right) + \mathcal{P}\rbr{ \cbr{\bU_\ell^{k}}, \cbr{\bV_m^{k}}},
\end{align*}
which is equivalent to $\mathcal{L} \rbr{\Theta^{k+1}} \leqslant \mathcal{L} \rbr{\Theta^{k}}$. 
\end{proof}

\clearpage

\section{Additional Simulation Results}
\label{app:simres}
 {To further highlight the advantage of ROTOT, we also compare it with OnlyCase-TOT and OnlyCell-TOT, which correspond to ROTOT with $\rho_1(z)=z^2$ and $\rho_2(z)=z^2$, respectively. These variants are therefore designed to handle only casewise or only cellwise outliers. We consider the simulation settings described in Section~\ref{sec:simu} with $\text{SNR}=5$.}

\begin{figure}[H]
\centering
\begin{tabular}{cccc}
 &\textbf{Cellwise}  & \textbf{Casewise} &{\textbf{Casewise \& Cellwise}}\\
\rotatebox{90}{\normalsize {\parbox{5cm}{\centering Variation in $\cbr{\cy_n}$}}} & 
\includegraphics[width=.30\textwidth]{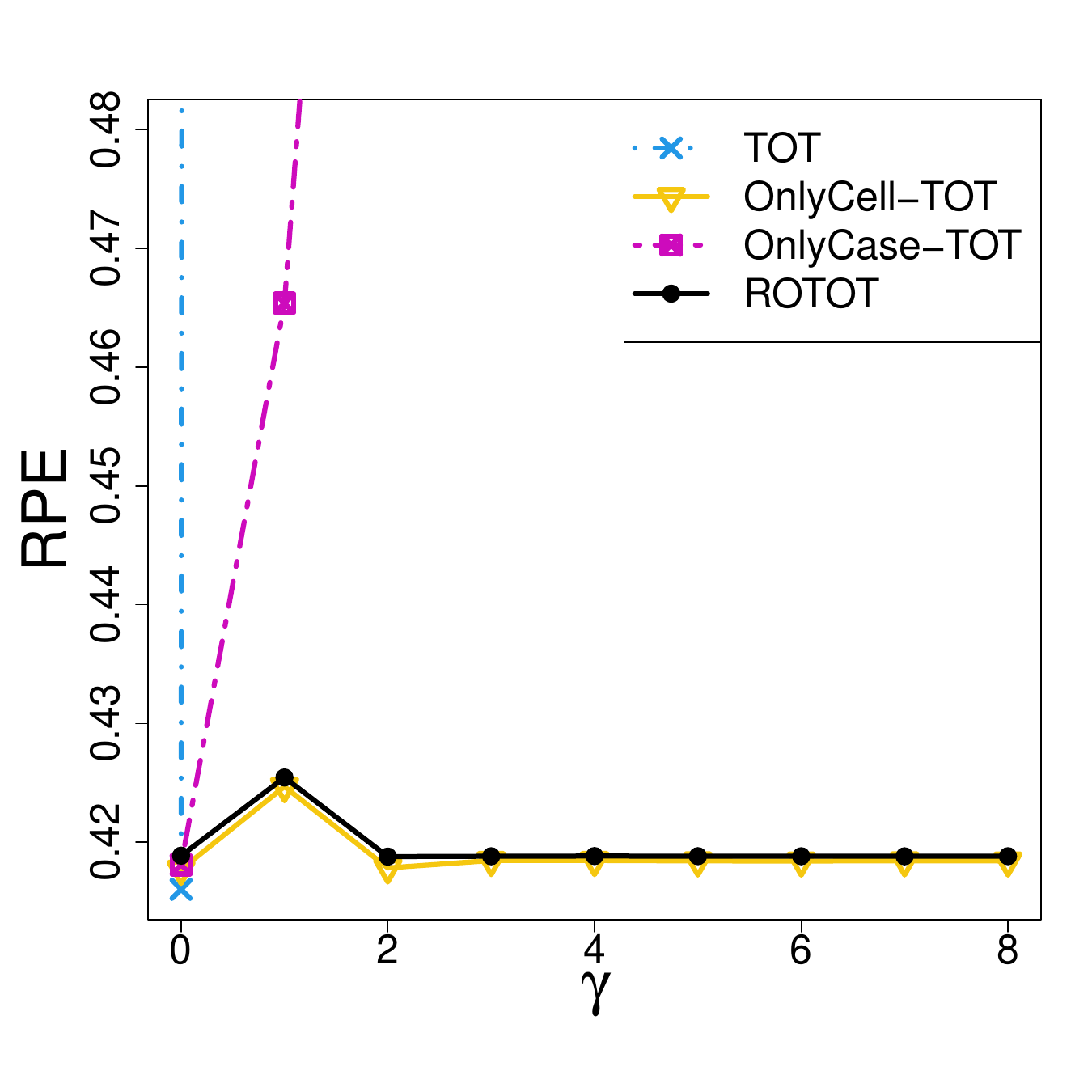} &%
\includegraphics[width=.30\textwidth]{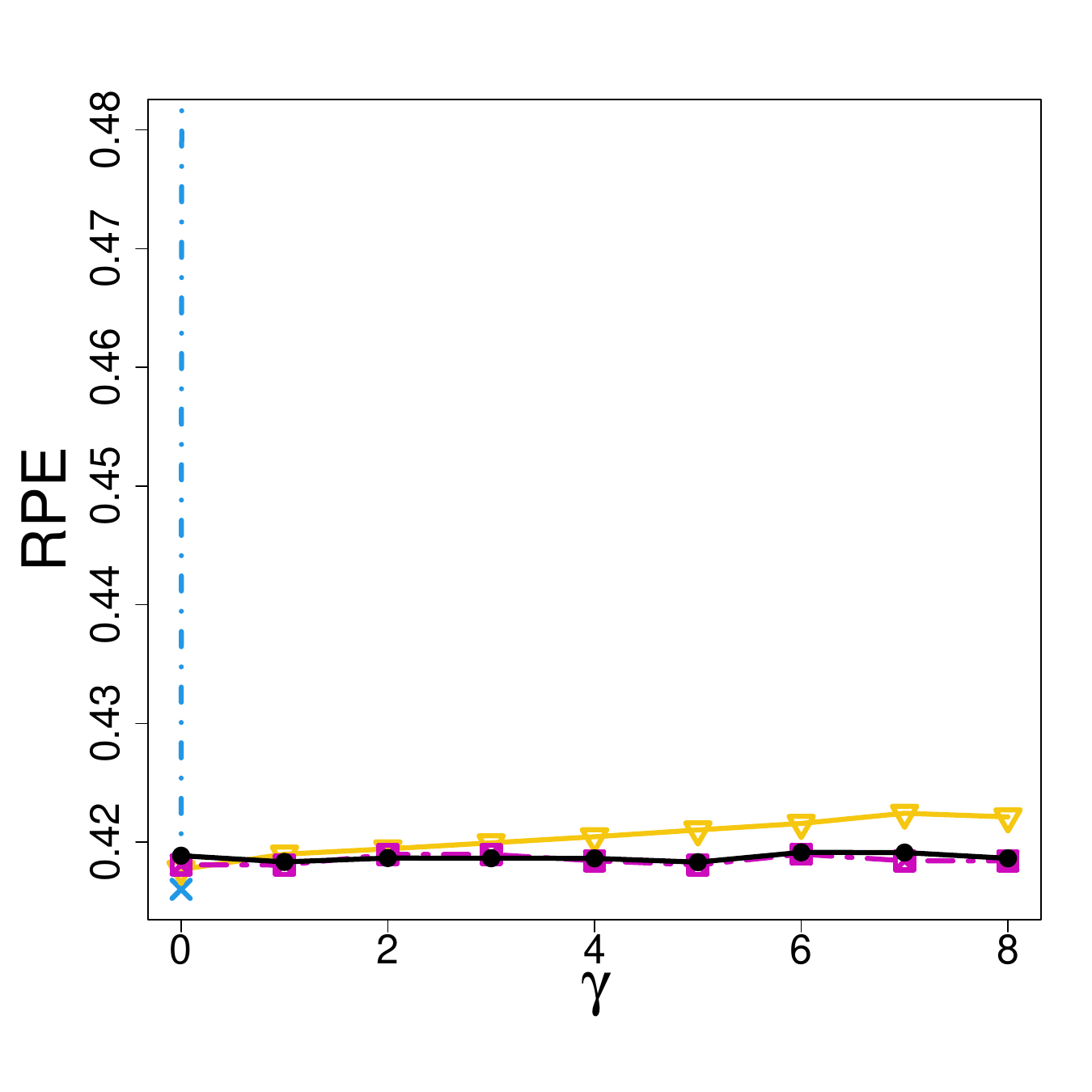}& %
\includegraphics[width=.30\textwidth]{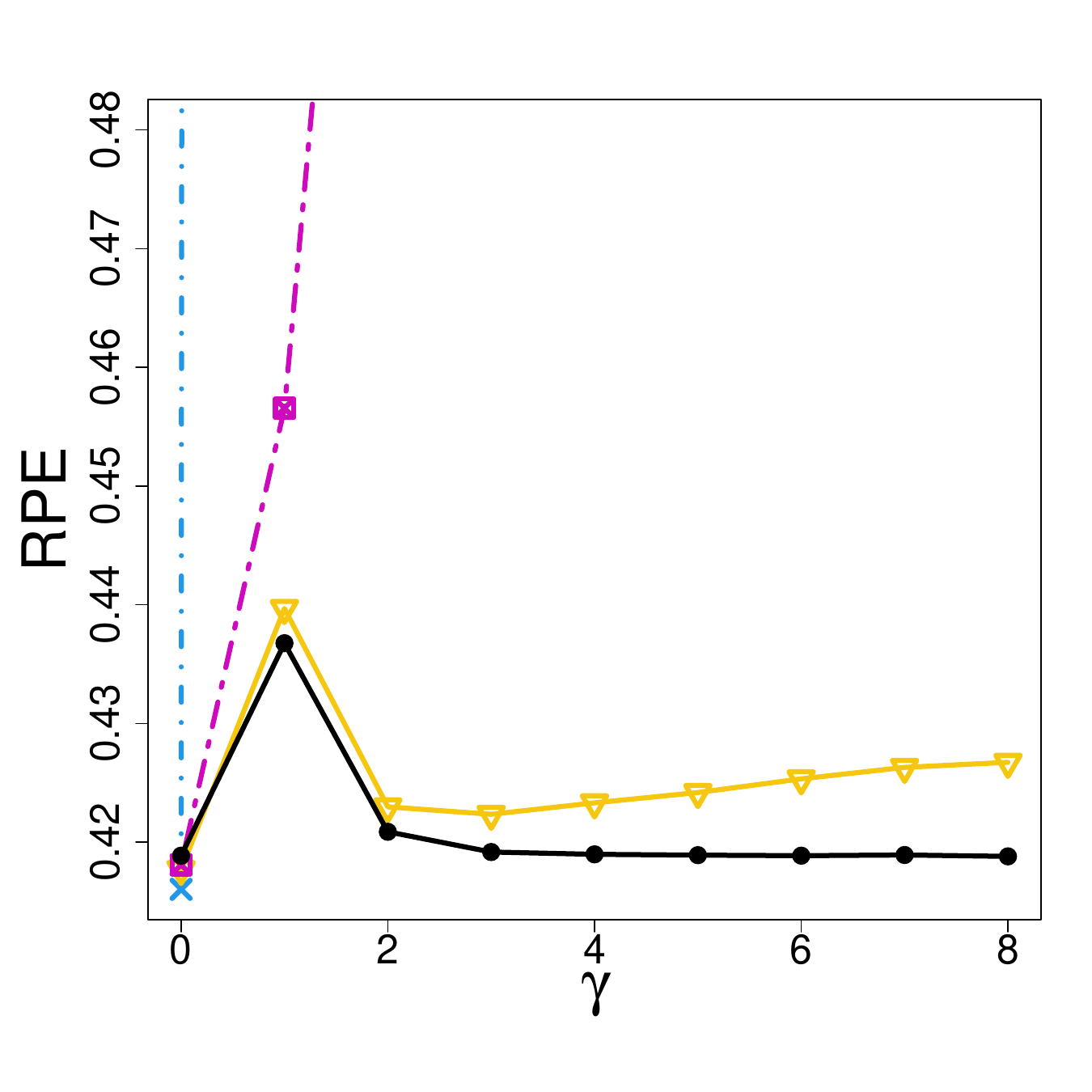}\\ 
\rotatebox{90}{\normalsize  {\parbox{5cm}{\centering Variation in $\cbr{\cx_n}$}}} &  
\includegraphics[width=.30\textwidth]{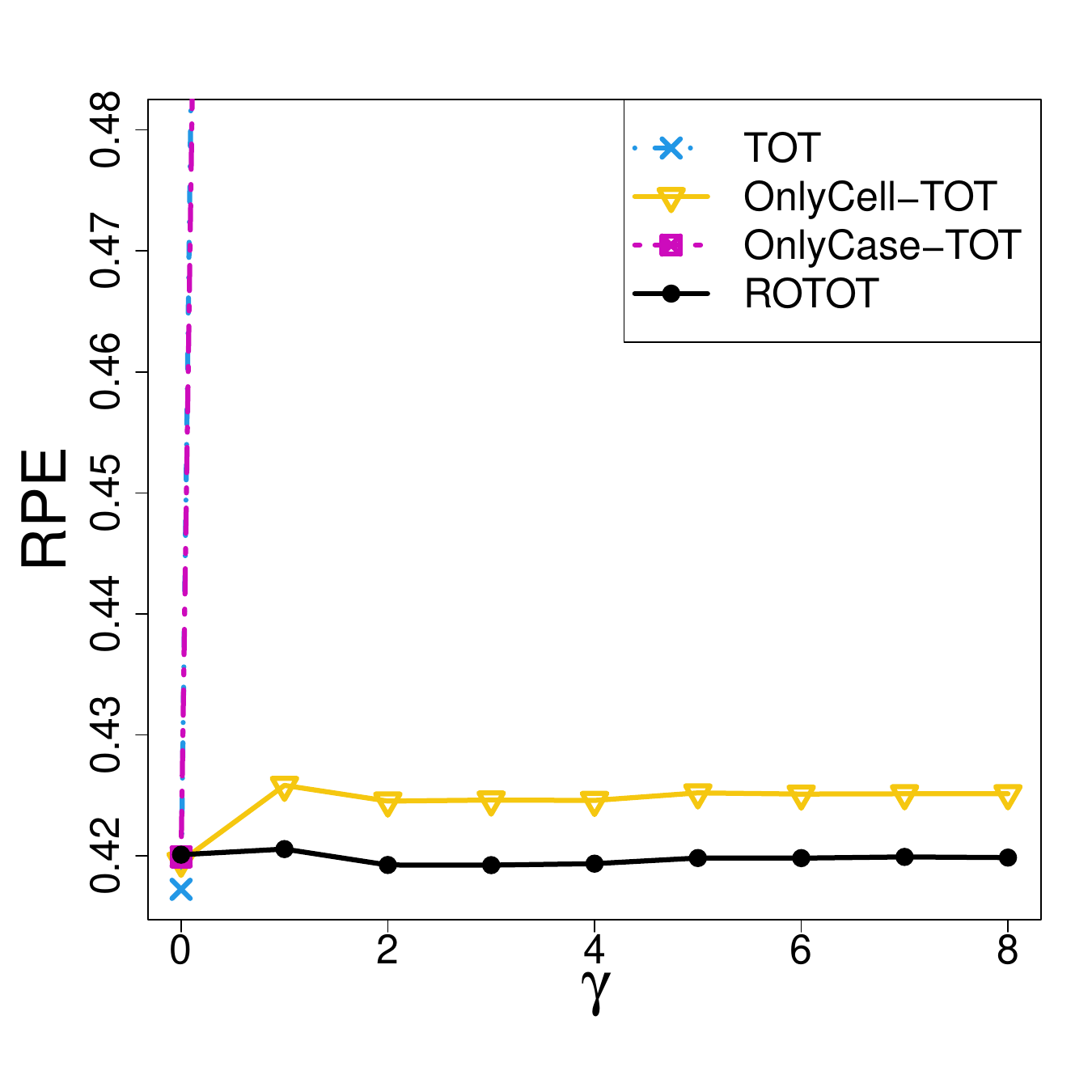} &%
\includegraphics[width=.30\textwidth]{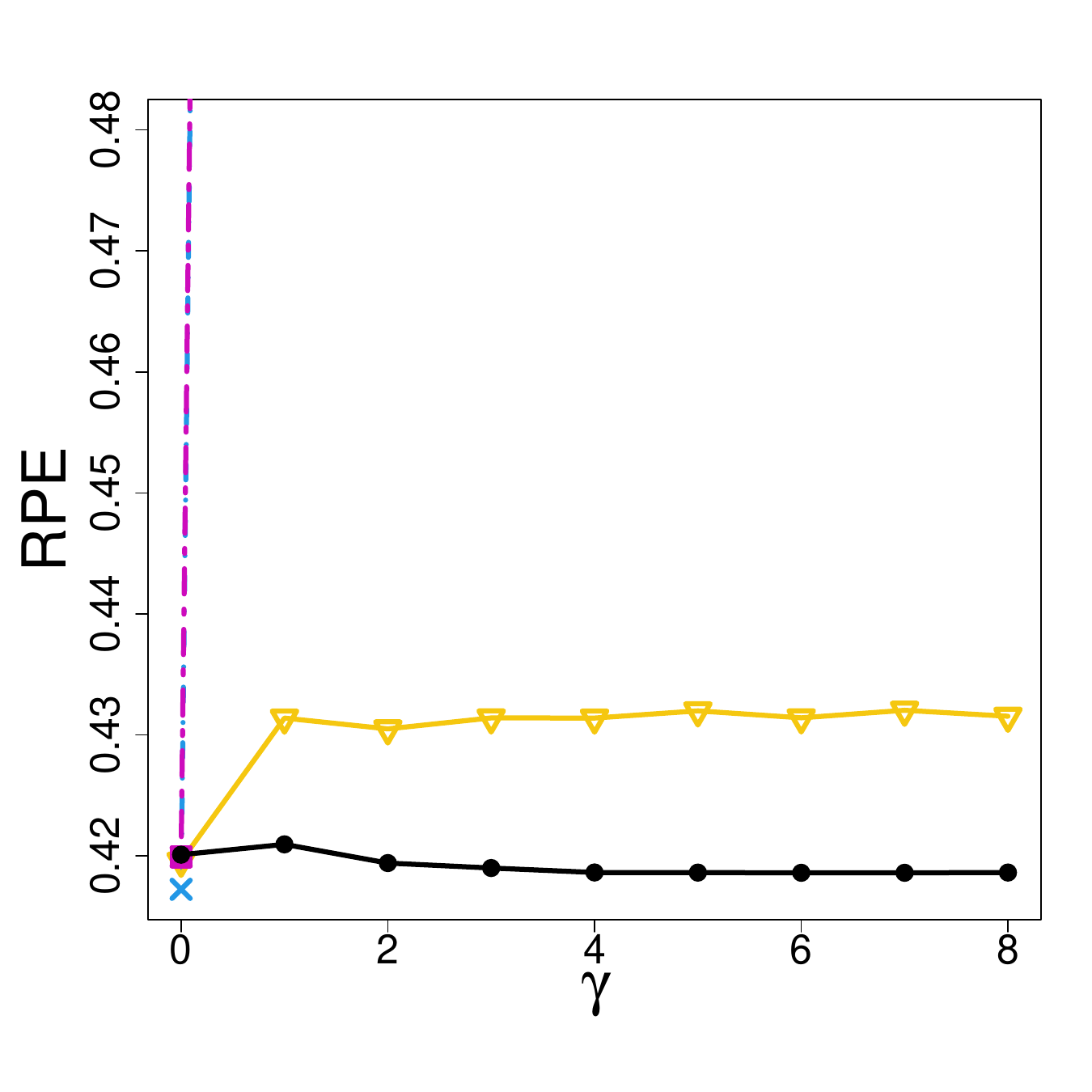} &%
\includegraphics[width=.30\textwidth]{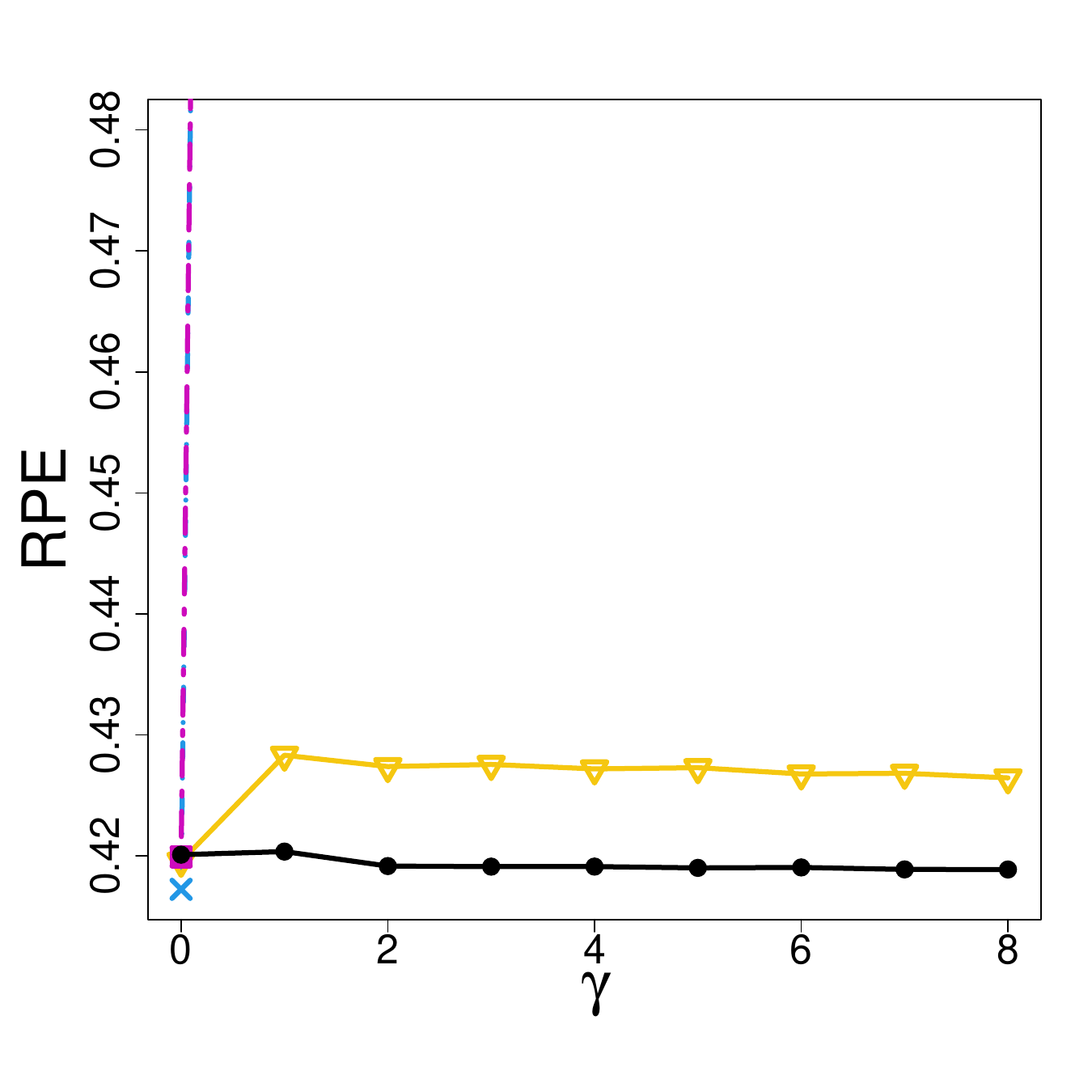} 
\end{tabular}
\caption{ {Median RPE attained by TOT, OnlyCell-TOT, OnlyCase-TOT, and ROTOT across the contamination scenarios for varying contamination magnitude in $\cbr{\cy_n}$ and $\cbr{\cx_n}$, with $\text{SNR} = 5$.}}%
\label{fig:RPEvarYX_zoomed}%
\end{figure}

 {Figure~\ref{fig:RPEvarYX_zoomed} presents the median RPE of the methods for both scenarios across the different contamination settings and contamination magnitudes. When contamination in the predictor is fixed, TOT and OnlyCase-TOT are outperformed by OnlyCell-TOT and ROTOT under cellwise contamination. Under casewise contamination, OnlyCase-TOT and ROTOT achieve the best performance, whereas ROTOT performs best when both types of contamination are present. When contamination in the response is fixed, ROTOT consistently outperforms the other methods. TOT and OnlyCase-TOT are adversely affected, mainly due to the presence of cellwise contamination in the response, while ROTOT and OnlyCell-TOT maintain good performance across the different settings.
  {This indicates again that the combination of imputation of $\cbr{\cx_n}$ and the downweighting of casewise outliers in $\cbr{\cx_n}$ is effective to handle outliers in the predictor tensors.}}

Additional simulations to those presented in Section~\ref{sec:simu} are conducted to assess the performance of ROTOT in the presence of missing values. We again consider two simulation scenarios. 
In the first scenario, the predictors $\cbr{\cx_n}$ are contaminated with $\gamma_{\text{\tiny cell}} = 30$, $\gamma_{\text{\tiny case}} = 10$, and $\varepsilon_{\text{\tiny cell}} = \varepsilon_{\text{\tiny case}} = \varepsilon_{\text{\tiny obs}} = 5\%$, while the contamination in the responses $\cbr{\cy_n}$ varies.
This includes a setting with only cellwise contamination, where $\varepsilon_{\text{\tiny cell}} = 10\%$ of outliers with $c^{\text{\tiny cell}} = 4.5$ and $\varepsilon_{\text{\tiny obs}} = 5\%$ missing values are introduced, and a setting with only casewise contamination, where $\varepsilon_{\text{\tiny case}} = 10\%$ of outliers with $c^{\text{\tiny case}} = 0.5$ and $\varepsilon_{\text{\tiny obs}} = 5\%$ missing values are introduced. 
In the last setting, the responses are contaminated with $\varepsilon_{\text{\tiny cell}} = 10\%$ cellwise, $\varepsilon_{\text{\tiny case}} = 10\%$ casewise outliers, and $\varepsilon_{\text{\tiny obs}} = 5\%$ missing values, using  $c^{\text{\tiny cell}} = 3.5$ and $c^{\text{\tiny case}} = 0.5$. 
We consider two signal-to-noise ratios: $\text{SNR} = 1$ and $\text{SNR} = 5$. The parameter $\gamma$ ranges from 0 to 8,  data are uncontaminated when $\gamma = 0$.

In the second scenario, the responses $\cbr{\cy_n}$ are contaminated with $\gamma_{\text{\tiny cell}} = 20$ and $\gamma_{\text{\tiny case}} = 3.5$ for $\text{SNR} = 1$ and $\gamma_{\text{\tiny case}} = 4$ for $\text{SNR} = 5$, with $\varepsilon_{\text{\tiny cell}} = \varepsilon_{\text{\tiny case}} = 10\%$ and $\varepsilon_{\text{\tiny obs}} = 5\%$, while the contamination in the predictors $\cbr{\cx_n}$ varies.
This includes a setting with only cellwise contamination, where $\varepsilon_{\text{\tiny cell}} = 10\%$ of outliers with $c^{\text{\tiny cell}} = 1.5$ and $\varepsilon_{\text{\tiny obs}} = 5\%$ missing values are introduced, and a setting with only casewise contamination, where $\varepsilon_{\text{\tiny case}} = 10\%$ of outliers with $c^{\text{\tiny case}} = 1$ and $\varepsilon_{\text{\tiny obs}} = 5\%$ missing values are introduced.
In the last setting, the responses are contaminated with $\varepsilon_{\text{\tiny cell}} = 5\%$ cellwise, $\varepsilon_{\text{\tiny case}} = 5\%$ casewise outliers, and $\varepsilon_{\text{\tiny obs}} = 5\%$ missing values, using  $c^{\text{\tiny cell}} = 1.5$ and $c^{\text{\tiny case}} = 1$.

\begin{figure}[H]
\centering
\begin{tabular}{cccc}
 &\textbf{Cellwise in $\brc{\cy_n}$}  &\textbf{Casewise in $\brc{\cy_n}$} &{\textbf{Casewise \& Cellwise in $\brc{\cy_n}$}}\\
\rotatebox{90}{\normalsize {\parbox{5cm}{\centering $\text{SNR} = 1$ }}} & 
\includegraphics[width=.30\textwidth]{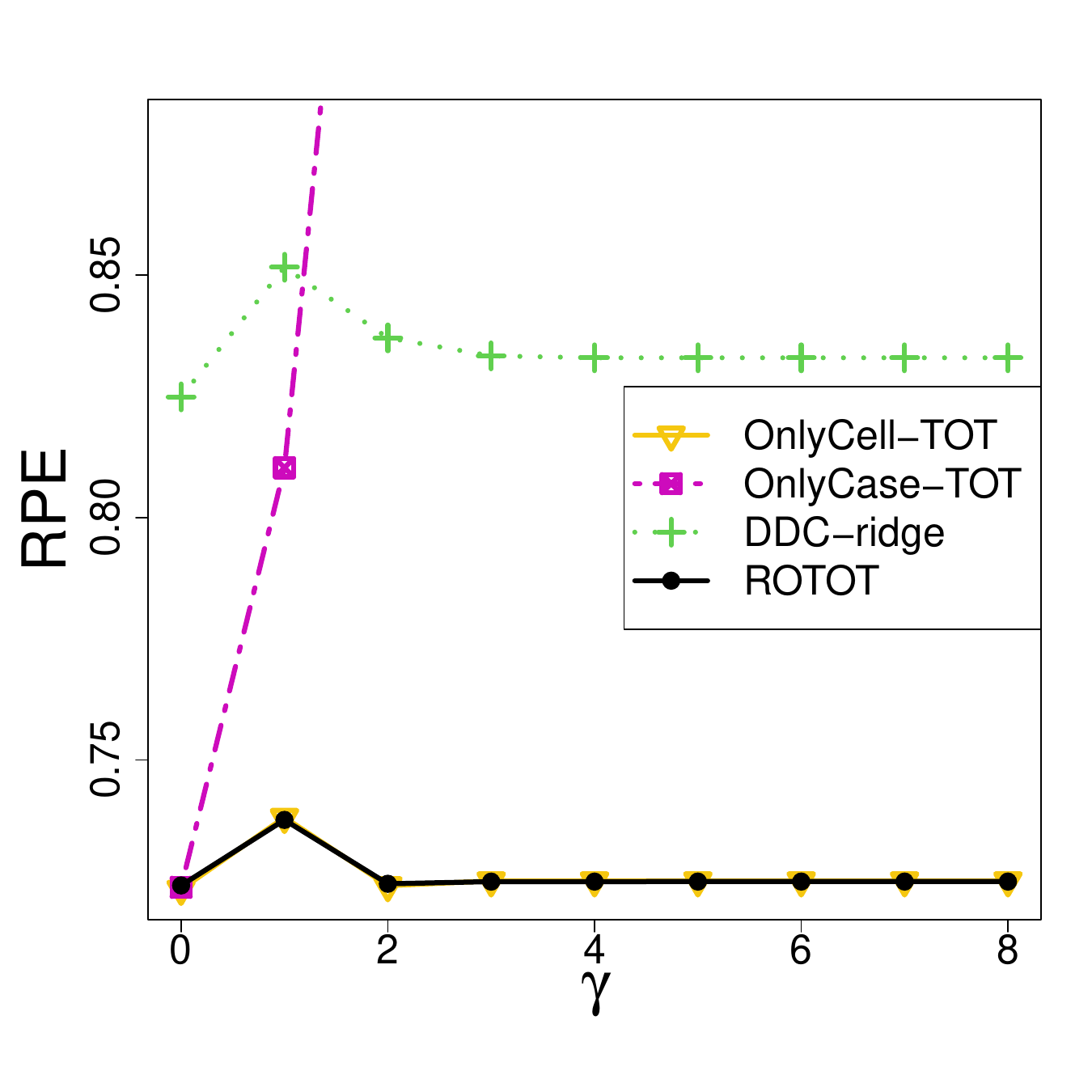} &%
\includegraphics[width=.30\textwidth]{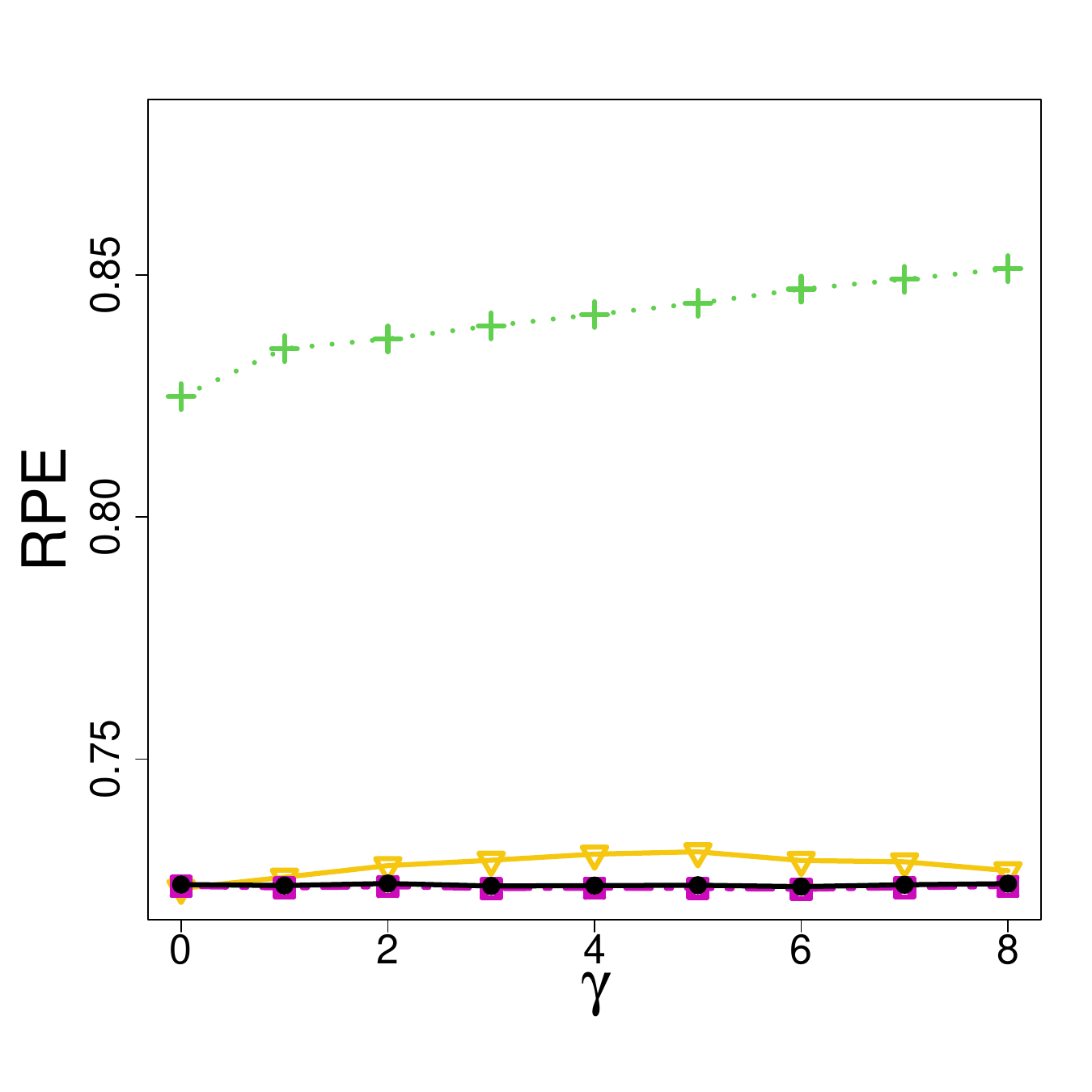}& %
\includegraphics[width=.30\textwidth]{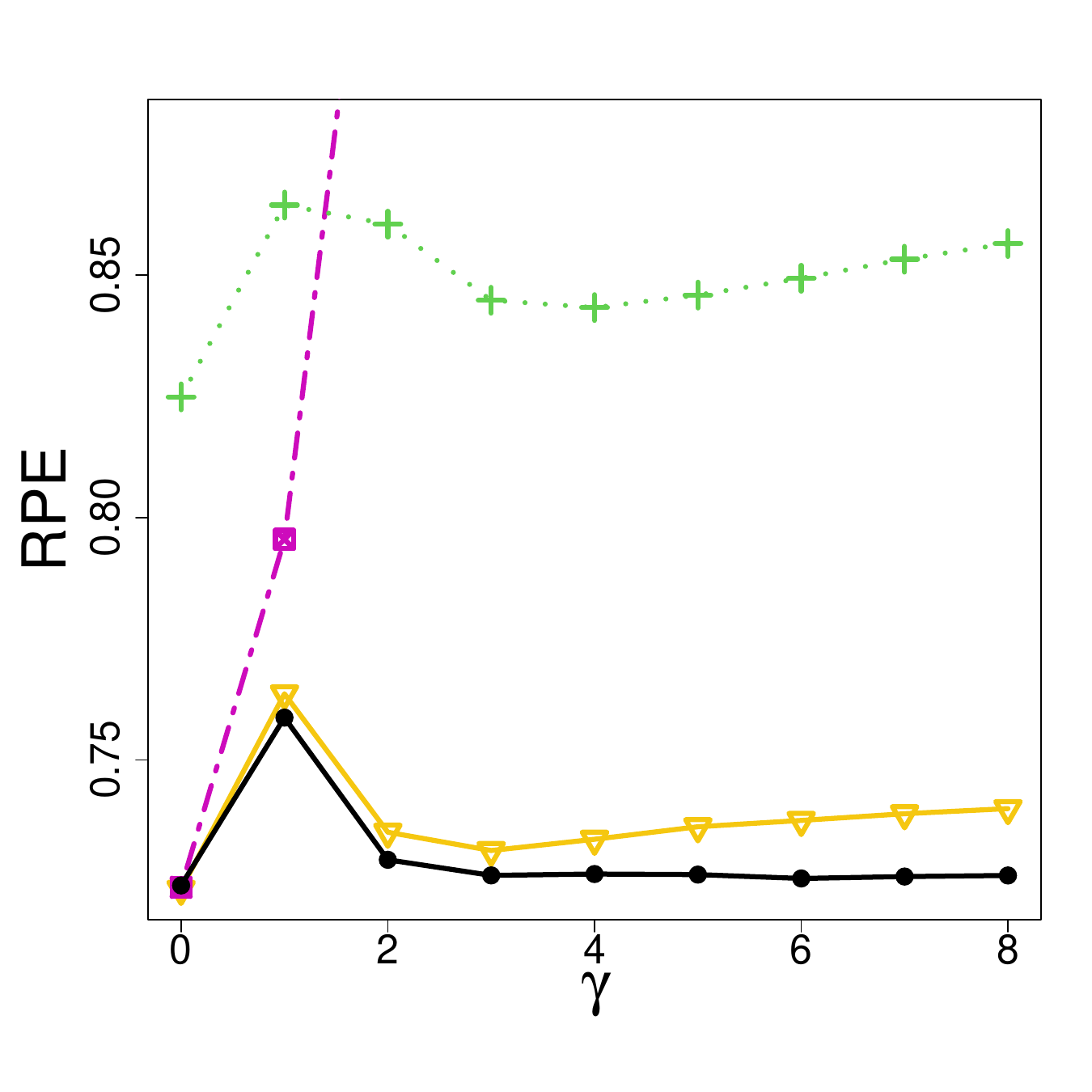}\\ 
\rotatebox{90}{\normalsize  {\parbox{5cm}{\centering $\text{SNR} = 5$}}} & 
\includegraphics[width=.30\textwidth]{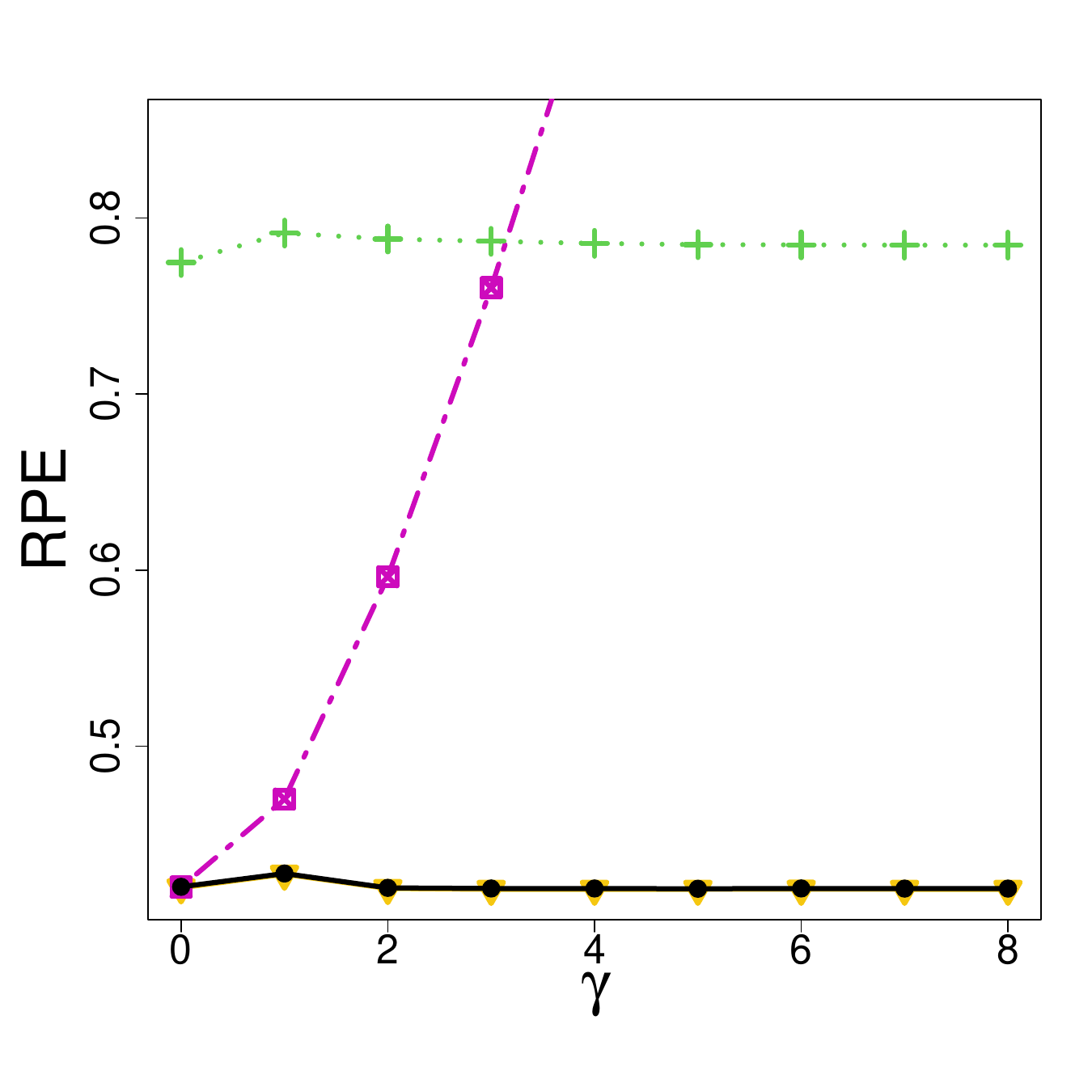} &%
\includegraphics[width=.30\textwidth]{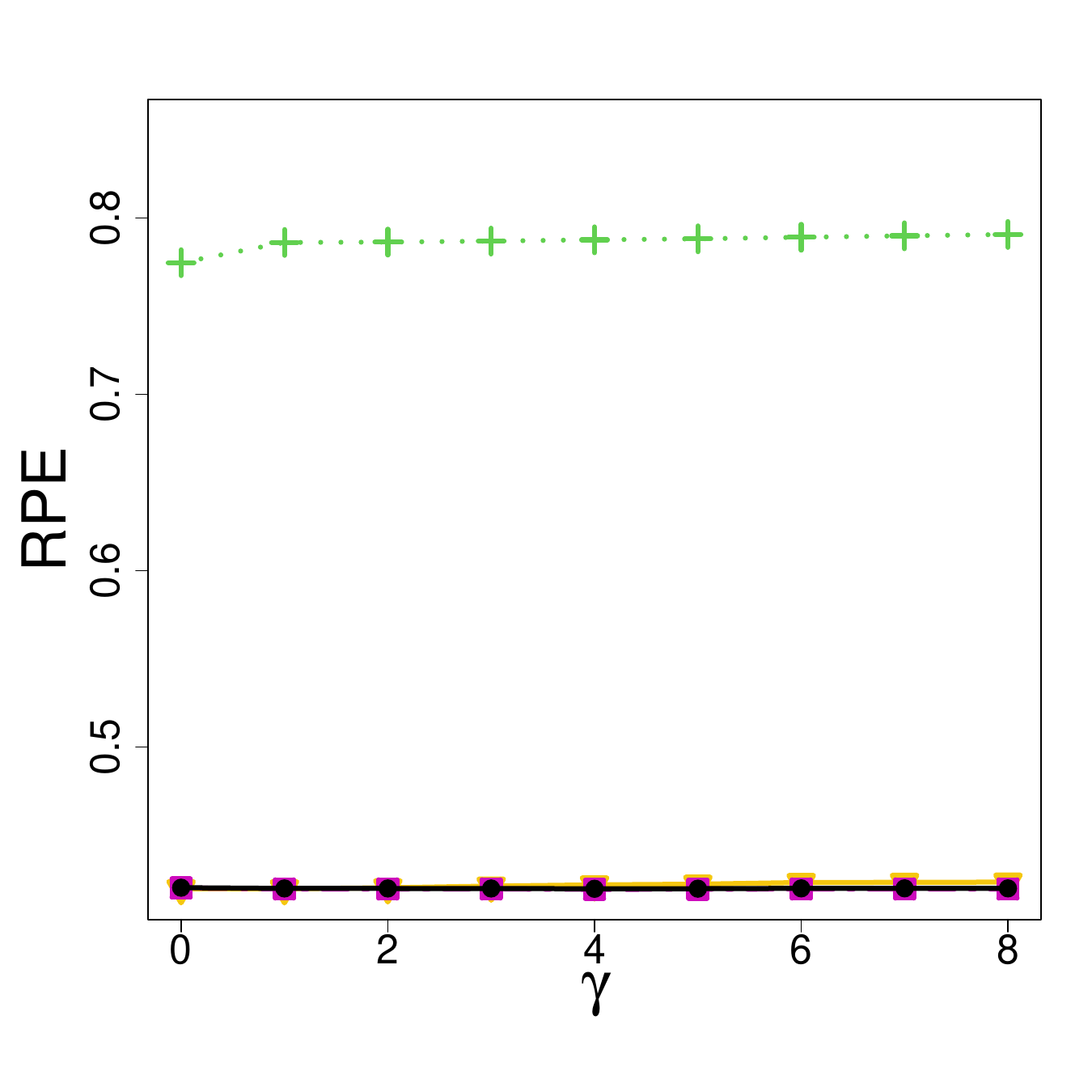} &%
\includegraphics[width=.30\textwidth]{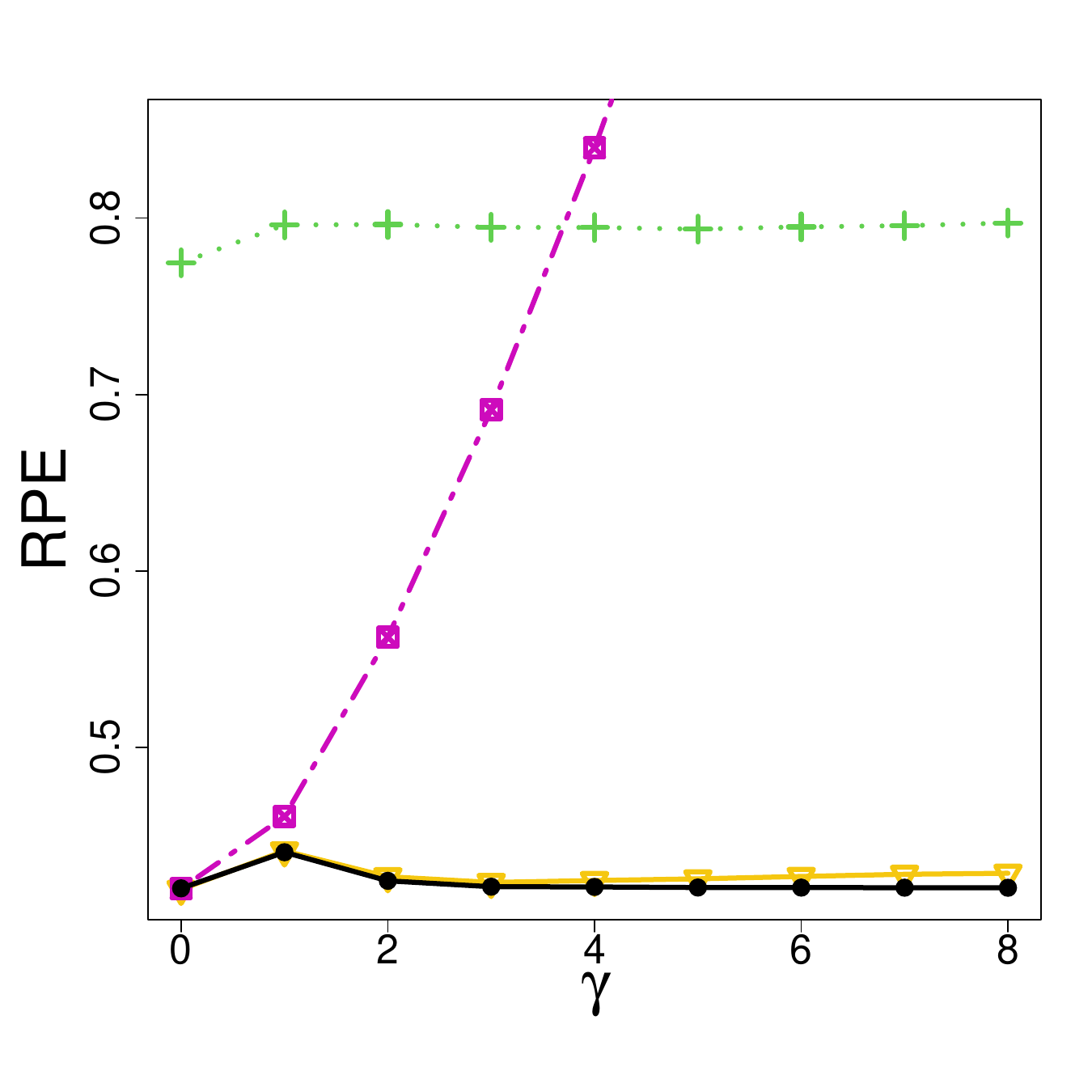} 
\end{tabular}
\caption{ {Median RPE attained by OnlyCell-TOT, OnlyCase-TOT, DDC-ridge and ROTOT across the different contamination settings, with fixed contamination in the predictors and varying contamination magnitudes in the responses, with missing values and $\text{SNR} = \{1,5\}$.}}%
\label{fig:RPEvarYNA}%
\end{figure}

\begin{figure}[H]
\centering
\begin{tabular}{cccc}
  &\textbf{Cellwise in $\brc{\cx_n}$}  &\textbf{Casewise in $\brc{\cx_n}$} &{\textbf{Casewise \& Cellwise in $\brc{\cx_n}$}}\\
\rotatebox{90}{\normalsize {\parbox{5cm}{\centering $\text{SNR} = 1$ }}} & 
\includegraphics[width=.30\textwidth]{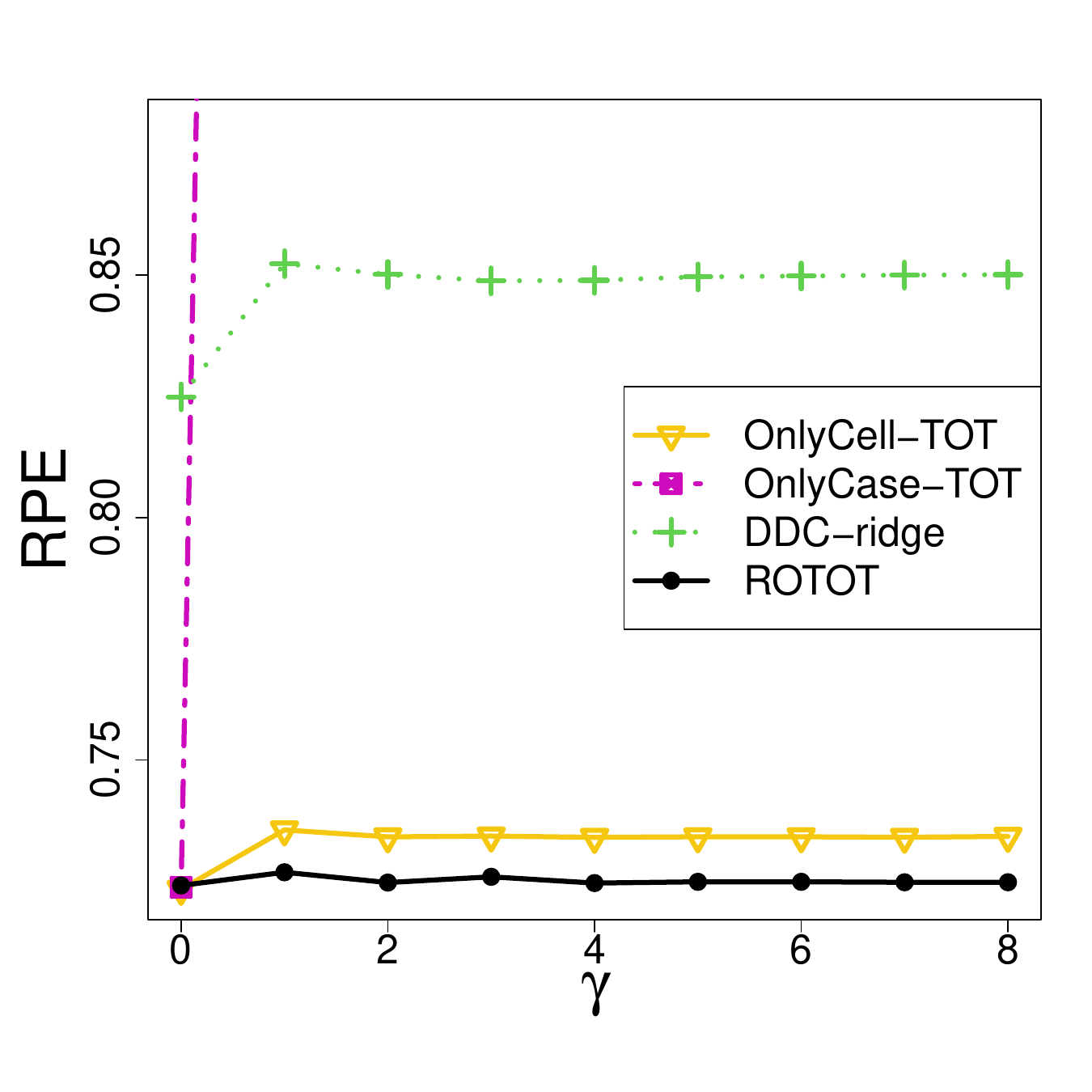} &%
\includegraphics[width=.30\textwidth]{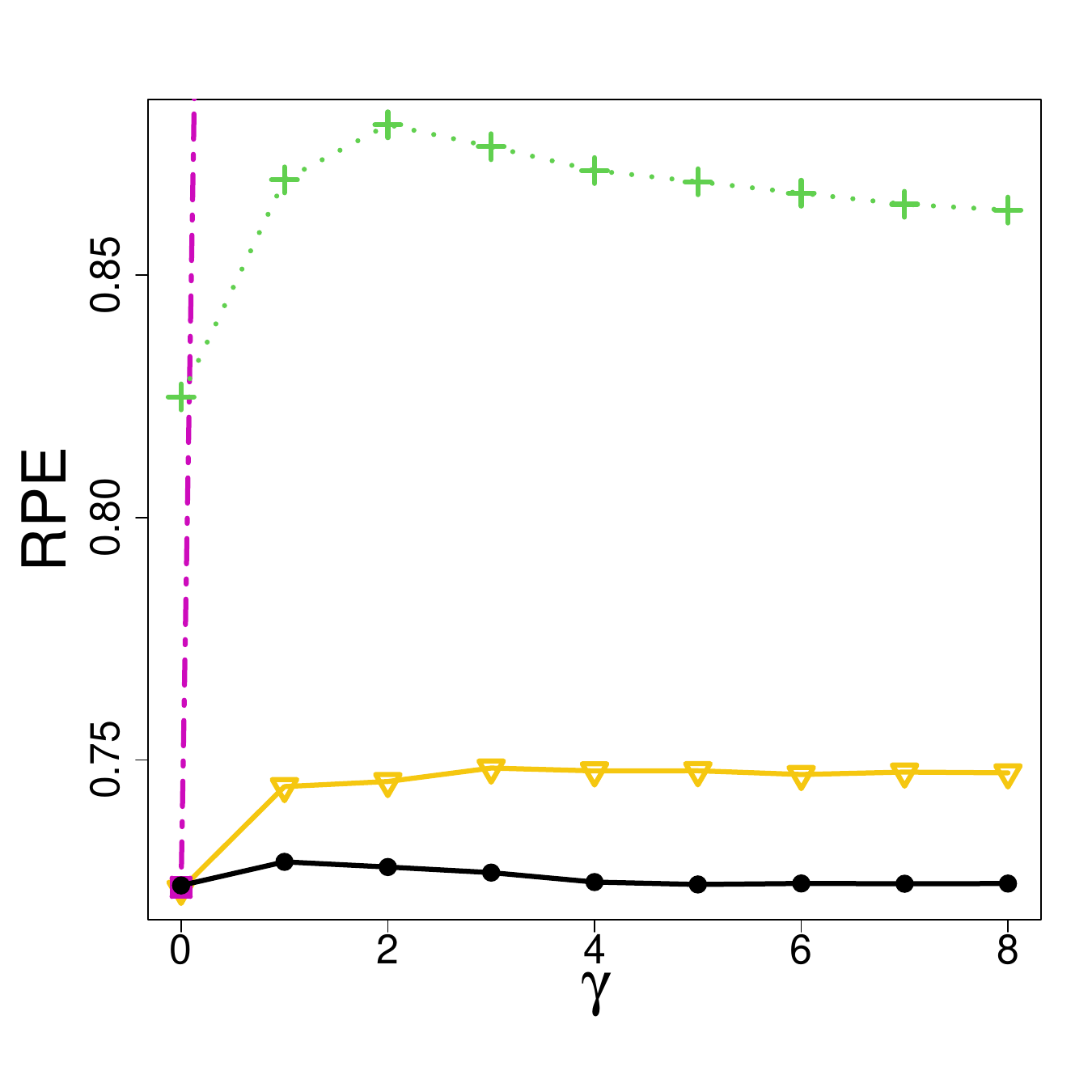}& %
\includegraphics[width=.30\textwidth]{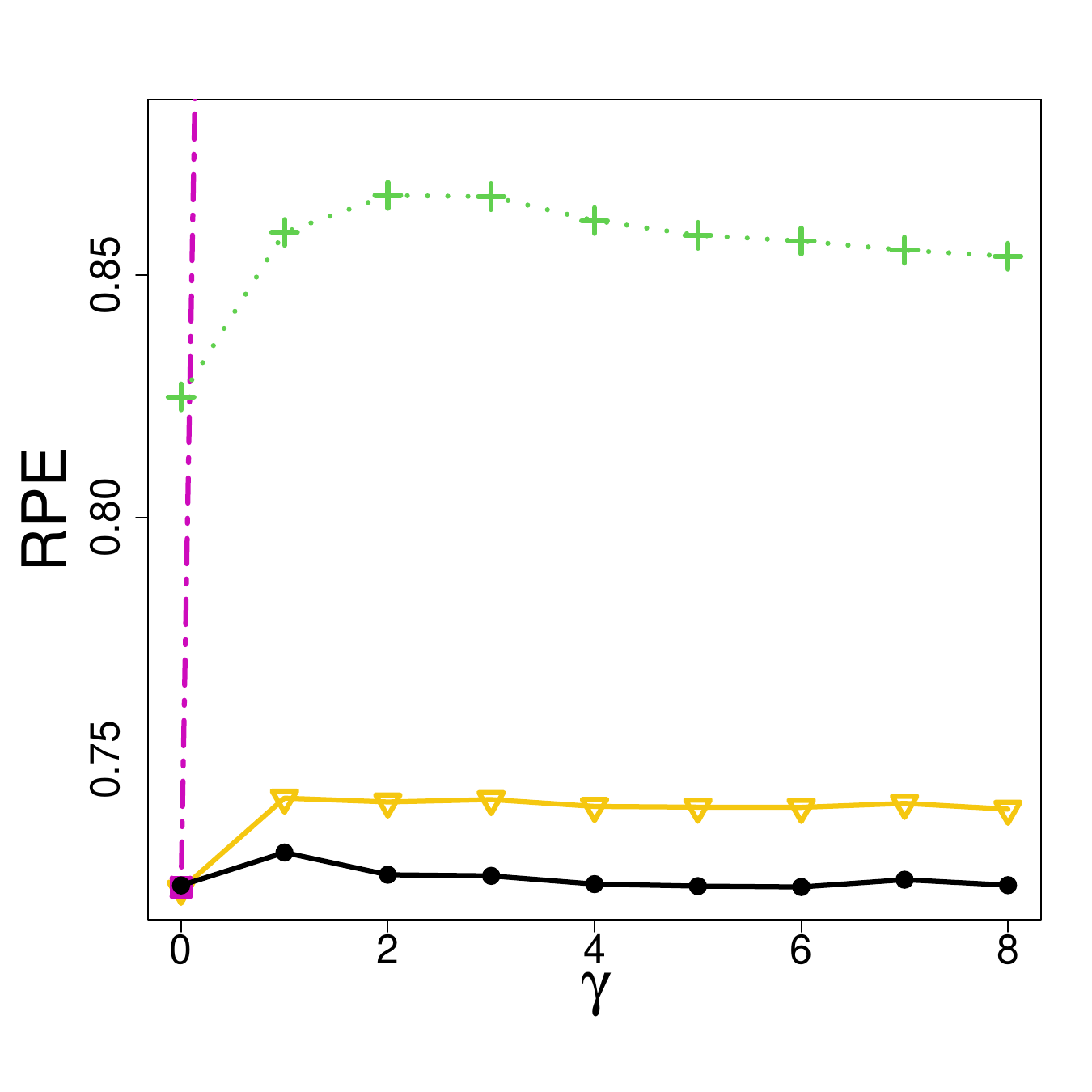}\\ 
\rotatebox{90}{\normalsize  {\parbox{5cm}{\centering $\text{SNR} = 5$}}} &  
\includegraphics[width=.30\textwidth]{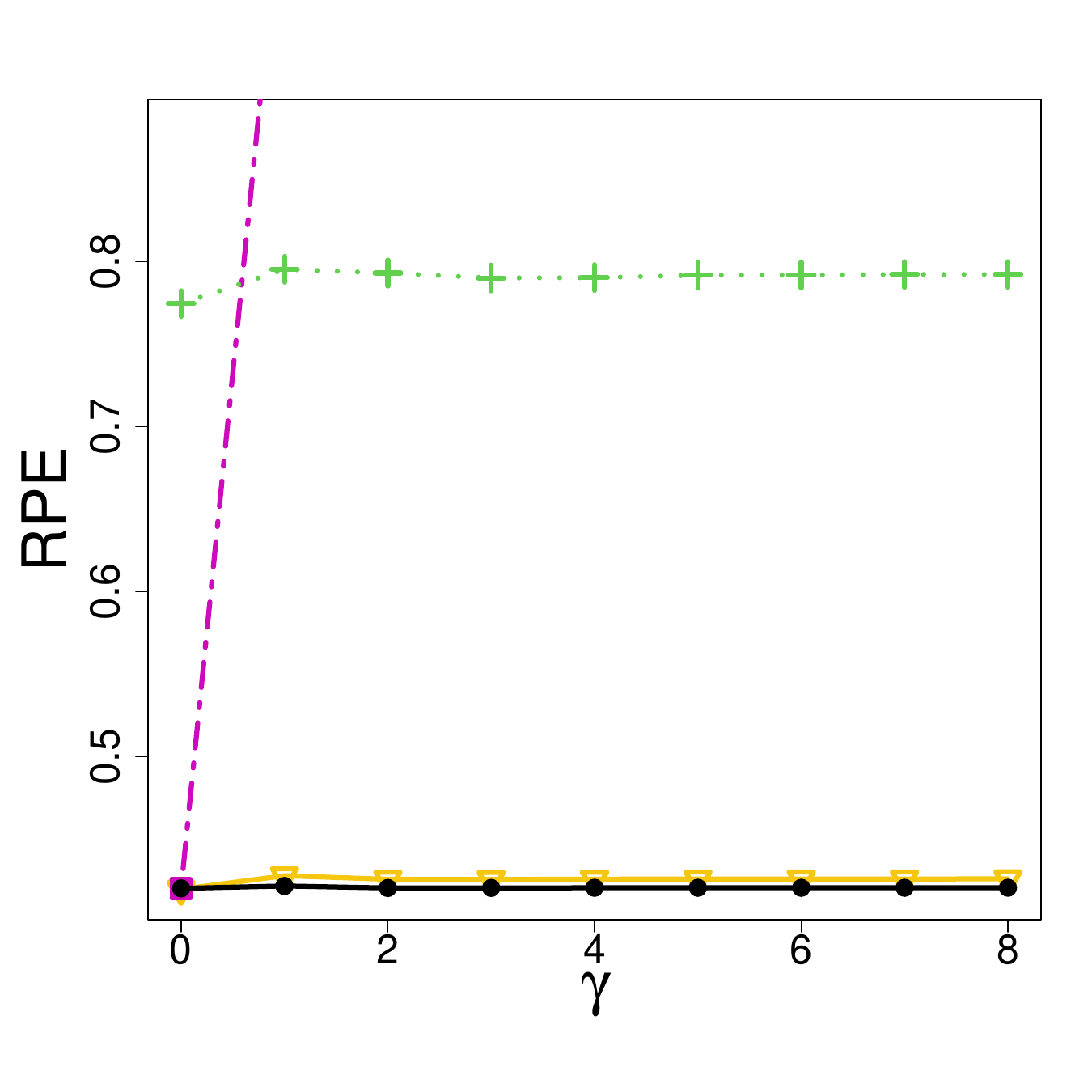} &%
\includegraphics[width=.30\textwidth]{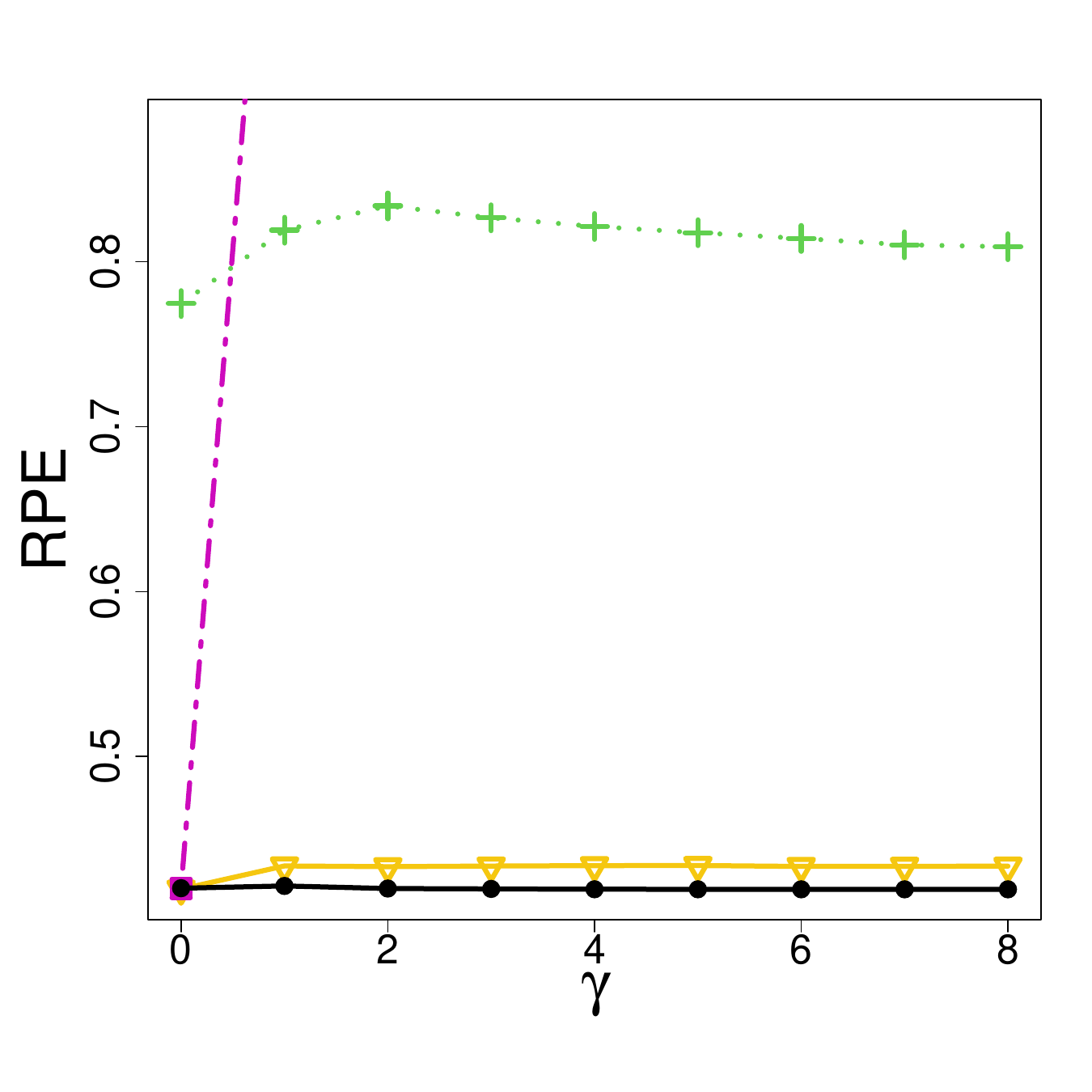} &%
\includegraphics[width=.30\textwidth]{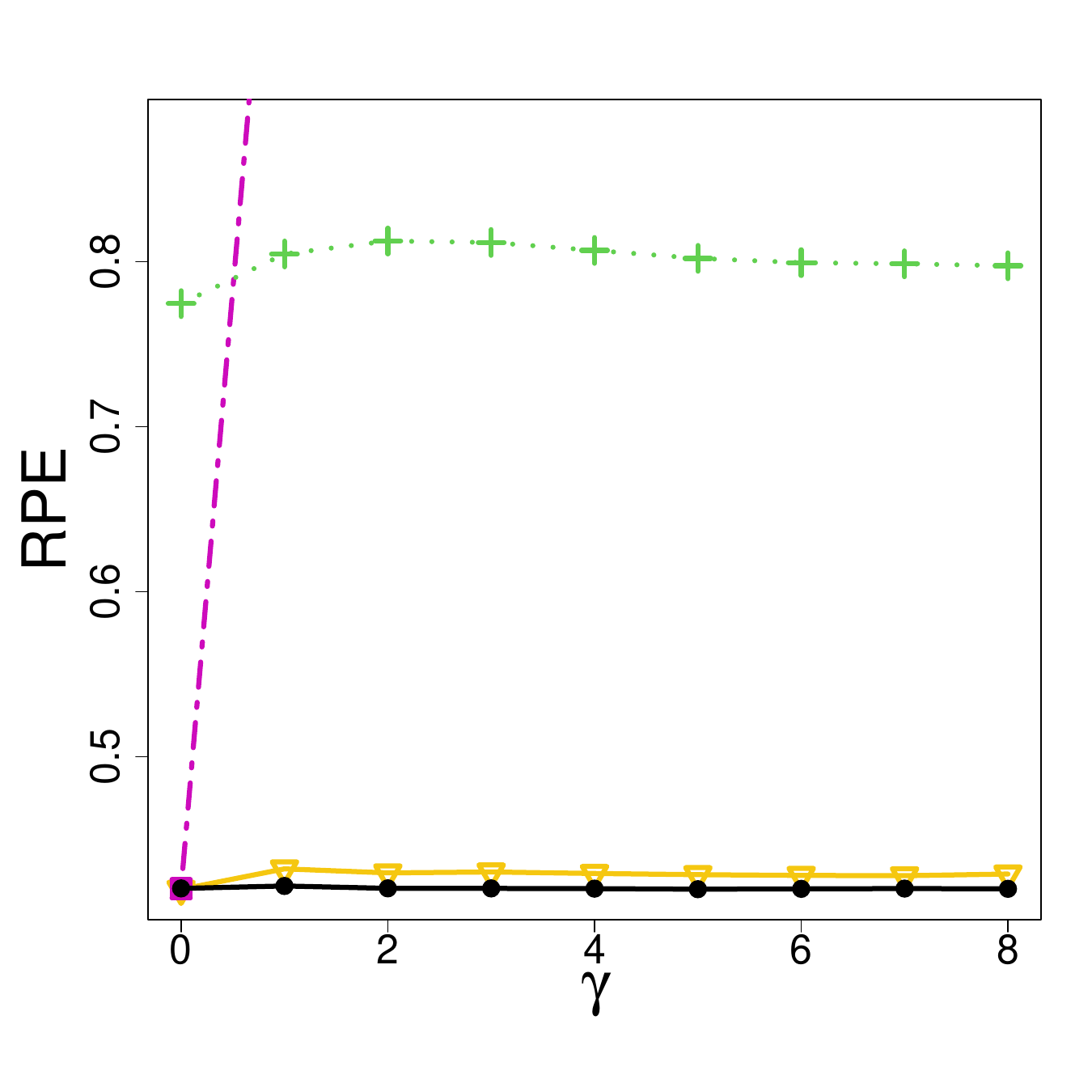} 
\end{tabular}
\caption{ {Median RPE attained by OnlyCell-TOT, OnlyCase-TOT, ROTOT, and DDC-ridge across the different contamination settings, with fixed contamination in the responses and varying contamination magnitudes in the predictors, with missing values and $\text{SNR} = \cbr{1,5}$.}}%
\label{fig:RPEvarXNA}%
\end{figure}

Figure~\ref{fig:RPEvarYNA} and \ref{fig:RPEvarXNA} show the median RPE for all methods that can handle missing values across different contamination scenarios and settings for varying contamination magnitude in the response and predictor, respectively, with missing values and   $\text{SNR} = \cbr{1,5}$. 
The conclusion remains the same as in the case where there are no missing values.

\begin{figure}[H]
\begin{center}
\includegraphics[width=0.35\textwidth]{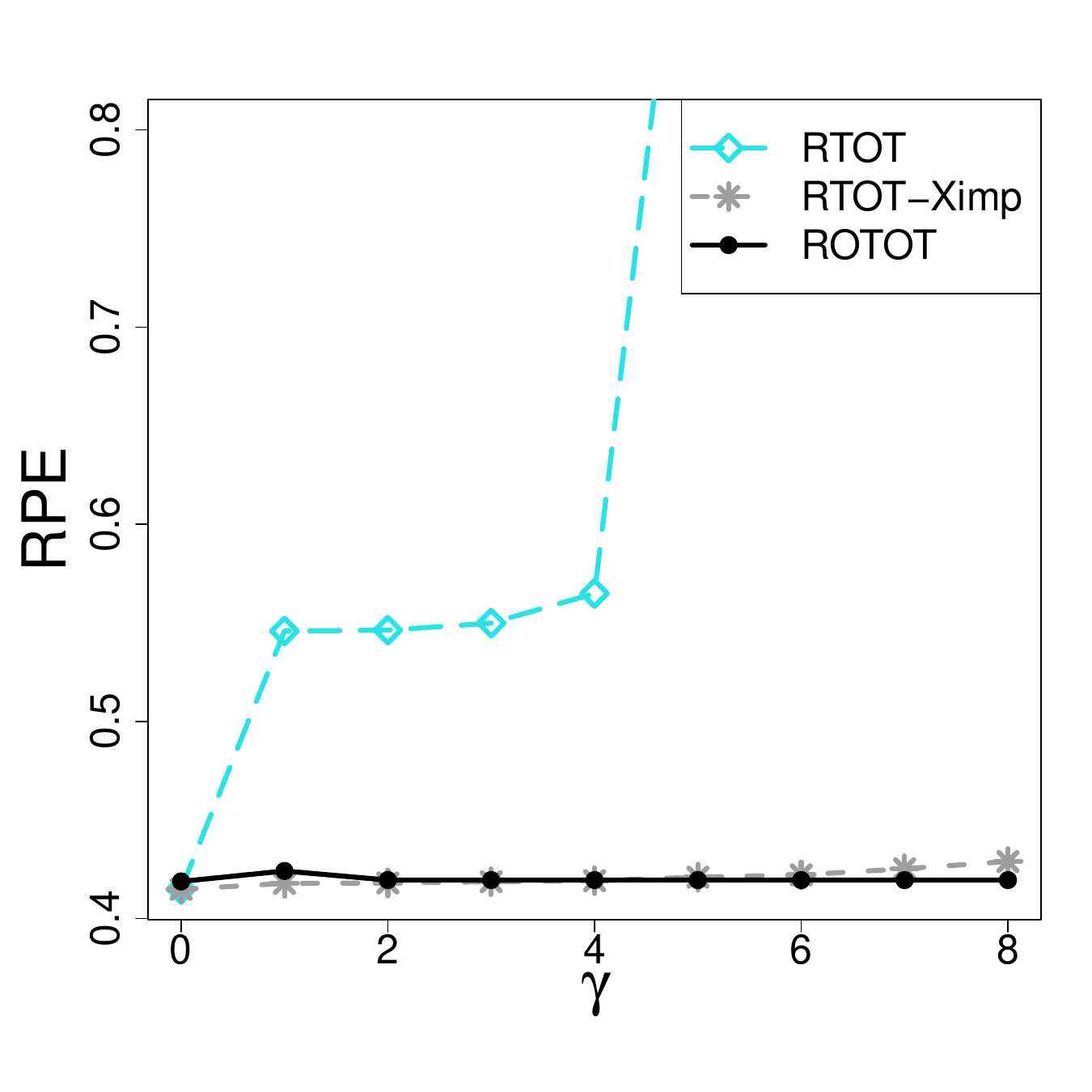}
\end{center}
\vspace{-0.5cm}
\caption{ {Median RPE attained by RTOT, RTOT-Ximp, and ROTOT under cellwise contamination in both the response and predictor for varying contamination magnitude in the response.}}
\label{fig:Simulation_RTOT_comp}
\end{figure}

 {We further compare RTOT-Ximp and ROTOT in the contamination setting for which RTOT-Ximp is specifically designed. Figure~\ref{fig:Simulation_RTOT_comp} presents the median RPE of RTOT, RTOT-Ximp, and ROTOT under cellwise contamination in both the response and predictor tensors, with varying contamination magnitudes in the response and $\text{SNR}=5$, following the first simulation scenario for the cellwise contamination setting described in Section~\ref{sec:simu}. The predictor tensors are contaminated with a fixed severe level of cellwise contamination, with $\gamma_{\text{\tiny cell}}=30$ and $\varepsilon_{\text{\tiny cell}}=5\%$.}
 {As expected, ROTOT performs well in this setting. RTOT performs poorly because it does not account for contamination in the predictor tensors. In contrast, RTOT-Ximp achieves good performance, as the cellwise outliers in the predictor tensors are handled through imputation.}

\begin{figure}[!ht]
\centering
\begin{tabular}{cccc}
 & \textbf{Cellwise in $\brc{\cy_n}$}  & \textbf{Casewise in $\brc{\cy_n}$} &{\textbf{Casewise \& Cellwise in $\brc{\cy_n}$}}\\
\rotatebox{90}{\normalsize {\parbox{5cm}{\centering $\text{SNR} = 5$ }}} & 
\includegraphics[width=.30\textwidth]{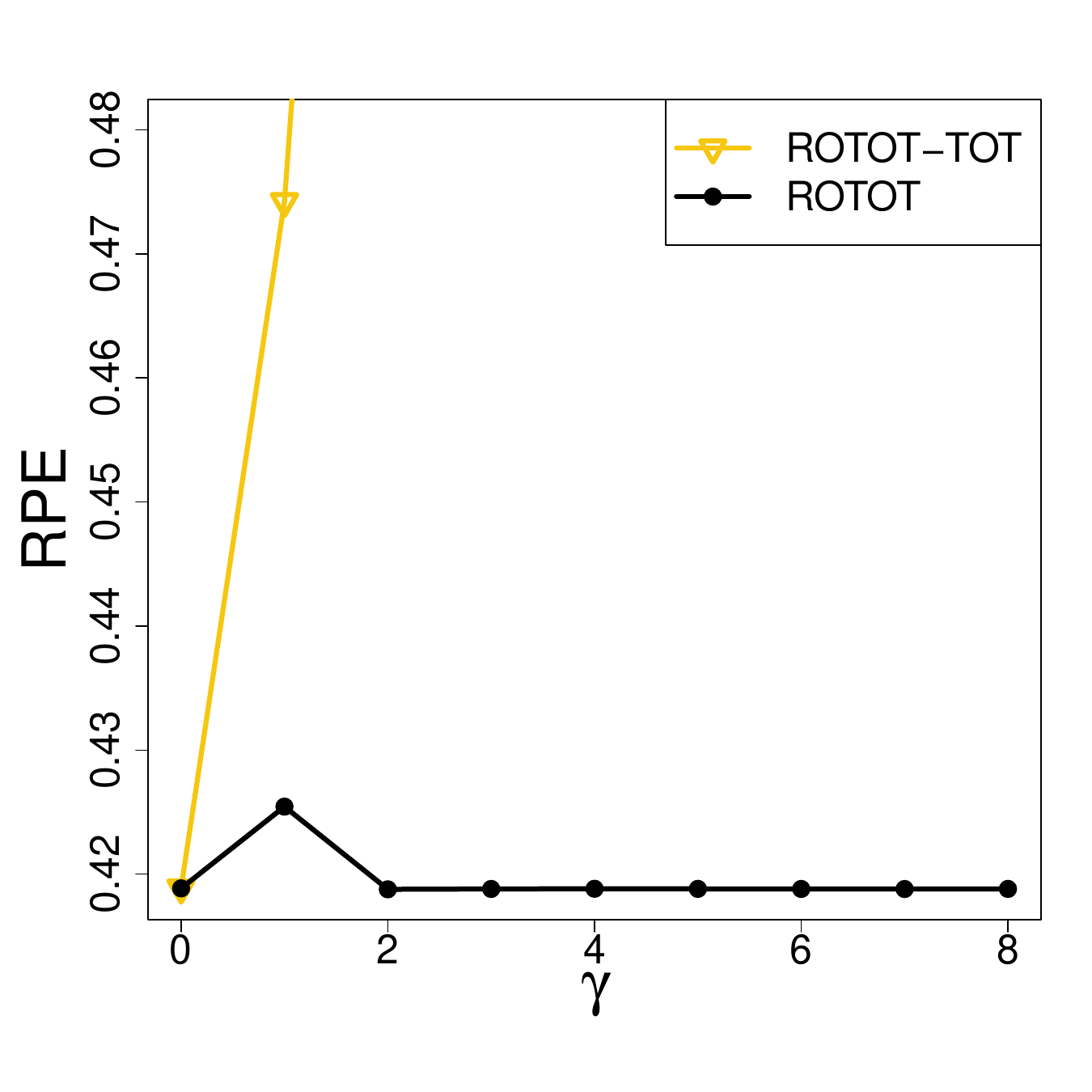} &%
\includegraphics[width=.30\textwidth]{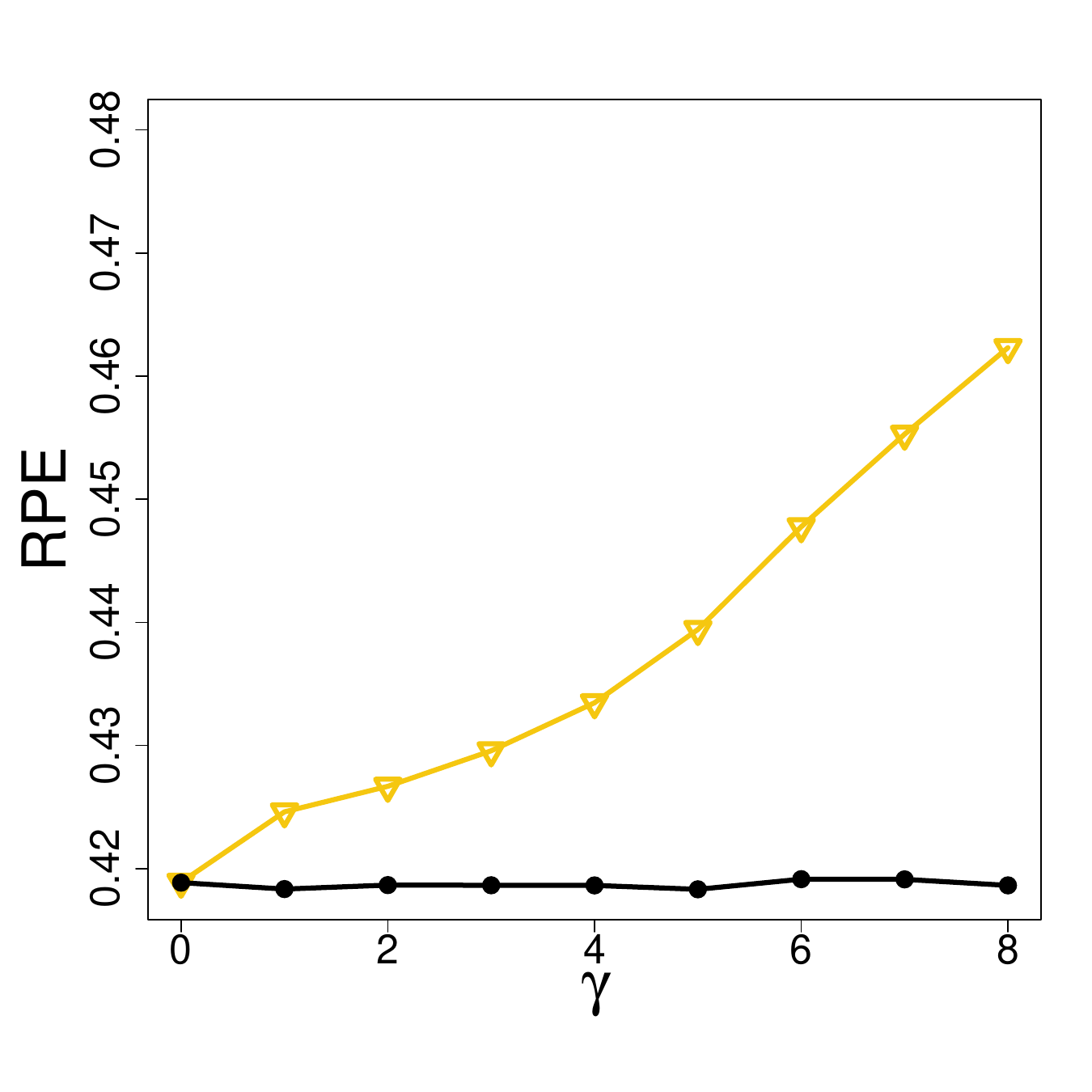}& %
\includegraphics[width=.30\textwidth]{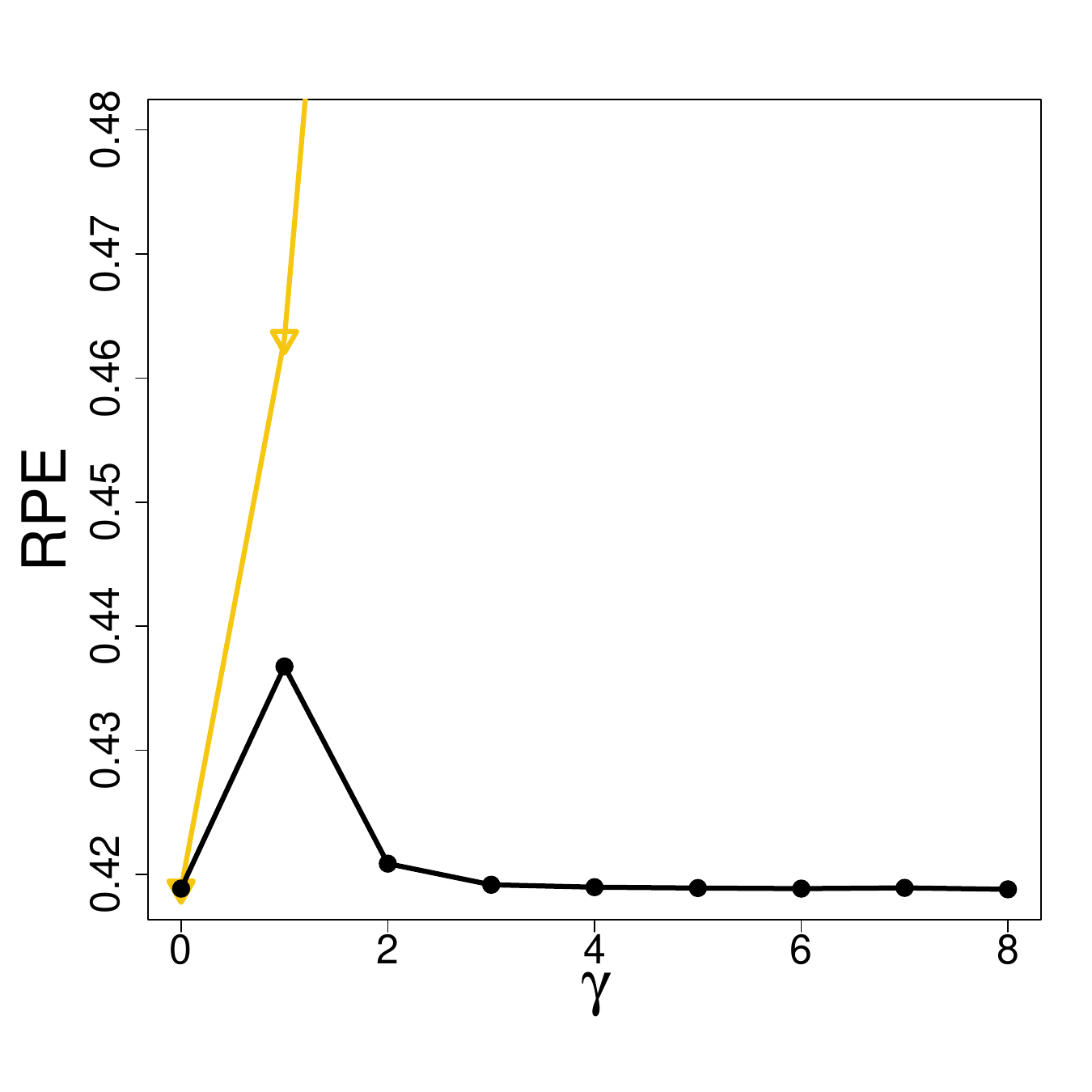}\\ 
\end{tabular}
\caption{ {Median RPE attained by ROTOT-TOT and ROTOT across the different contamination settings, with fixed contamination in the predictors and varying contamination magnitudes in the responses.}}%
 \label{fig:RPEvarYinitchoice}%
\end{figure}

 {Figure~\ref{fig:RPEvarYinitchoice} illustrates the sensitivity of ROTOT to the choice of initialization strategy. 
  {The plots report the median RPE across the different contamination settings and contamination magnitudes in the response, for $\text{SNR} = 5$ under the first contamination scenario as described in Section~\ref{sec:simu}.}
ROTOT-TOT denotes ROTOT initialized with  TOT, whereas ROTOT uses the initialization strategy described in Section~\ref{sec:setup}.
As expected, ROTOT-TOT performs worse, since the  TOT initialization is not robust to outliers. In contrast, ROTOT maintains good performance, consistent with the preceding simulation results. These findings highlight the importance of using a robust initialization for ROTOT.}

\begin{figure}[H]
\begin{center}
\includegraphics[width=0.35\textwidth]{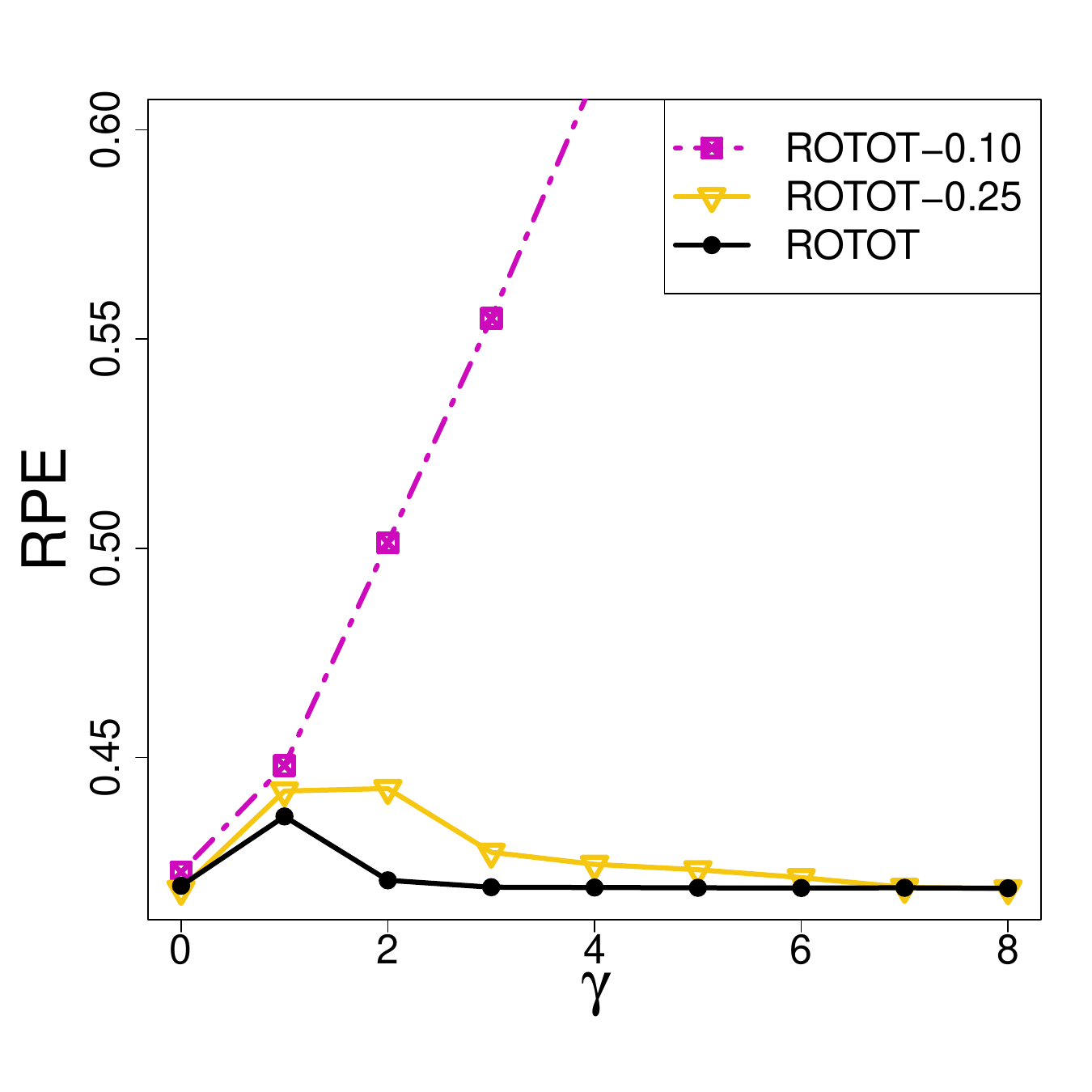}
\end{center}
\vspace{-0.5cm}
\caption{ {Median RPE attained by ROTOT-0.10, ROTOT-0.25, and ROTOT under casewise and cellwise contamination in both responses and predictors for varying contamination magnitude in the responses.}}
\label{fig:Simulation_BV}
\end{figure}

 {Figure~\ref{fig:Simulation_BV} illustrates the sensitivity of the proposed ROTOT method to the choice of the tuning constants $\delta$ and $a$ in the M-scale defined in~\eqref{eq:Mscale}. 
  {The plots report the median RPE under casewise and cellwise contamination in both the response and predictor tensors, without missing values, for different choices of these tuning constants and varying contamination magnitudes in the response, with $\text{SNR} = 5$ under the first contamination scenario described in Section~\ref{sec:simu}.}
We compare the default values, $\delta=1.8811$ and $a=0.34308$, which correspond to the maximal breakdown value of $50\%$, with two alternative calibrations. The first corresponds to a breakdown value of $25\%$, obtained with $\delta=0.94055$ and $a=0.6492$, while the second corresponds to a breakdown value of $10\%$, obtained with $\delta=0.37622$ and $a=1.1291$. We denote these three settings by ROTOT, ROTOT-0.25, and ROTOT-0.10, respectively.}

 {ROTOT performs well, as already shown in the previous simulations. ROTOT-0.25 also maintains good performance, since the overall proportion of contamination in the response tensors remains below $25\%$ in the considered simulation setting. In contrast, ROTOT-0.10 breaks down completely because the contamination level exceeds the $10\%$ breakdown value for which its M-scale is calibrated.}

\begin{figure}[H]
\begin{center}
\includegraphics[width=0.35\textwidth]{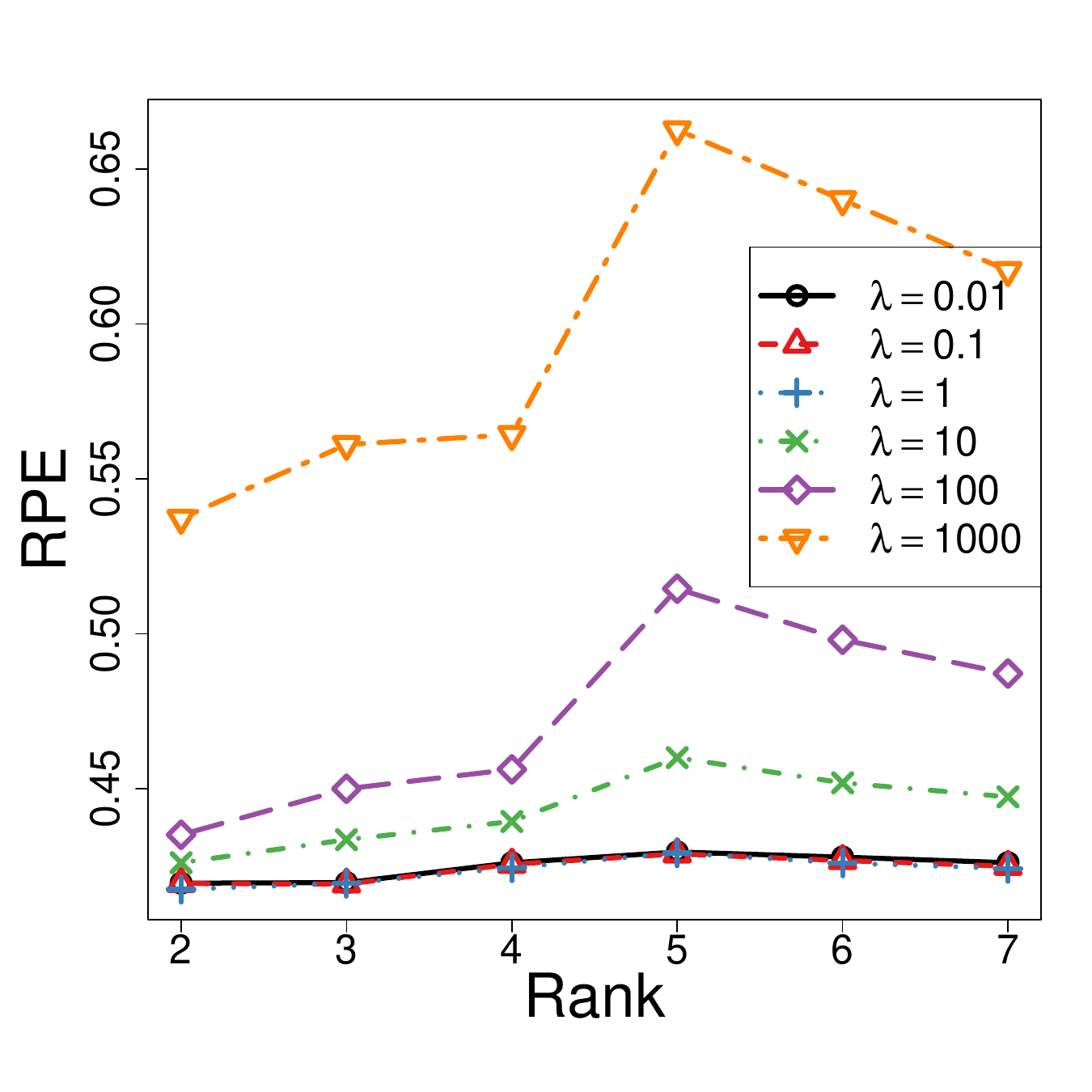}
\end{center}
\vspace{-0.5cm}
\caption{ {ROTOT median RPE for varying $(\lambda, R)$ at contaminated data.}}
\label{fig:Simulation_GridLambdaRank}
\end{figure}

 {We further investigate the sensitivity of ROTOT to the choice of the rank $R$. Figure~\ref{fig:Simulation_GridLambdaRank} reports the median RPE over $50$ simulation runs for different combinations of the rank $R$ and the regularization parameter $\lambda$. We consider $R \in \cbr{2,3,4,5,6,7}$ and $\lambda \in \cbr{0.01, 0.1, 1, 10, 100, 1000}$. 
   {The data are generated under casewise and cellwise contamination in both the response and predictor tensors, without missing values, with $\text{SNR} = 5$, following the first contamination scenario described in Section~\ref{sec:simu}. The contamination magnitude in $\brc{\cy_n}$ is fixed at the highest value $\gamma = 8$.}
The true rank used to generate the data is $R = 2$ and the optimal value of $\lambda$ under uncontaminated data is $0.1$.}
 {
 Figure~\ref{fig:Simulation_GridLambdaRank} shows that the median RPE remains stable across a broad range of $(\lambda,R)$ combinations. In particular, moderate overestimation of the rank does not lead to a deterioration in predictive performance when a suitable amount of regularization is applied.
 }

\begin{figure}[H]
\begin{center}
\includegraphics[width=0.35\textwidth]{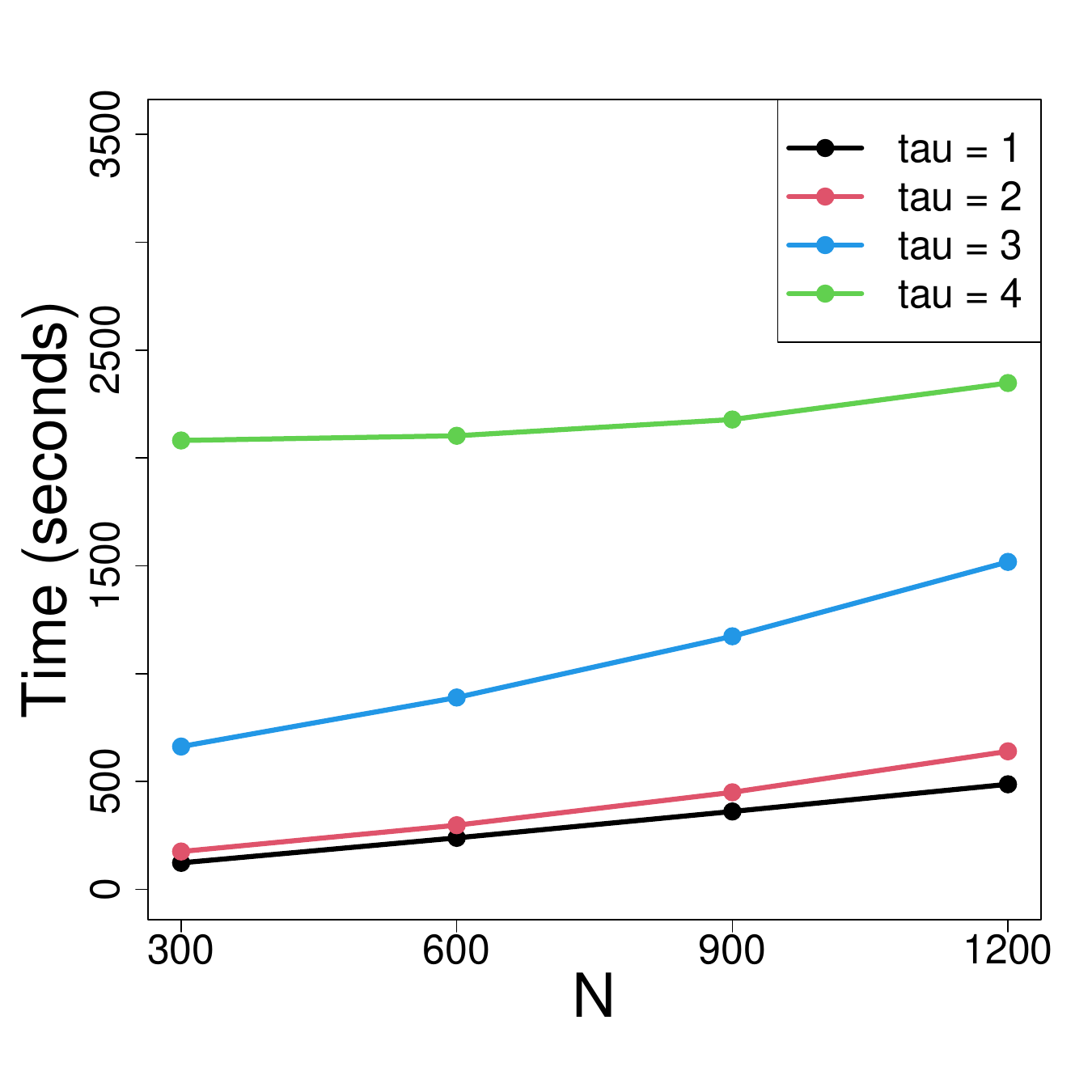}
\end{center}
\vspace{-0.5cm}
\caption{ {Computing time in seconds of ROTOT for varying sample sizes and tensor dimensions.}}
\label{fig:Simulation_comptime}
\end{figure}

 {Figure~\ref{fig:Simulation_comptime} reports the computing times of ROTOT for different sample sizes and tensor dimensions. The simulated data are generated as in Section~\ref{sec:simu}, with $\text{SNR}=5$ and no missing values. The sample size varies over $N \in \brc{300,600,900,1200}$, while each line corresponds to a scaling factor $\tau \in \brc{1,2,3,4}$ applied to the baseline tensor dimensions. Specifically, the response tensors $\cy_n$ have dimensions $(5\tau,10\tau)$ and the predictor tensors $\cx_n$ have dimensions $(15\tau,20\tau)$. Hence, for each value of $\tau$, the computing time is evaluated at all four sample sizes. It can be seen that the method scales reasonably well with both the sample size and the tensor dimensions.}

\clearpage

\section{Structured Outliers}
\label{app:stroutl}
 {The ROTOT method can simultaneously handle casewise and cellwise outliers in both the predictor and response tensors. In practice, the contamination structure may lie between these two extremes. We refer to this intermediate form of contamination as structured outliers, an example is discussed in \cite{Mou:StrData}.}
 {
A structured outlier is a form of contamination that affects only a subset of the cells within an observation, while the remaining cells are uncontaminated. Such outliers typically exhibit a recurring pattern across observations and share a similar structure. Here, we focus on slicewise contamination, a particular type of structured contamination in which an entire hyper-slice of a tensor is contaminated. 
 {Following the terminology of~\cite{BallardKolda:bookTensor}, we use the term hyper-slice to denote the subtensor obtained by fixing a single index of an arbitrary-order tensor $\cy_n$. Thus, under slicewise contamination, a hyper-slice of the response tensor $\cy_n$ may be contaminated while the rest of the observation remains clean.}
  }

\begin{figure}[H]
\centering
\begin{tabular}{cccc} 
&\textbf{Contamination in $\brc{\cy_n}$}  & \textbf{Contamination in $\brc{\cx_n}$} &{\textbf{Contamination in $\brc{\cy_n}$ and $\brc{\cx_n}$}}\\
&
\includegraphics[width=.30\textwidth]{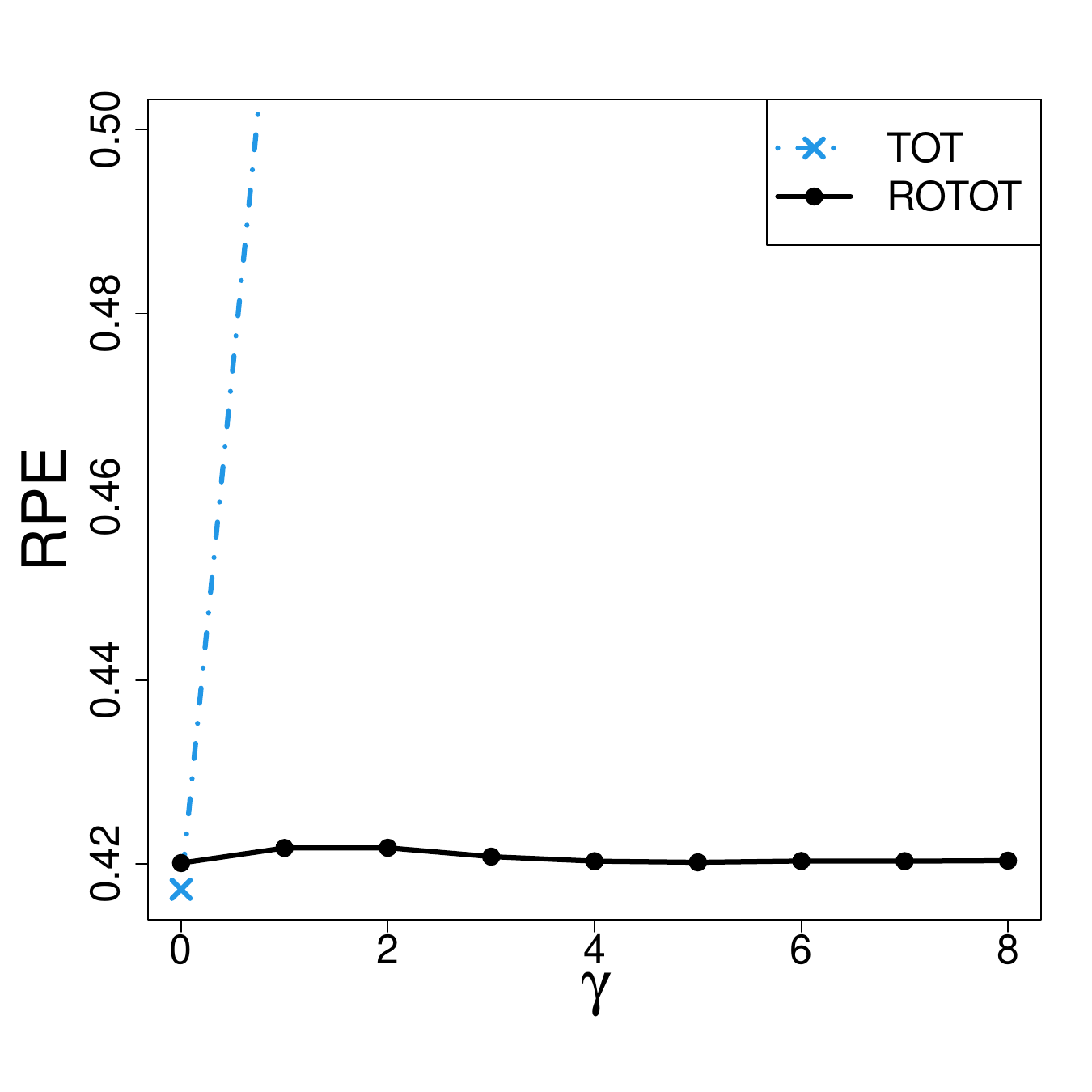} &%
\includegraphics[width=.30\textwidth]{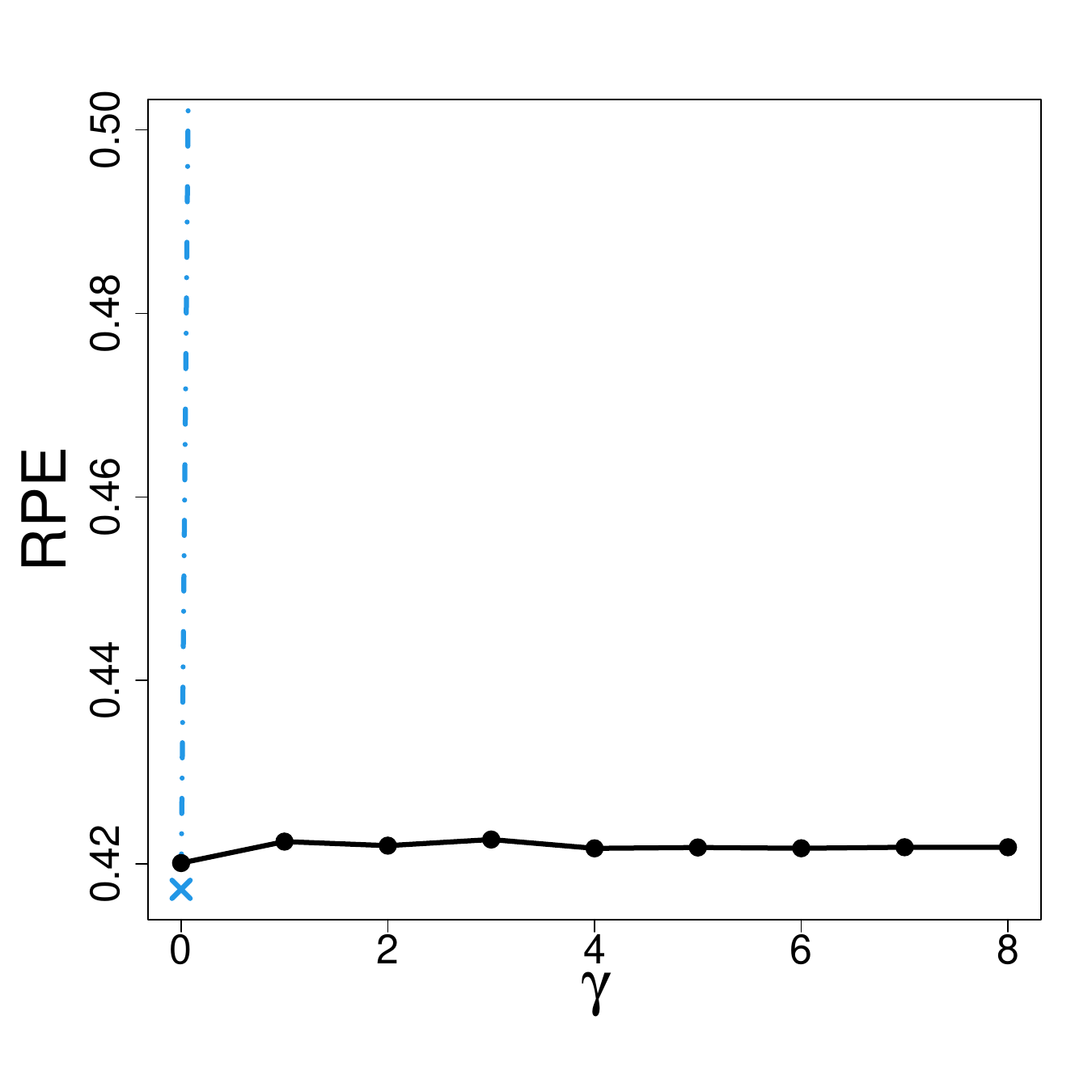}& %
\includegraphics[width=.30\textwidth]{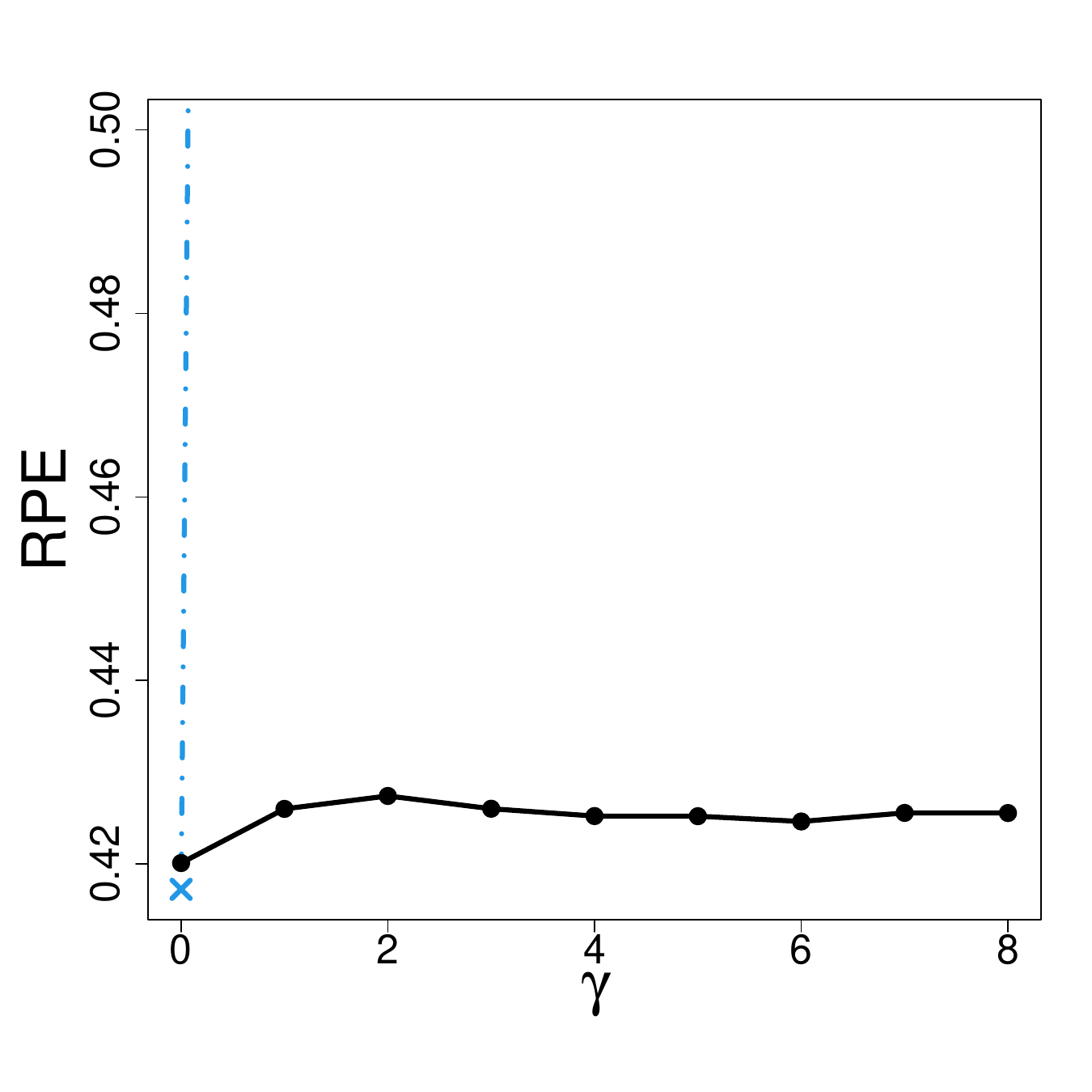}\\ 
\end{tabular}
\caption{ {Median RPE attained by TOT and ROTOT under slicewise contamination for varying contamination magnitudes in the response (left), predictor (middle), and both response and predictor (right).}}%
 \label{fig:RPEslice}%
\end{figure}

 {We assess the performance of ROTOT under slicewise contamination. The outliers are introduced into the response and predictor tensors in the same way as casewise contamination described in Section~\ref{sec:simu}. However, rather than shifting an entire observation, only a subset of its slices is contaminated. More specifically, slicewise outliers in $\cbr{\cy_n}$ and $\cbr{\cx_n}$  are generated by shifting $\varepsilon_{\text{\tiny slice}} = 10 \%$ of the slices by the tensor $\ce_n^{\text{\tiny slice}}$. The elements of $\ce_n^{\text{\tiny slice}}$ are independently drawn from $N(\gamma_{\text{\tiny slice}},2)$, where $\gamma_{\text{\tiny slice}} = \gamma \cdot c^{\text{\tiny slice}}$ and $c^{\text{\tiny slice}} = 3.5$. The parameter $\gamma$ ranges from 0 to 8, data are uncontaminated when $\gamma = 0$.
We generate $N=200$ observations, the dimensions of the coefficient tensor $\cb$ are set to $\rbr{P_1=15, P_2=20, Q_1=5, Q_2 = 10}$, and the SNR is fixed at $5$. The core tensors $\core^x_n$ of the predictors have rank $K_1=4$ and $K_2=6$, and $\cb$ is assumed to have a low rank representation with $R = 2$. We consider three different settings: varying the contamination magnitude when only the response tensors are contaminated, when only the predictor tensors are contaminated, and when both the predictor and response tensors are contaminated.}
 {Figure~\ref{fig:RPEslice} presents the median RPE of TOT, and ROTOT under slicewise contamination for the three contamination settings and varying contamination magnitudes in the predictor and response tensors. As expected, ROTOT remains robust in the presence of slicewise outliers and still clearly outperforms TOT.}

\begin{figure}[H]
\begin{center}
\includegraphics[width=0.5 \textwidth]{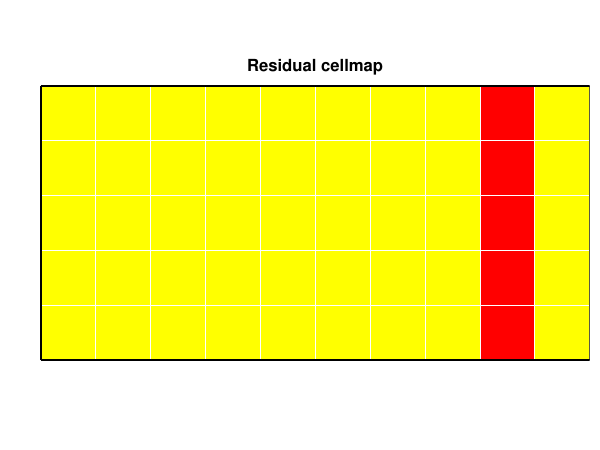}
\end{center}
\vspace{-1cm}
\caption{ {Residual cellmap corresponding to a randomly selected observation from the simulated data under slicewise contamination.}}
\label{fig:Simulation_slicewiseResmap}
\end{figure}

 {Structured outliers can also be visualized using the residual cellmap introduced in Section~\ref{sec:outdetect}. Figure~\ref{fig:Simulation_slicewiseResmap} displays the residual cellmap corresponding to a randomly selected observation from the simulated data under slicewise contamination. ROTOT correctly flags the contaminated cells within each structured outlier, allowing the corresponding outlying slice to be identified. 
 }

\clearpage
\bibliographystyle{chicago}
\setlength{\bibsep}{5pt plus 0.2ex}
{\small
\spacingset{1}
\bibliography{mybibliography}
}